\documentstyle[floats,prd,aps,epsf]{revtex}

\draft
\begin{document}

%
\makeatletter
\@ifundefined{lesssim}{\def\lesssim{\mathrel{\mathpalette\vereq<}}}{}
\@ifundefined{gtrsim}{\def\gtrsim{\mathrel{\mathpalette\vereq>}}}{}
\def\vereq#1#2{\lower3pt\vbox{\baselineskip1.5pt \lineskip1.5pt
\ialign{$\m@th#1\hfill##\hfil$\crcr#2\crcr\sim\crcr}}}
\makeatother
\newcommand{\beq}{\begin{equation}}
\newcommand{\eeq}{\end{equation}}
\newcommand{\beqn}{\begin{eqnarray}}
\newcommand{\eeqn}{\end{eqnarray}}
\newcommand{\pa}{\partial}
\newcommand{\vp}{\varphi}
\newcommand{\varep}{\varepsilon}
\def\zero{\hbox{$_{(0)}$}}
\def\bL{\hbox{$\,{\cal L}\!\!\!$--}}
\def\bI{\hbox{$\,I\!\!\!$--}}
\def\bm#1{{\hbox{\boldmath $#1$}}}
\def\riso{\hbox{$r_{\rm ISCO}$}}
\def\jisom{\hbox{$j_{\rm ISCO:max}$}}

\begin{center}
{\large\bf{Axisymmetric collapse simulations of rotating
    massive stellar cores in full general relativity: Numerical study for
    prompt black hole formation}}
 ~\\
~\\
 Yu-ichirou Sekiguchi and Masaru Shibata \\
{\em Graduate School of Arts and 
Sciences, University of Tokyo, Tokyo, 153-8902, Japan}\\
\end{center}
\begin{abstract}
We perform axisymmetric simulations for gravitational collapse of a
massive iron core to a black hole in full general relativity.
The iron cores are modeled by $\Gamma = 4/3$ equilibrium polytrope for
simplicity.
The hydrodynamic equations are solved using a high-resolution shock
capturing scheme with a parametric equation of state. The Cartoon
method is adopted for solving the Einstein equations.  Simulations are
performed for a wide variety of initial conditions changing the mass
($\approx 2.0$--$3.0M_{\odot}$), the angular momentum, the
rotational velocity profile of the core, and the parameters of the
equations of state which are chosen so that the maximum mass of the
cold spherical polytrope is $\approx 1.6M_{\odot}$. Then, the
criterion for the prompt black hole formation is clarified in terms of
the mass and the angular momentum for several rotational velocity 
profile of the core and equations of state. It is found that (i) with 
the increase of the thermal energy generated by shocks, the threshold 
mass for the prompt black hole formation is increased by 20--40$\%$, (ii) the 
rotational centrifugal force increases the threshold mass by $\alt 
25\%$, (iii) with the increase of the degree of differential rotation, 
the threshold mass is also increased, and (iv) the amplification
factors shown 
in the results (i)--(iii) depend sensitively on the equation of state. 
We also find that the collapse dynamics and the structure of the shock 
formed at the bounce depend strongly on the stiffness of the 
adopted equation of state. In particular, as a new feature, a strong 
bipolar explosion is observed for the collapse of
rapidly rotating iron cores with 
an equation of state which is stiff in subnuclear density and soft in 
supranuclear density. Gravitational waves are computed in terms of a 
quadrupole formula. It is also found that the waveform depends 
sensitively on the equations of state.
\end{abstract}

\section{Introduction}

The black hole is one of the most striking and intriguing objects
predicted in general relativity. A wide variety of recent observations
have shown that black holes actually exist in the universe
\cite{Rees}. Among several types of the black holes,
the existence of the stellar-mass black holes has been tightly
confirmed. So far, about 20 stellar-mass black holes for which
the mass is determined within a fairly small error have been
observed in binary systems of our Galaxy and the Large Magelanic
Clouds \cite{McClintock}. Such black holes are believed to be formed
from stellar core collapse of massive stars. This fact stimulates
the theoretical study for clarifying the
physics of gravitational collapse and formation mechanism of the
stellar-mass black holes. The formation of 
the black hole through the gravitational collapse is a highly nonlinear and
dynamical phenomena. Therefore,
numerical simulation in full general relativity is the unique 
approach to this problem.

Realistic simulations for the formation of stellar-mass black holes
in massive rotating stellar core collapse is also increasingly
important due to its possible association with gamma-ray bursts \cite{GRB}. 
Recent observations have indicated that at least long-duration
gamma-ray bursts are of cosmological origin \cite{GRB2} and
associated with rotating stellar core collapse \cite{GRB3,Wang1}, 
probably to a black hole surrounded by a massive disk 
as suggested by \cite{Woosley,Paczynski,MW}. 
Recent numerical analyses have also shown that if a progenitor of the
collapse is massive and the angular momentum is large enough, a black
hole surrounded by a massive disk will be indeed formed 
\cite{ShibaSA,Shapiro,SS3}. However detailed simulations have not yet
been done. 

Stellar-mass black holes can be formed through the gravitational
collapse of a degenerate iron core in excess of the Chandrasekhar
mass \cite{ST,WHW}. It is well known that stars whose initial mass are
larger than $\approx 10M_{\odot}$ evolve to form a core mainly
composed of iron group elements \cite{WHW}. 
Since the iron is the most stable nuclei and it does not generate
energy by nuclear burning, the iron core contracts
gradually. Accordingly, the central temperature, $T_{c}$, 
and central density, $\rho_{c}$, rise to be $T_{c} \agt 10^{10}$ K and
$\rho_{c} > 10^{9}$ g$/$cm$^{3}$, resulting in 
the photo-dissociation of iron
to lighter elements and subsequent electron capture that reduce the
entropy of electrons. 
As a result, the adiabatic index $\Gamma_{s}$ decreases below $4/3$ and
the iron core is destabilized to collapse. 
If the mass of the iron core is much larger than the maximum neutron star mass,
a black hole will be formed soon after the collapse. 
However, the threshold mass of the iron core
for prompt formation of a black hole has not been clarified yet. 
Note that due to the contribution of the thermal pressure and
rotation, the iron core can be much larger than the maximum allowed
neutron star mass of $\sim 2M_{\odot}$ for a sufficiently massive and
rotating star.

Since massive stars in nature are rapidly rotating in general
\cite{Fukuda}, it is necessary to explore the gravitational
collapse of a rotating star in full general relativity for
clarifying the black hole formation. 
Since Nakamura \cite{Nak} and his collaborators \cite{NOK} 
first presented a series of preliminary numerical simulations, 
a number of simulations for rotating stellar collapse to a
black hole have been performed in full general relativity
\cite{SP,ShiSha,SBS,ShiApJ03,Baiotti,Duez}. 
However, the initial conditions and equations of state in the previous
works are not very realistic for modeling the stellar core
collapse. Thus, the simulations with a realistic setting
remain to be an unsolved issue in general relativity. 

Realistic simulations of rotating stellar core collapse have been
performed intensively in the framework of Newtonian gravity
\cite{Muller,FE,Monch,BM,Yamada,Zweg,Rammp,FH,Kotake,Ott}.
Most of these studies mainly aim at clarifying the effect of the rotation
on the dynamics of {\it neutron star formation} and gravitational
waveforms from it. 
In particular, a comprehensive parameter study sweeping through
various values of the stiffness of a parametric equation of state
as well as rotational parameters was performed in \cite{Zweg}. It was
shown that the dynamics of the collapse and resulting gravitational
waveforms depend strongly not only on rotation but also on the
stiffness of equations of state. 
Dimmelmeier et al. \cite{HD} extended the aforementioned study to
general relativistic case using a conformal flatness formalism \cite{WM}.
Fully general relativistic numerical studies of neutron star formation
have been recently performed \cite{SS2,Siebel}. 
As shown in \cite{HD,SS2}, the general relativistic effects
significantly modify the dynamics of the collapse even in the formation 
of neutron stars. 

Taking into account the present status described above, in this paper,
we study a criterion for prompt black hole formation in the iron core
collapse performing fully general relativistic simulations. 
The iron cores are modeled by $\Gamma = 4/3$ polytropes in
equilibrium for simplicity. 
The major purpose of this paper is to clarify the threshold mass of
the iron core for the prompt black 
hole formation and its dependence on the angular momentum, the
rotational velocity profiles of the iron core, and the equations of
state.
To clarify the dependence on the equations of state in a clear manner, 
we adopt a parametric equation of state following previous papers
\cite{SS2,SS4}. 

The simulations are performed assuming that the collapse proceeds in an
axisymmetric manner. This assumption is reasonable 
as far as the progenitor of the collapse is not very rapidly and 
highly differentially rotating (e.g., \cite{SS4}). In this paper,
we do not adopt such progenitor that are likely to be dynamical
unstable against nonaxisymmetric deformation (cf. Sec. \ref{Sec-rot}
for discussion). 
Although for several models, the rotational kinetic energy
is so large that the outcome formed in the 
collapse may be secularly unstable against nonaxisymmetric
deformation, the secular time scale is much longer than the dynamical
time scale of the core collapse. Hence, the collapse will proceed
in an approximately axisymmetric manner in the time scale of interest.
On the other hand, rapidly and differentially rotating stellar core collapse
has to be studied in the three-dimensional simulation. Such
simulation was recently performed and the detailed 
results are shown in a companion paper \cite{SS4}. 

This paper is organized as follows. 
In Sec. \ref{Numerical implementation}, we briefly review our
formulation for general relativistic hydrodynamic simulations,
equations of state, and a quadrupole formula adopted in the present
paper. In Sec. \ref{secInitial}, we describe the initial conditions. 
A detail of computational setting
is also described. Sec. \ref{Numerical result} presents the numerical
results, emphasizing the threshold mass for the
prompt black hole formation and 
its dependence on the angular momentum and the adopted equations of state.
Gravitational waveforms emitted in the neutron star
formation are also shown. Sec. \ref{Summary} is devoted to a summary. 
Throughout the paper, we adopt the geometrical units $G=c=1$ 
where $G$ and $c$ are the gravitational constant and speed of light,
respectively. 
The Latin indices $i, j, k, \cdots$ denote the spatial components
of $x, y$, and $z$, and the Greek indices $\mu \cdots$ denote 
the spacetime components. 

\section{Numerical implementation} \label{Numerical implementation}

\subsection{Brief summary of formulation and numerical method}

We perform fully general relativistic simulations for rotating 
stellar core collapse in axial symmetry using the same formulation and 
numerical method as those presented in~\cite{S2003}, to which the 
reader may refer for details of basic equations and successful test
simulations. 

In the 3+1 formulation, the metric can be written in the form 
\beq
ds^{2} = (-\alpha^{2} + \beta_{k}\beta^{k})dt^{2} + 2\beta_{k}dtdx^{k} 
+ \gamma_{ij}dx^{i}dx^{j},
\eeq
where $\alpha$, $\beta^k$, and $\gamma_{ij}$ are the lapse
function, the shift vector and metric in 3D spatial
hypersurface, respectively. The extrinsic curvature is defined by 
\beq
(\partial _{t} - \bL _{\beta})\gamma_{ij} = -2 \alpha K_{ij},
\eeq
where $\bL _{\beta}$ is the Lie derivative with respect to $\beta^{k}$.

As in the series of our papers, we evolve 
$\phi \equiv \log(\det \gamma_{ij})/12$, 
$\tilde \gamma_{ij} \equiv e^{-4 \phi }\gamma_{ij}$,  
$\tilde A_{ij} \equiv e^{-4\phi}(K_{ij}-\gamma_{ij} K_k^{~k})$,
and trace of the extrinsic curvature $K_k^{~k}$ 
together with three auxiliary functions
$F_i\equiv \delta^{jk}\pa_{j} \tilde \gamma_{ik}$ with an
unconstrained free evolution code
as in \cite{AMS,SBS,bina1,bina2,SN,Shiba2000,S2003}. 
The Einstein equations are solved in Cartesian coordinates. 
To impose axisymmetric boundary conditions, the Cartoon method \cite{alcu}
is used with the grid size $N \times 3 \times N$ in $(x, y, z)$ which
covers  a computational domain as 
$0 \leq x \leq L$, $0 \leq z \leq L$, 
and $-\Delta \leq y \leq \Delta$. 
Here, $N$ and $L$ are constants and $\Delta = L/N$. 

The fundamental variables for the hydrodynamics are
$\rho$ : rest mass density, $\varep$ : specific internal energy,
$P$ : pressure, $u^{\mu}$ : four velocity, and
\beq
v^i ={dx^i \over dt}={u^i \over u^t}.
\eeq
As the variables to be evolved in the numerical
simulations, we define a weighted density, $\rho_{\ast}$, 
a weighted four-velocity $\hat{u}_{\mu}$, and a specific energy
density as
\beqn
&&\rho_* \equiv \rho w e^{6\phi}, \nonumber \\
&&\hat u_i = h u_i, \nonumber \\
&&\hat{e} \equiv h w - \frac{P}{\rho w},
\eeqn
where $w \equiv \alpha
u^{t}$, and $h \equiv 1 + \varepsilon + P/\rho$. 
From these variables, the total baryon rest mass and angular momentum 
of the system, which are conserved quantities in an axisymmetric
spacetime, can be defined as 
\beqn 
M_*&=&\int d^3 x \rho_*, \\
J  &=&\int d^3 x \rho_*\hat u_{\varphi} . 
\eeqn
The general relativistic hydrodynamic equations are solved using 
a so-called high-resolution shock-capturing scheme \cite{Font} 
on the $y=0$ plane with the cylindrical coordinates $(x, z)$
(in Cartesian coordinates with $y=0$). Details about our numerical
scheme are described in \cite{S2003} 

We neglect effects of viscosity and magnetic fields. The time scale of
dissipation and angular momentum transport due to these effects are
much longer than the time scale of collapse $\sim 100$ ms, unless the
magnitude of viscosity or magnetic fields is extremely large
\cite{BSS}. Thus neglecting them is an appropriate assumption.

As the slicing condition we impose 
an ``approximate'' maximal slicing condition in which 
$K_k^{~k} \approx 0$ is required\cite{AMS}. 
As the spatial gauge, 
we adopt a dynamical gauge condition \cite{DynGauge} in which
the equation for the shift vector is written as 
\beq
\pa_t \beta^k = \tilde \gamma^{kl} (F_l +\Delta t \pa_t F_l).
\label{dyn}
\eeq
Here, $\Delta t$ denotes the time step in numerical
computation\cite{ShiApJ03}.  Note that in this gauge condition,
$\beta^{i}$ obeys a hyperbolic-type equation for a sufficiently small
value of $\Delta t$ because the right-hand side of the evolution
equation for $F_i$ contains a vector Laplacian term \cite{SN}. 
It has already been illustrated that stable
simulations for rotating stellar collapse and merger of binary neutron
stars are feasible in this gauge \cite{ShiApJ03,bina2}.

An outgoing-wave boundary condition 
is imposed for $h_{ij}(\equiv \tilde \gamma_{ij}-\delta_{ij})$, 
$\tilde A_{ij}$, and $F_{i}$
at the outer boundaries of the computational domain. 
The condition adopted is the same as that described in \cite{SN}. 
On the other hand, for $\phi$ and $K_k^{~k}$,
other types of outer boundary conditions are imposed as
$r\, \phi = {\rm const}$ and $K_k^{~k}=0$, respectively. 

Existence of a black hole is determined using an apparent horizon
finder developed in \cite{ShibaAH}. We compute the apparent horizon
mass $M_{\rm AH}$ which is defined as \cite{YP}
\beq
M_{\rm AH} = \sqrt{\frac{A}{16\pi}} ,
\eeq
where $A$ denotes area of an apparent horizon.

During the numerical simulations, conservation of 
the Arnowitt-Deser-Misner (ADM) mass, $M_{\rm ADM}$,
and the angular momentum are monitored as code
checks. The ADM mass is defined by 
\beq
M_{\rm ADM} = \int d^{3}\left[ 
\psi^{5}(\rho h w^{2} - P) + \frac{\psi^{5}}{16\pi}\left(
\tilde{A}_{ij}\tilde{A}^{ij} - \frac{2}{3}(K_k^{~k})^{2} 
\right) - \frac{1}{16\pi}\tilde{R}\psi
\right],
\eeq
where $\psi \equiv e^{\phi}$.

For the analysis of numerical results, 
we define a rest-mass distribution $m_{\ast}(j)$ \cite{ShiSha}, 
which is the integrated baryon rest mass of fluid elements with the
specific angular momentum less than a given value of $j=\hat u_{\varphi}$:
\beq
m_{\ast}(j) \equiv 2\pi \int_{j'< j} \rho_{\ast} r^{2}dr d(\cos \theta).
\eeq
Similarly, a specific angular momentum distribution is defined according to
\beq
J(j) \equiv 2\pi \int _{j'< j} \rho_{\ast} j' r^{2}dr d(\cos \theta).
\eeq
These distribution functions are preserved in axisymmetric spacetimes
of ideal fluid. 
Gauge independence and preservation of these distribution functions
in axial symmetry can be proven by the hydrodynamical equations
\beqn
&&\frac{\partial \rho_{\ast}}{\partial t} 
+ \frac{\partial(\rho_{\ast}v^{I})}{\partial x^{I}} = 0, \\
&& \frac{\partial (\rho_{\ast}j)}{\partial t} 
+ \frac{\partial (\rho_{\ast}j v^{I})}{\partial x^{I}} = 0,
\eeqn
where the index $I$ denotes the component of $\varpi$ and $z$.

From these distribution functions, we define a spin parameter
distribution as
\beq \label{spin_distri}
q_{\ast}(j) \equiv \frac{J(j)}{m_{\ast}(j)^{2}}.  
\eeq
This may be approximately regarded as a spin 
parameter of the inner region of the core composed of
fluid elements with the specific angular momentum less than $j$. 

\subsection{Equations of state}\label{secEOS}
\begin{table}[hbt]
  \vspace{-4mm}
  \begin{center}
    \begin{tabular}{cccccc}
      Model & $\Gamma_{1}$ & $\Gamma_{2}$ & $\rho_{\rm nuc}$(g/cm$^{3}$) &
      $\Gamma_{\rm th}$ & $M_{\rm max}$ 
      \\ \hline
      a & 1.32 & 2.25 & $2.0 \times 10^{14}$ & 1.32 & 1.623 \\
      b & 1.30 & 2.5  & $2.0 \times 10^{14}$ & 1.30 & 1.600 \\
      c & 1.30 & 2.22 & $1.0 \times 10^{14}$ & 1.30 & 1.599 \\
      d & 1.28 & 2.75 & $2.0 \times 10^{14}$ & 1.28 & 1.597
    \end{tabular}
  \end{center}
  \caption{
Adopted sets of $(\Gamma_{1}, \Gamma_{2}, \rho_{\rm nuc},
    \Gamma_{\rm th})$. The values of $\Gamma_{1}$, $\Gamma_{2}$, 
    and $\rho_{\rm nuc}$ are chosen so that the maximum ADM mass of a
    cold spherical polytrope for each set becomes $\approx
    1.6M_{\odot}$.}\label{Table1}
\end{table}
\begin{figure}[htb]
  \vspace{-4mm}
  \begin{center}
    \epsfxsize=3.in
    \leavevmode
    \epsffile{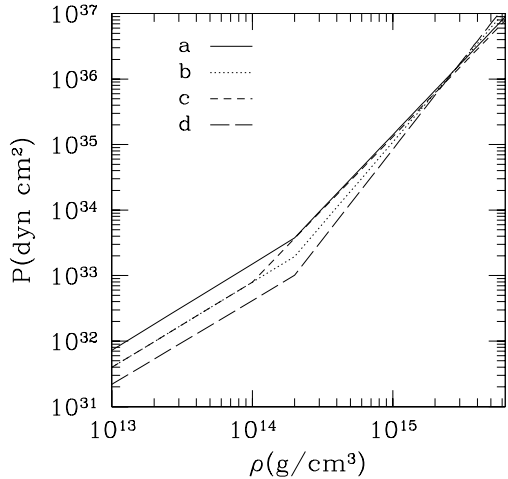} 
  \end{center}
  \vspace{-8mm}
  \caption{The pressure, $P$, as a function of the density, $\rho$,
  for cold equations of state with the parameters listed in Table
  \ref{Table1}.}\label{figure1} 
\end{figure}

During dynamical evolution, a parametric equation of state is adopted
following M\"uller and his collaborators \cite{Janka,Zweg,HD}. In this
equation of state, one assumes that the pressure consists of the sum
of polytropic and thermal parts as
\beq
P=P_{\rm P}+P_{\rm th}. \label{EOSII}
\eeq
$P_{\rm P}$ denotes the cold (zero temperature) nuclear
equation of state and is given by 
$P_{\rm P}=K_{\rm P}(\rho) \rho^{\Gamma(\rho)}$
where $K_{\rm P}$ and $\Gamma$ are functions of $\rho$.
In this paper, we follow \cite{HD} for the choice of 
$K_{\rm P}(\rho)$ and $\Gamma(\rho)$: 
For the density smaller than the nuclear density,
$\rho_{\rm nuc}$, we set $\Gamma=\Gamma_1 < 4/3$, and 
for $\rho \geq \rho_{\rm nuc}$, $\Gamma=\Gamma_2 > 2$.
Namely,
\beqn
P_{\rm P}=
\left\{
\begin{array}{ll}
  K_1 \rho^{\Gamma_1}, & \rho \leq \rho_{\rm nuc}, \\
  K_2 \rho^{\Gamma_2}, & \rho \geq \rho_{\rm nuc}, \\
\end{array}
\right.\label{P12EOS}
\eeqn
where $K_1$ and $K_2$ are constants. 
Since $P_{\rm P}$ should be continuous at $\rho = \rho_{\rm nuc}$, the
relation, $K_2=K_1\rho_{\rm nuc}^{\Gamma_1-\Gamma_2}$, is required. 
Following \cite{Zweg,HD}, the value of $K_1$ is fixed to 
be $5\times 10^{14}$ cgs. With this choice, 
the polytropic part of the equation of state for 
$\rho < \rho_{\rm nuc}$, in which the degenerate pressure of electrons is 
dominant, is approximated well. Since the specific internal energy 
should be also continuous at $\rho=\rho_{\rm nuc}$, the polytropic part of
the specific internal energy, $\varepsilon_{\rm P}$, is defined as 
\beqn
\varepsilon_{\rm P}=
\left\{
\begin{array}{ll}
  \displaystyle
      {K_1 \over \Gamma_1-1} \rho^{\Gamma_1-1}, & \rho \leq \rho_{\rm nuc}, \\
      \displaystyle 
	  {K_2 \over \Gamma_2-1} \rho^{\Gamma_2-1}
	  +{(\Gamma_2-\Gamma_1)K_1 \rho_{\rm nuc}^{\Gamma_1-1}
	    \over (\Gamma_1-1)(\Gamma_2-1)},  & \rho \geq \rho_{\rm nuc}. \\
\end{array}
\right.
\eeqn
With this setting, a realistic equation of state for high-density,
cold nuclear matter is mimicked. 

For realistic simulations of stellar core collapse, it would be better
to adopt realistic equations of state \cite{Latti,Shen}. However, in
the realistic equations of state, many micro-physical processes are
simultaneously taken into account together. In such cases, it is not
easy to extract an important element responsible for an output of
numerical simulations.  With the parametric equations of state, on the
other hand, one can systematically investigate dependence of the
dynamics of stellar core collapse on the equations of state by changing
their own parameters. Therefore, as a first step to the more realistic
simulations of rotating stellar core collapse, we adopt the parametric
equations of state. We now plan to perform simulations with realistic
equations of state. Some discussion comparing the
parametric equations of state and a realistic equation of state
is presented in appendix \ref{compareEOS}.

There is four parameters in the
parametric equations of state, namely 
$(\Gamma_{1}, \Gamma_{2},\rho_{\rm nuc}, \Gamma_{\rm th})$. 
In this paper, we choose sets of $(\Gamma_{1},\Gamma_{2}, \rho_{\rm nuc})$ 
so that the maximum allowed ADM mass of the cold spherical polytrope 
becomes an approximately identical value as
$M_{\rm ADM,max} \approx 1.6M_{\odot}$. 
Note that this value is larger than the mass 
of neutron stars in binary neutron stars accurately determined, and 
thus, a reasonable choice \cite{Stairs}. 
Following \cite{Zweg,HD}, we typically set 
$\rho_{\rm nuc} = 2.0 \times 10^{14}$g/cm$^{3}$ 
($\rho_{14} \equiv \rho_{\rm nuc}/(10^{14}~{\rm g/cm^{3}}) = 2.0$). 
For $\Gamma_{1}$, we choose the three values of 1.32, 1.30, and 1.28. 
Requiring that the maximum allowed ADM mass should be 
$\approx 1.6M_{\odot}$, the values of $\Gamma_{2}$ are determined to
be 2.25, 2.50, and 2.75, respectively.
To investigate the dependence of the output physics
on the value of $\rho_{\rm nuc}$, we also pick up a case with 
$\rho_{\rm nuc} = 1.0 \times 10^{14}~{\rm g/cm^{3}}$ ($\rho_{14} = 1.0$) 
and set $(\Gamma_{1}, \Gamma_{2}) = $ $(1.30,2.22)$
for comparison.
 
To summarize, the parameter sets adopted in this paper are
$(\Gamma_{1},\Gamma_{2},\rho_{14})$ 
$ = (1.32, 2.25, 2.0)$, $(1.30, 2.50, 2.0)$,
$(1.30, 2.22, 1.0)$, and $(1.28, 2.75, 2.0)$ which are referred to
as equations of state 'a', 'b', 'c', and 'd', respectively 
(cf. Table \ref{Table1}). Although the maximum allowed mass of the cold
spherical neutron star in equilibrium is approximately identical for
all of the equations of state, the difference in the values of 
$(\Gamma_{1},\Gamma_{2}, \rho_{\rm nuc})$ yields a significant 
variation in the collapse dynamics and in the criterion for prompt
black hole formation. 

Figure \ref{figure1} shows relations between the pressure and the
density for each set of $(\Gamma_{1}, \Gamma_{2},\rho_{\rm nuc})$. 
It is found that for the smaller value of $\Gamma_{1}$, the depletion
fraction of the pressure for $\rho \le \rho_{\rm nuc}$ is increased by a 
large factor. It is also worthy to note that the equation of state 'c' is
stiffer than 'b' in the density range between $10^{14}$ and 
$\approx 2 \times 10^{15}$g/cm$^{3}$. These result in significant
difference in the collapse dynamics and the threshold mass for 
the prompt black hole formation. 

$P_{\rm th}$ is related to the thermal energy
density, $\varepsilon_{\rm th}\equiv \varepsilon-\varepsilon_{\rm P}$, as 
\beq
P_{\rm th}=(\Gamma_{\rm th}-1)\rho \varepsilon_{\rm th}. 
\eeq
In this paper we set $\Gamma_{\rm th} = \Gamma_{1}$ for simplicity.
As shown in \cite{SS2}, the collapse
dynamics depends rather weakly on $\Gamma_{\rm th}$ as far as it is 
in the range $4/3 \alt \Gamma_{\rm th} \alt 5/3$, besides the fact that 
the oscillation amplitude of a formed protoneutron star depends
slightly on it. 

To prepare rotating stars in equilibrium as initial conditions, we use the
polytropic equations of state with $\Gamma=4/3$ 
\beq
P= K_{0} \rho^{4/3} = (K_{\rm deg} + K_{\rm T}) \rho^{4/3},\label{EOS43}
\eeq
where following \cite{HD}, 
$K_{\rm deg}$ is set to be $5 \times 10^{14}~{\rm cm^3/s^2/gr^{1/3}}$, 
with which
a soft equation of state governed by the electron degenerate pressure
is approximated well \cite{ST}. Here, $K_{\rm deg}$ and $K_1$ are related
by $K_1=K_{\rm deg}\rho_0^{4/3-\Gamma_1}$ 
where we set $\rho_0=1~{\rm g/cm^3}$. 
The extra pressure of $P_{\rm T} \equiv K_{\rm  T}\rho^{4/3}$ denotes 
the pressure generated by the ideal gas pressure and the radiation
pressure that are nonzero only for the finite temperature. 
The values of $K_{0}$ and $K_{\rm T}$ adopted in the present
paper are described in Sec. \ref{secInitial}.
To induce the collapse, we slightly decrease the value of
the adiabatic index from $\Gamma=4/3$ to $\Gamma_1 < 4/3$ at $t=0$. 
This implies that at $t=0$, $P_{\rm P} = K_{\rm deg}\rho^{\Gamma_{1}}$
and $P_{\rm th} = K_{\rm T}\rho^{\Gamma_{1}}$, respectively.

\subsection{Quadrupole formula}\label{quad}

In the present work, gravitational waveforms are computed 
using a quadrupole formula described in \cite{SS,SS2}.
In quadrupole formulas, only
the $+$-mode of gravitational waves with 
$l=2$ is nonzero in axisymmetric spacetime and it is written as 
\beq
h_+^{\rm quad} = {\ddot I_{zz}(t_{\rm ret}) - \ddot I_{xx}(t_{\rm ret})
\over r}\sin^2\theta, \label{quadr}
\eeq
where $I_{ij}$ denotes a quadrupole moment, $\ddot I_{ij}$ 
its second time derivative, and $t_{\rm ret}$ a retarded time. 

In fully general relativistic and dynamical spacetime, 
there is no unique definition for the quadrupole moment and 
nor is for $\ddot I_{ij}$. Following a previous paper \cite{SS}, 
we choose the simplest definition as 
\beq
I_{ij} = \int \rho_* x^i x^j d^3x. 
\eeq
Then, using the continuity equation of the form 
\beq
\pa_t \rho_* + \pa_i (\rho_* v^i)=0, 
\eeq
the first time derivative can be written as 
\beq
\dot I_{ij} = \int \rho_* (v^i x^j +x^i v^j)d^3x.
\eeq
To compute $\ddot I_{ij}$, the finite differencing of the numerical result 
for $\dot I_{ij}$ is carried out. 

In the following, we present
\beq
A_{2}(t_{\rm ret}) =
\ddot{I}_{zz}(t_{\rm ret}) - \ddot{I}_{xx}(t_{\rm ret}),
\eeq
in the quadrupole formula. This provides the amplitude of $l=2$ mode 
measured by an observer located in the most optimistic direction
(i.e., in the equatorial plane).

The energy power spectrum of gravitational waves 
is given in \cite{Thorne} as 
\beq
\frac{dE}{df} = \frac{\pi}{2}(4\pi r^{2})f^{2} 
\langle | \tilde{h}_{+}(f) |^{2} \rangle  \ \ (f > 0), 
\eeq
where $\tilde{h}_{+}(f)$ denotes the Fourier transform
\beq
\tilde{h}_{+}(f) = \int h_{+}^{\rm quad}(t)e^{2\pi i f t} dt,
\eeq
and the bracket denotes the angle-averaged value. 
This can be expressed in the present case as (e.g. \cite{Zhuge})
\beq
\langle | \tilde{h}_{+}(f) |^{2} \rangle 
= \frac{8}{15} \frac{1}{r^{2}}\left| \tilde{A}_{2}(f) \right|^{2},
\eeq
where 
\beq
\tilde{A}_{2}(f) = \int A_{2}(t) e^{2\pi i f t} dt.
\eeq
Thus, the effective amplitude of gravitational waves observed
in the most optimistic direction is denoted by 
\beqn
h_{\rm eff} \equiv {|f A_2(f)| \over r}=5.0 \times 10^{-20}
\biggl({dE/df \over 10^{46}~{\rm erg/Hz}}\biggr)^{1/2}
\biggl({10~{\rm kpc} \over r}\biggr). 
\eeqn

As indicated in \cite{SS}, it is possible to compute gravitational
waves from oscillating and rapidly rotating neutron stars of high
values of compactness fairly accurately with the present choice of
$I_{ij}$, besides possible systematic errors for the amplitude of
order $M/R$.  For the stellar core collapse in which the outcomes are
protoneutron stars of $M/R \sim 0.1$--0.2, it is likely that the wave
amplitude is computed within an error of $\sim 10$--$20\%$. The
phase of gravitational waves will be computed very accurately as indicated
in \cite{SS}. For the stellar core collapse to a black hole, on the
other hand, the quadrupole formula will be no longer valid because the
value of $M/R$ is high and, moreover, quasi-normal mode ringing of the
formed black hole is not taken into account.

\section{Initial conditions and computational setting}\label{secInitial}
%
\begin{table}[p]
 \begin{center}
  \begin{tabular}{c|cccccccc}
    Model & $K_{0}$ (cgs) & $R_{p}/R_{e}$ 
    & $M_{\rm ADM}$($M_{\odot}$) & $R_{\rm ce}$(km) & $T_{\rm rot}/W$ 
    & $q$ & $\Omega_{a}$(1/s) 
    \\ \hline
    A0 & $6.0\times 10^{14}$ & 1 & 1.920 & 1717
    & 0.0 & 0.0 & 0.0 \\
    A1 & $6.0\times 10^{14}$ & 119/120 & 1.921 & 1718
    & $3.98 \times 10^{-4}$  & 0.234 & 0.904 \\
    A2 & $6.0\times 10^{14}$ & 110/120 & 1.940 & 1836
    & $3.68 \times 10^{-3}$  & 0.716 & 2.71  \\
    A3 & $6.0\times 10^{14}$ & 100/120 & 1.957 & 1999
    & $6.51 \times 10^{-3}$  & 0.959 & 3.56  \\
    A4 & $6.0\times 10^{14}$ & 90/120 & 1.967 & 2207
    & $8.28 \times 10^{-3}$  & 1.086 & 3.98  \\
    A5 & $6.0\times 10^{14}$ & 80/120 & 1.971 & 2478
    & $8.88 \times 10^{-3}$  & 1.128 & 4.11
    \\ \hline
    B0 & $6.5\times 10^{14}$ & 1 & 2.163 & 1793 
    & 0.0 & 0.0 & 0.0 \\ 
    B1 & $6.5\times 10^{14}$ & 119/120 & 2.163 & 1793 
    & $3.98 \times 10^{-4}$  & 0.224 & 1.01 \\
    B2 & $6.5\times 10^{14}$ & 110/120 & 2.185 & 1914 
    & $3.67 \times 10^{-3}$  & 0.688 & 2.71 \\ 
    B3 & $6.5\times 10^{14}$ & 100/120 & 2.203 & 2080
    & $6.50 \times 10^{-3}$  & 0.921 & 3.56 \\
    B4 & $6.5\times 10^{14}$ & 90/120 & 2.215 & 2300 
    & $8.27 \times 10^{-3}$  & 1.044 & 3.98 \\
    B5 & $6.5\times 10^{14}$ & 80/120 & 2.219 & 2582
    & $8.87 \times 10^{-3}$  & 1.084 & 4.11 
    \\ \hline
    C0 & $6.75 \times 10^{14}$ & 1 & 2.286 & 1812
    & 0.0 & 0.0 & 0.0 \\
    C1 & $6.75 \times 10^{14}$ & 119/120 & 2.289 & 1824
    & $ 3.97 \times 10^{-4}$ & 0.221 & 0.901 \\
    C15 & $6.75 \times 10^{14}$ & 115/120 & 2.298 & 1876
    & $ 1.92 \times 10^{-3}$ & 0.487 & 1.97 \\
    C2 & $6.75 \times 10^{14}$ & 110/120 & 2.310 & 1948
    & $ 3.67 \times 10^{-3}$ & 0.675 & 2.70 \\
    C25 & $6.75 \times 10^{14}$ & 105/120 & 2.321 & 2030
    & $ 5.20 \times 10^{-3}$ & 0.807 & 3.20 
    \\ \hline
    D0 & $7.0\times 10^{14}$ & 1 & 2.412 & 1845
    & $0.0$ & 0.0 & 0.0 \\
    D1 & $7.0\times 10^{14}$ & 119/120 & 2.414 & 1859 
    & $3.97 \times 10^{-4}$  & 0.217 & 0.902 \\ 
    D15 & $7.0\times 10^{14}$ & 115/120 & 2.425 & 1909
    & $1.92 \times 10^{-3}$  & 0.478 & 1.97 \\
    D2 & $7.0\times 10^{14}$ & 110/120 & 2.438 & 1986 
    & $3.67 \times 10^{-3}$  & 0.663 & 2.70 \\
    D25 & $7.0\times 10^{14}$& 105/120 & 2.449 & 2065
    & $5.20\times 10^{-3}$ & 0.792 & 3.20 & \\
    D3 & $7.0\times 10^{14}$ & 100/120 & 2.459 & 2162
    & $6.49 \times 10^{-3}$  & 0.888 & 3.55 \\
    D4 & $7.0\times 10^{14}$ & 90/120 & 2.472 & 2387
    & $8.26 \times 10^{-3}$  & 1.006 & 3.97 \\
    D5 & $7.0\times 10^{14}$ & 80/120 & 2.476 & 2679
    & $8.86 \times 10^{-3}$  & 1.045 & 4.10 
    \\ \hline
    E2 & $7.25\times 10^{14}$ & 110/120 & 2.568 & 2019 
    & $3.66 \times 10^{-3}$ & 0.652 & 2.70 \\
    E25 & $7.25\times 10^{14}$ & 105/120 & 2.580 & 2103 
    & $5.20 \times 10^{-3}$ & 0.780 & 3.19 \\
    E3 & $7.25\times 10^{14}$ & 100/120 & 2.590 & 2198 
    & $6.49 \times 10^{-3}$ & 0.873 & 3.55 \\
    E4 & $7.25\times 10^{14}$ & 90/120 & 2.604 & 2428 
    & $8.25 \times 10^{-3}$ & 0.989 & 3.97 \\
    E5 & $7.25\times 10^{14}$ & 80/120 & 2.608 & 2726 
    & $8.86 \times 10^{-3}$ & 1.03 & 4.09 
    \\ \hline
    F0 & $7.5\times 10^{14}$ & 1 & 2.672 & 1909
    & 0.0 & 0.0 & 0.0 \\
    F1 & $7.5\times 10^{14}$ & 117/120 & 2.680 & 1951 
    & $1.17 \times 10^{-3}$  & 0.361 & 1.55 \\ 
    F2 & $7.5\times 10^{14}$ & 110/120 & 2.700 & 2056 
    & $3.66 \times 10^{-3}$  & 0.641 & 2.71 \\
    F3 & $7.5\times 10^{14}$ & 100/120 & 2.723 & 2238
    & $6.49 \times 10^{-3}$  & 0.858 & 3.55 \\
    F4 & $7.5\times 10^{14}$ & 90/120 & 2.738 & 2471
    & $8.26 \times 10^{-3}$  & 0.972 & 3.97 \\
    F5 & $7.5\times 10^{14}$ & 80/120 & 2.742 & 2773
    & $8.86 \times 10^{-3}$  & 1.010 & 4.10 
    \\ \hline
    G0 & $7.75\times 10^{14}$ & 90/120 & 2.874 & 2510 
    & $8.24 \times 10^{-3}$ & 0.957 & 3.96 \\
    G5 & $7.75\times 10^{14}$ & 80/120 & 2.879 & 2818
    & $8.85 \times 10^{-3}$  & 0.993 & 4.09
    \\ \hline
    H0 & $8.0\times 10^{14}$ & 1 & 2.940 & 1972 
    & 0.0 & 0.0 & 0.0 \\
    H5 & $8.0\times 10^{14}$ & 80/120 & 3.016 & 2864
    & $8.84 \times 10^{-3}$  & 0.978 & 4.10 
  \end{tabular}
 \end{center}
 \caption{The values of $K_{0}$, axial ratio $R_{p}/R_{e}$,
   ADM mass $M_{\rm ADM}$, equatorial circumferential
   radius $R_{\rm ec}$, ratio of the rotational kinetic energy,
   $T_{\rm rot}$, to the gravitational potential energy, $W$,
   nondimensional spin parameter 
   $q \equiv J/M_{\rm ADM}^{2}$, and angular velocity $\Omega_{a}$ for
   rigidly rotating iron cores in equilibrium used as the initial
   condition for numerical simulation.
 }\label{Table2}
\end{table}
\begin{table}[htb]
  \begin{center}
    \begin{tabular}{c|cccccccc}
      Model & $K_{0}$ (cgs) & $A$ 
      & $M_{\rm ADM}$($M_{\odot}$) & $R_{\rm e}$(km) & $T_{\rm rot}/W$ 
      & $q$ & $\Omega_{a}$(1/s) 
      \\ \hline
      D20d1 & $7.0\times 10^{14}$ & 1.0 & 2.437 & 1912
      & $3.54 \times 10^{-3}$ & 0.647 & 3.06 \\
      D22d1 & $7.0\times 10^{14}$ & 1.0 & 2.441 & 1924
      & $4.12 \times 10^{-3}$ & 0.698 & 3.29 \\
      D23d1 & $7.0\times 10^{14}$ & 1.0 & 2.445 & 1936
      & $4.69 \times 10^{-3}$ & 0.746 & 3.50 \\
      D25d1 & $7.0\times 10^{14}$ & 1.0 & 2.450 & 1950 
      & $5.26 \times 10^{-3}$ & 0.791 & 3.69 \\
      D15d05 & $7.0\times 10^{14}$ & 0.5 & 2.424 & 1858
      & $1.75 \times 10^{-3}$ & 0.440 & 2.84 \\
      D17d05 & $7.0\times 10^{14}$ & 0.5 & 2.430 & 1866
      & $2.63 \times 10^{-3}$ & 0.541 & 2.84 \\
      D20d05 & $7.0\times 10^{14}$ & 0.5 & 2.437 & 1874
      & $3.52 \times 10^{-3}$ & 0.626 & 3.47 \\
      D25d05 & $7.0\times 10^{14}$ & 0.5 & 2.444 & 1881
      & $4.42 \times 10^{-3}$ & 0.702 & 4.21 
      \\ \hline
      H5d1 & $8.0\times 10^{14}$ & 1.0 & 3.019 & 2190
      & $9.03 \times 10^{-3}$ & 0.978 & 4.75 \\
      H5d05 & $8.0 \times 10^{14}$ & 0.5 & 3.020 & 2054
      & $8.98 \times 10^{-3}$ & 0.941 & 6.31
    \end{tabular}
  \end{center}
  \caption{The values of $K_{0}$, 
   $A \equiv \varpi_{d}/R_{e}$,  $M_{\rm ADM}$, $R_{e}$, $T_{\rm rot}/W$, 
   $q \equiv J/M_{\rm ADM}^{2}$, and angular velocity at the rotational
   axis $\Omega_{a}$ for differentially rotating
   initial models.}\label{Table3}
\end{table}

\subsection{Initial conditions}

Recent numerical study \cite{ET,UmeNom} for stellar evolution of 
very massive and low metallicity stars from main-sequence to the
pre-iron-core-collapse suggests that initially massive stars evolve to
form an iron core of mass $\agt 2M_{\odot}$--$3M_{\odot}$. 
Taking these fact into account, we consider a wide range of the mass
for the progenitor of the collapse as 
$2M_{\odot} \alt M \alt 3M_{\odot}$. 

It should be addressed that such large mass of the iron core may not
be a special product of low metallicity or large progenitor mass.
Recently, Hirschi et al. \cite{HMM} study 
presupernova evolution of {\it rotating} massive star of solar
metallicity and show that mass of the iron 
core at the onset of the collapse depends strongly on the treatment
of both convection and rotation. 
They find that the rotation significantly increases (by a factor of
$\sim 1.5$) the core mass: An iron core of mass of 
$\approx 2.0M_{\odot}$ is formed from the ZAMS (zero-age main sequence
) of $15M_{\odot}$ by
including the rotation while its non-rotating counterpart yields an iron
core of mass $\approx 1.5M_{\odot}$. This result is quite
different from that in \cite{HLW}.
This suggests that mass of the iron core at the
onset of collapse may depend sensitively on the chemical abundance, 
and the stellar rotation
as well as the magnitude of the viscosity and the mixing
length which are not well understood. 

The central density and the central temperature of 
such very massive iron cores are $\rho_{c} \agt 5 \times 10^{9}$
g/cm$^{3}$ and $T_{c} \sim 10^{10}$ K, respectively \cite{UmeNom}. 
Thus, we set the central density of the initial conditions to be
$10^{10}$ g/cm$^{3}$. 
The electrons under such a high density are extremely
degenerate even with $T \approx 10^{10}$ K, 
since ratio of the Fermi energy of free electrons, $\varepsilon_{F}$,
to thermal energy, $k_{B}T$, is much larger than unity \cite{ST}
according to 
\beq
\frac{\varepsilon_{F}}{k_{\rm B}T} 
\approx 10 \left(\frac{Y_{e}}{0.45}\right)^{1/3}
           \left(\frac{\rho}{10^{10}{\rm ~g/cm}^{3}}\right)^{1/3}
	   \left(\frac{T}{10^{10}{\rm ~K}}\right)^{-1}.
\eeq
Therefore, the main contribution of the pressure comes from the electron
degenerate pressure that is denoted by the polytropic form as
$P_{\rm deg} = K_{\rm deg}\rho^{4/3}$. 

In addition to the electron degenerate pressure, the thermal and
radiation pressure should be taken into account. 
The adiabatic index relevant for them in the iron core 
may be close to $4/3$ at the onset of the collapse since the
photo-dissociation of the irons are accelerated with increasing the
density to reduce the pressure. Hence, we put the gas and radiation 
pressure together into the polytropic form as 
\beq
P_{T} \equiv  P_{\rm gas} + P_{\rm rad} = K_{T}\rho^{\frac{4}{3}}.
\eeq
With these assumptions, rotating polytropes in equilibrium with the
polytropic index $\Gamma=4/3$ and the polytropic constant 
$K_{0} = K_{\rm deg} + K_{T}$ [see Eq. (\ref{EOS43})] are
given as the initial models of rotating iron cores.

The iron core model adopted in this paper is nothing more than a
simplified one. The nature is more complicated and the iron core of the
evolved massive star are not simple polytropes \cite{Ott}. 
We will nevertheless use the term 'iron core' in the
following, to emphasizing that we aim
at clarifying the threshold mass of the {\it iron core} for the prompt
black hole formation.

We adopt the values of $K_{14}\equiv K_{0}/10^{14}$ (cgs) to be 6.0,
6.5, 6.75, 7.0, 7.25, 7.5, 7.75, and 8.0, which we refer to models A,
B, C, D, E, F, G, and H, respectively (cf. Table \ref{Table2}).
In the Newtonian case, mass of the $\Gamma=4/3$ spherical polytrope is 
given by $M \approx 4.555 (K_{0}/G)^{3/2}$. 
Thus, for the adopted values of $K_{0}$, the ADM mass of the initial 
spherical iron cores are $M_{\rm ADM}/M_{\odot}\approx 2.0$, 2.2, 2.3
2.4, 2.6, 2.7, 2.9, and 3.0, respectively (see Table \ref{Table2}). 
These values are larger than the maximum allowed ADM mass, $\approx 1.6 
M_{\odot}$, of the spherical cold polytrope adopted in this paper. 
Namely, in the absence of shock heating and rotation, the 
core collapses to a black hole. 

For our choice of the iron core mass, the corresponding helium core mass  
is $M_{\rm He}/M_{\odot}\approx 17$, 22, 25, 30, 32, 35, 40, and 42
for models A, B, C, D, E, F, G, and H, respectively,
according to the calculation by Umeda and Nomoto \cite{UmeNom}. 
Note that these values are larger than a critical value of helium
core mass for direct black hole formation $\approx 15M_{\odot}$
estimated in \cite{Fryer,HFWLH}. 

The velocity profiles of equilibrium rotating cores are given 
according to a popular relation \cite{KEH,Ster} 
\beq
u^t u_{\varphi} = \varpi_d^2( \Omega_a - \Omega ), 
\eeq 
where $\Omega_a$ denotes the angular velocity
along the rotational axis, and $\varpi_d$ is a constant. 
In the Newtonian limit, the rotational profile is written as 
\beq
\Omega = \Omega_a{\varpi_d^2 \over \varpi^2 + \varpi_d^2}. 
\eeq
Thus, $\varpi_d$ controls the steepness of differential rotation.
In this paper, we pick up mainly the rigidly rotating models in which
$\varpi_d \rightarrow \infty$. For illustration of the effect of
differential rotation, we select differentially rotating models
with $A \equiv \varpi_{d}/R_{e} = 1$ and 0.5 (see table \ref{Table3}).

For rigidly rotating initial models, 
we choose the axial ratio $R_{p}/R_{e}$ of polar radius, $R_{p}$, to
equatorial radius, $R_{e}$, as $1$ (spherical configuration), $119/120$,
$115/120$, $110/120$, $105/120$, $100/120$, $90/120$, and $80/120$. 
The models with these axial ratios are referred to as models X0, X1, X15,
X2, X25, X3, X4, and X5, where X denotes A--H. For example, a
model with $K_{14} = 7.0$ and $R_{p}/R_{e} = 80/120$ is
abbreviated as D5. Note that for 
models with $R_{p}/R_{e} = 80/120$, the angular velocity at the
equatorial stellar surface is nearly equal to the Keplerian velocity;
namely, a rapidly rotating initial condition near the mass shedding
limit is chosen for this case.
On the other hand, differentially rotating initial models are chosen
only for a selected set of parameters with
$K_{14}=7.0$ and 8.0, since the purpose in this
paper is to clarify the effect of differential rotation, comparing
the results with those for the rigidly rotating initial models of
nearly identical values of mass and angular momentum. 

In Tables \ref{Table2} and \ref{Table3}, several fundamental
quantities for the models adopted in the present numerical computation
are listed. 
Here, $q$ is a nondimensional spin parameter defined by 
$J/M_{\rm ADM}^{2}$. 
$T_{\rm rot}$ and $W$ are the rotational kinetic energy and the
gravitational potential energy, and defined according to \cite{FIP,CST}
\beqn
T_{\rm rot}&&
\equiv {1 \over 2} \int d^3x \rho_{\ast} \hat u_{\varphi} \Omega,\\
W&&\equiv  M_{\ast} + \int \rho_{\ast} \varepsilon d^{3}x - M_{\rm
  ADM} + T_{\rm rot},
\eeqn
where $W$ is defined to be positive. 
Note that for the rigidly rotating case, the maximum value 
of $T_{\rm rot}/W$ is $\approx 0.009$.

For the rigidly or weakly differentially rotating stars, the stability
against nonaxisymmetric perturbation is determined by $T_{\rm rot}/W$
\cite{Karino}. A star with $T_{\rm rot}/W \agt 0.27$ and $T_{\rm
rot}/W \agt 0.14$ will become unstable against the nonaxisymmetric
dynamical and secular instabilities, respectively. Here, the secular
timescale is much longer than the collapse timescale, so that the
dynamical instabilities are only relevant during collapse and bounces. 
Since we assume that the rotating iron core collapse proceeds in an 
axisymmetric manner, it is important to ensure that the iron core should
not spin up to be dynamically unstable against nonaxisymmetric
deformation.  We will discuss the spin-up and the stabilities of the
inner cores formed after the bounce in Sec. \ref{Sec-rot}.

For the differentially rotating case with a small value of $A$,
it is possible to make equilibrium states with $T_{\rm rot}/W \gg 0.009$. 
With such an initial condition, the collapsing core may 
form a differentially rotating object
of a highly nonspherical shape and of a high value of $T_{\rm rot}/W$
\cite{SS4,Rammp}. 
It is also known that rapidly rotating neutron stars of a high degree of 
differential rotation is dynamically unstable against nonaxisymmetric 
deformation even for $T_{\rm rot}/W=O(0.01)$ (e.g., \cite{SKS} and 
references therein). To follow such collapse, 
a nonaxisymmetric simulation will be necessary. In this paper, 
we do not choose such initial conditions and focus only on 
dynamically stable cases against nonaxisymmetric deformation. 
The collapse of highly differentially rotating initial conditions
is studied in three-dimensional simulations in \cite{SS4}. 

\subsection{Computational Settings}

The central density increases from
$10^{10}~{\rm g/cm^3}$ to $\agt 10^{15}~{\rm g/cm^3}$ 
during the collapse.
This implies that the characteristic length scale of the system varies by
a factor of $\sim 100$. 
To compute such a collapse accurately saving the CPU time efficiently,
a regridding technique as described in \cite{ShiSha,SS2} is helpful. 
The regridding is carried out whenever the characteristic radius
of the collapsing star decreases by a factor of a few. At each regridding, 
the grid spacing is decreased by a factor of 2. 
All the quantities in the new grid are calculated
using the cubic interpolation. 
To avoid discarding the matter in the outer region, we also increase the 
grid number at the regridding.

For the regiridding, we define 
a relativistic gravitational potential
$\Phi_c \equiv 1 -\alpha_c~ (\Phi_c>0)$.
Since $\Phi_c$ is approximately proportional to $M/R$, 
$\Phi_c^{-1}$ can be used as a measure of the characteristic
length scale of the core for the regridding. 
We typically choose $N$ at each regridding in the following manner. 
From $t=0$ to the time at which
$\Phi_c = 0.025$, we set $N=620$ with the grid spacing $\Delta
R_{e}/600$. At $\Phi_c = 0.025$, the characteristic stellar radius
becomes approximately one fourth of the initial value. 
Then, the first regridding is performed; the grid 
spacing is changed to the half of the previous one
and the grid number is increased to $N=1020$. 
Subsequently, the value of $N$ is chosen in the following manner; 
for $0.025 \leq \Phi_c \leq 0.05$, we set $N=1020$; 
for $0.05 \leq \Phi_c \leq 0.1$, we set $N=1700$; and 
for $0.1 \leq \Phi_c \leq 0.25$, we set $N=2500$, and keep this number
until the termination of the simulations. 
For the typical cases, the physical
size of the grid spacing is $\Delta \sim 4$ km at the beginning of
simulations and $\sim 0.5$ km at the end. 

In the case of black hole formation, $\Phi_{c}$ approaches to 1. In
this case, we carry out one more regridding at the time of
$\Phi_{c}=0.25$. In this final regridding, the grid spacing is made
half while keeping $N=2500$. For the typical cases in which a black
hole is formed, the physical
size of the grid spacing is $\Delta \sim 4$ km at the beginning of
simulations and $\sim 0.25$ km at the end.
In this treatment, the total discarded fraction of the baryon rest-mass
which is located outside the new regridded domains is $\alt 4\%$.

Simulations for each model with the higher grid resolution 
are performed for 
40,000--100,000 time steps.
The required CPU time for one model is about 30--90 hours using 8
processors of FACOM VPP 5000 at the data processing center
of National Astronomical Observatory of Japan. 

\section{Numerical Results}\label{Numerical result}
\begin{figure}[htb]
  \vspace{-6mm}
  \begin{center} 
    \epsfxsize=3.in
    \leavevmode
    \epsffile{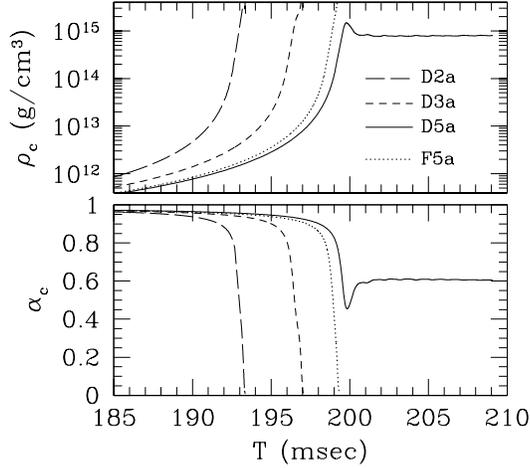}
  \end{center}
  \vspace{-8mm}
  \caption{Evolution of the central density $\rho_{c}$ (upper
    panel) and the central value of 
    the lapse function
    $\alpha_{c}$ (lower panel) for models D2a
    (long dashed curve), D3a (dashed curve), D5a (solid curve), and
    F5a (dotted curve).
  }\label{figure2}
\end{figure}
\begin{figure}[htb]
  \vspace{-6mm}
  \begin{center} 
    \epsfxsize=3.in
    \leavevmode
    \epsffile{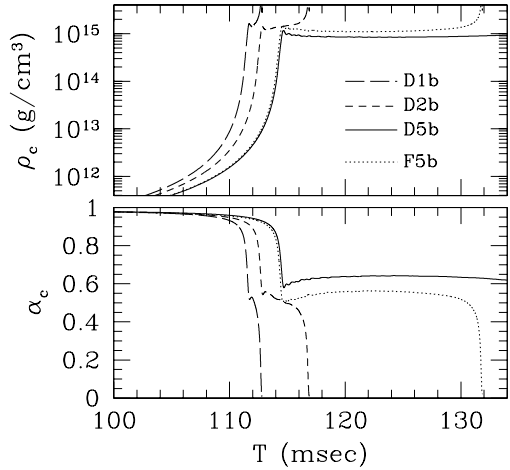}
  \end{center}
  \vspace{-8mm}
  \caption{The same as Fig. \ref{figure2} but for models D1b
    (long dashed curve), D2b (dashed curve), D5b (solid curve), and
    F5b (dotted curve).
  }\label{figure3}
\end{figure}
\begin{figure}[htb]
  \vspace{-6mm}
  \begin{center} 
    \epsfxsize=3.in
    \leavevmode
    \epsffile{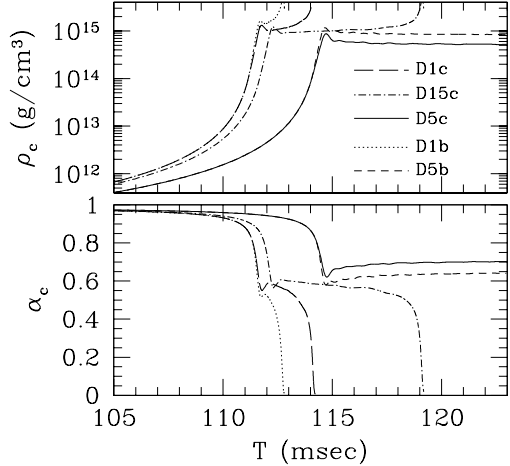}
  \end{center}
  \vspace{-8mm}
  \caption{The same as Fig. \ref{figure2} but for models D1c
    (long dashed curve), D15c (dotted dashed curve), D5c (solid
    curve). Results for models D1b (dotted curve) and D5b (dashed
    curve) are also displayed for comparison.
  }\label{figure4}
\end{figure}
\begin{figure}[htb]
  \vspace{-6mm}
  \begin{center} 
    \epsfxsize=3.in
    \leavevmode
    \epsffile{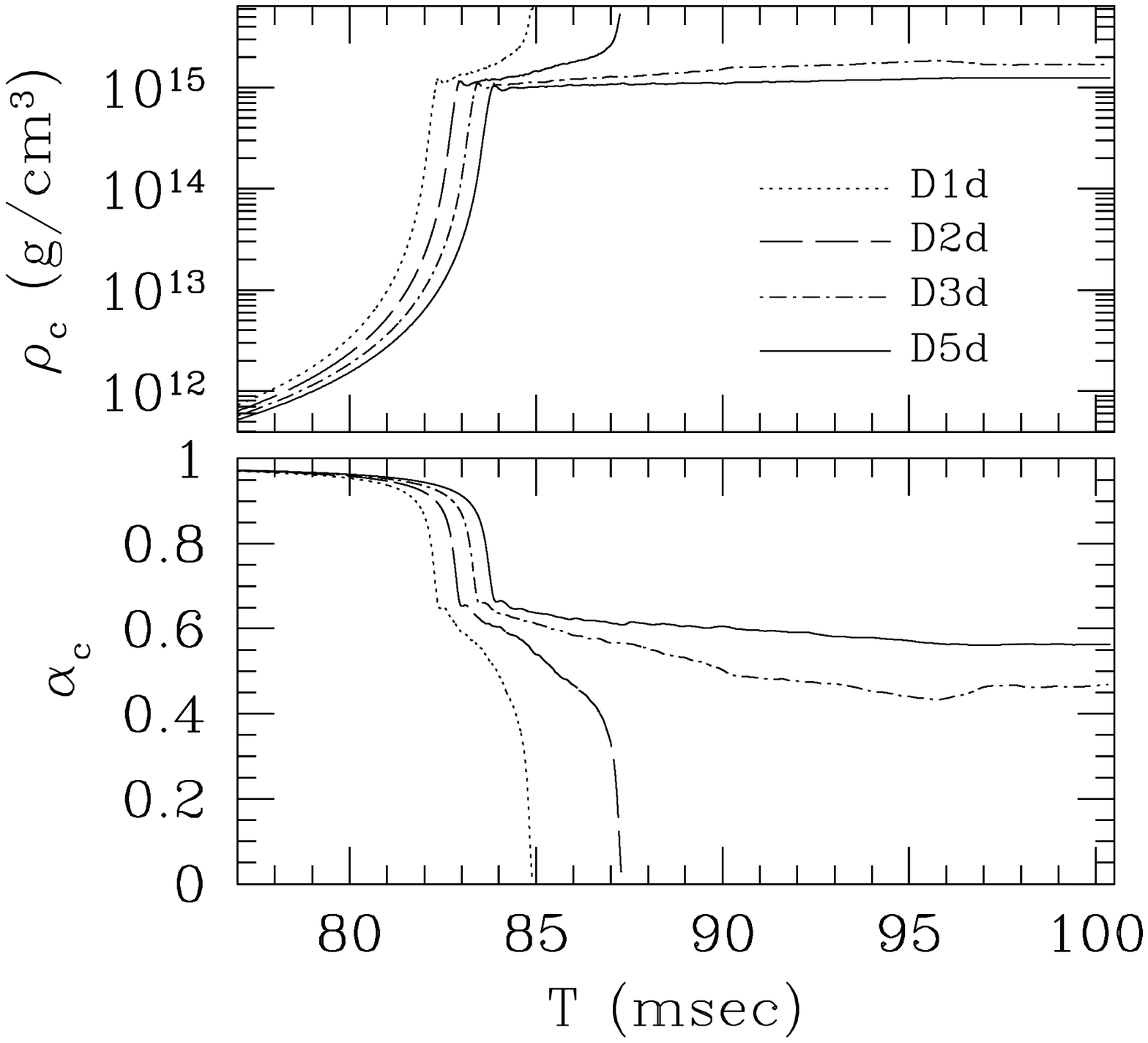}
  \end{center}
  \vspace{-8mm}
  \caption{ The same as Fig. \ref{figure2} but for models D1d
    (dotted curve), D2d (long dashed curve), D3d (dotted dashed
    curve), and D5d (solid curve).
  }\label{figure5}
\end{figure}

\subsection{General feature of the collapse}\label{general feature}

We first summarize the outline of the dynamics of stellar
core collapse. Detailed features of the collapse will be presented in
the subsequent subsections.
In Figs. \ref{figure2}--\ref{figure5}, we display 
evolution of the central density $\rho_{c}$ 
and the central value of the lapse function $\alpha_{c}$ for selected
models.
As indicated in these figures,
rotating stellar core collapse to a neutron star can be divided
into three phases; the infall phase, the bounce phase, and the ringdown
(or post-bounce oscillation) phase \cite{Monch,Zweg}. 

The infall phase sets in due to the onset of the gravitational instability
of the progenitor triggered by the sudden softening of the equation
of state which is associated with the reduction of the adiabatic index.
During this phase, the central
density, $\rho_{c}$, (the central value of the lapse function,
$\alpha_{c}$), monotonically
increases (decreases) until it reaches the nuclear density, 
provided that the core is not very rapidly rotating
initially. The inner part of the core, which collapses nearly
homologously with a subsonic infall velocity, constitutes the inner
core. On the other hand, the outer region in which the infall
velocity is supersonic constitutes the outer core \cite{Monch,Zweg}.

The bounce phase sets in when the density around
the central region exceeds the nuclear density.
At this phase, the inner core decelerates rapidly due to (a) the sudden
stiffening of the equation of state or (b) the strong centrifugal force.
Hereafter, we pay attention to the case (a) since the 
collapse is halted due to the sudden stiffening of the equation 
of state for all the models (cf. Figs. \ref{figure2}--\ref{figure5}).
Because of its large inertia and large kinetic energy induced by the
infall, the inner core overshoots its hypothetical equilibrium state. 
The degree of the overshooting depends on the mass,
the amount of rotational kinetic energy, and 
the stiffness of the equation of state. 

If its mass is not too large, the inner core experiences a bounce.
The stored internal energy of the inner
core at maximum compression is released through 
a strong pressure wave generated inside the inner core \cite{Monch}.
The pressure wave travels from the center to the outer region 
until it reaches the sonic point located at the edge of the inner
core. Since the sound cones tilt inward beyond the sonic point, the
pressure disturbance cannot travel further and forms a shock
just inside the sonic point. 
During the formation of the shock, the inner core transfers its
kinetic energy to the shock through compressional work that 
powers the shock \cite{vanRiper,Monch}. 
It is important to note that the shock is formed at the outer edge of the
inner core, and hence, the bulk of the inner core matter never
undergoes shock heating and acceleration. Therefore, the shock heating 
is relatively less efficient for models with larger mass of the inner core.

On the other hand, if the mass and inertia of the inner core at the bounce
are sufficiently large, the pressure supplied by the sudden stiffening of
the equation of state and the centrifugal force cannot halt the
collapse. Then, the inner core will promptly collapse to a 
black hole without any distinct bounce (cf. models
D2a, D3a, and D5a in Fig. \ref{figure2} in which a black hole is
formed). As illustrated in Sec. \ref{secGam},
shocks do not propagate outward in such cases.

In the ring-down phase, the inner core oscillates quasi-radially and
then settles down to a quasi-stationary state, since the compressional
work done by the inner core on the matter of the outer region leads to
damping of the oscillation.
In this phase, the amplitude of the oscillation and the strength of
shocks generated by the outward oscillation depend on the stiffness of the
equation of state for the formed protoneutron star.
In the outer region, 
on the other hand, shock waves propagate and are accelerated due to the
density gradient in the outer core.
They sweep materials of the outer envelopes, and convert the infall kinetic
energy into the thermal energy, which helps further driving the shock
outward. However, 
if the explosion is too weak to eject sufficient matter of the
progenitor star, the subsequent fallback of the matter into
the formed protoneutron star will trigger formation of black hole.

%
\begin{table}[p]
 \vspace{-5mm}
 \begin{center}
  \begin{tabular}{c|ccccc|cccc}
    Initial model & $K_{0}$ (cgs) & $R_{p}/R_{e}$ 
    & $M_{\rm ADM}(M_{\odot})$ & $q$ & 
    &\multicolumn{4}{c}{Adopted  equations of state} \\
     &   &   &   &  &  &  a   &   b   &   c   &  d  \\ \hline
    B0  & $6.5\times 10^{14}$ & 1       & 2.163 & 0.0 & 
    & BH &    BH & no BH & BH  \\
    B1  & $6.5\times 10^{14}$ & 119/120 & 2.163 & 0.217 & 
    & no BH & no BH & no BH & no BH  \\
    B2  & $6.5\times 10^{14}$ & 110/120 & 2.185 & 0.688 & 
    & no BH & no BH & no BH & no BH  \\ 
    B3  & $6.5\times 10^{14}$ & 100/120 & 2.203 & 0.921 & 
    & no BH & no BH & no BH & no BH  \\ 
    B4  & $6.5\times 10^{14}$ & 90/120  & 2.215 & 1.044 & 
    & no BH & no BH & no BH & no BH  \\
    B5  & $6.5\times 10^{14}$ & 80/120  & 2.219 & 1.084 & 
    & no BH & no BH & no BH & no BH
    \\ \hline
    C0 & $6.75 \times 10^{14}$ & 1       & 2.286 & 0.0 & 
    & BH & BH & BH & --- \\
    C1 & $6.75 \times 10^{14}$ & 119/120 & 2.289 & 0.221 & 
    & BH & BH & no BH & --- \\
    C15 & $6.75 \times 10^{14}$ & 115/120 & 2.298 & 0.487  &
    & BH & BH & no BH & --- \\
    C2 & $6.75 \times 10^{14}$ & 110/120 & 2.310 & 0.675  &
    & BH & no BH & no BH & --- \\
    C25 & $6.75 \times 10^{14}$ & 105/120 & 2.321 & 0.807  &
    & no BH & no BH & no BH & --- 
    \\ \hline
    D0  & $7.0\times 10^{14}$ & 1       & 2.412 & 0.0 & 
    &    BH &    BH &    BH &    BH  \\
    D1  & $7.0\times 10^{14}$ & 119/120 & 2.414 & 0.217 & 
    &    BH &    BH &    BH &    BH  \\
    D15 & $7.0\times 10^{14}$ & 115/120 & 2.425 & 0.478 & 
    &    BH &    BH &    BH &    BH  \\
    D2  & $7.0\times 10^{14}$ & 110/120 & 2.438 & 0.663 & 
    &    BH &    BH & no BH &    BH  \\
    D25 & $7.0\times 10^{14}$ & 105/120 & 2.449 & 0.792 & 
    &    BH &    BH & no BH &    BH  \\ 
    D3  & $7.0\times 10^{14}$ & 100/120 & 2.459 & 0.888 & 
    &    BH & no BH & no BH & no BH  \\
    D4  & $7.0\times 10^{14}$ & 90/120  & 2.472 & 1.006 & 
    & no BH & no BH & no BH & no BH  \\
    D5  & $7.0\times 10^{14}$ & 80/120  & 2.476 & 1.045 & 
    & no BH & no BH & no BH & no BH  
    \\ \hline
    E2 & $7.25\times 10^{14}$ & 110/120 & 2.568 & 0.652 &
    & BH &    BH &    BH & --- \\
    E25 & $7.25\times 10^{14}$ & 105/120 & 2.580 & 0.780 &
    & BH &    BH & no BH & --- \\
    E3 & $7.25\times 10^{14}$ & 100/120 & 2.590 & 0.873 &
    & BH &    BH & no BH & --- \\
    E4 & $7.25\times 10^{14}$ & 90/120 & 2.604 & 0.989 &
    & BH & no BH & no BH & --- \\
    E5 & $7.25\times 10^{14}$ & 80/120 & 2.608 & 1.03 &
    & BH & no BH & no BH & --- 
    \\ \hline
    F0  & $7.5\times 10^{14}$ & 1       & 2.672 & 0.0 & 
    & BH &    BH &    BH & BH  \\
    F1  & $7.5\times 10^{14}$ & 119/120 & 2.680 & 0.361 & 
    & BH &    BH &    BH & BH  \\
    F2  & $7.5\times 10^{14}$ & 110/120 & 2.700 & 0.641 & 
    & BH &    BH &    BH & BH  \\
    F3  & $7.5\times 10^{14}$ & 100/120 & 2.723 & 0.858 & 
    & BH &    BH &    BH & BH  \\
    F4  & $7.5\times 10^{14}$ & 90/120  & 2.738 & 0.972 & 
    & BH &    BH & no BH & BH  \\
    F5  & $7.5\times 10^{14}$ & 80/120  & 2.742 & 1.010 & 
    & BH &    BH & no BH & BH
    \\ \hline
    G4  & $7.75\times 10^{14}$ & 90/120 & 2.874 & 0.957 & 
    &  BH &    BH &    BH & ---  \\
    G5  & $7.75\times 10^{14}$ & 80/120 & 2.879 & 0.993 & 
    &  BH &    BH &    BH & --- 
    \\ \hline
    H0  & $8.0\times 10^{14}$ & 1      & 2.940 & 0.0 & 
    &  BH &    BH &    BH &  BH  \\
    H5  & $8.0\times 10^{14}$ & 80/120 & 3.016 & 0.978 & 
    &  BH &    BH &    BH &  BH 
  \end{tabular}
 \end{center}
\caption{Summary of the outcome in the iron core collapse with rigidly 
rotating initial models. The adopted equations of state are discriminated by a
single alphabet: 'a' for
  $(\Gamma_{1}, \Gamma_{2}, \rho_{14})$ 
  $= (1.32, 2.25,2.0)$, 
  'b' for $(1.3, 2.50,2.0)$, 
  'c' for $(1.3, 2.22, 1.0)$,
  and 'd' for $(1.28, 2.75, 2.0)$. ``BH'' implies that a black
hole is formed in a dynamical time scale of the collapse. ``no BH''
implies that a black hole is not formed promptly. ``---'' in the last
  column implies that we did not perform simulations for such models.}
\label{Table4}
\end{table}
%
%
\begin{table}[p]
 \vspace{-5mm}
 \begin{center}
  \begin{tabular}{c|ccccc|c}
    \ \ \ \ \ model \ \ \ \ \ & $K_{0}$ (cgs) & $A$ 
    & $M_{\rm ADM}/M_{\odot}$ & $q$ & & outcome 
    \\ \hline
    D20d1b & $7.0\times 10^{14}$ & 1.0 & 2.437 & 0.647 & & BH \\
    D22d1b & $7.0\times 10^{14}$ & 1.0 & 2.441 & 0.698 & & BH \\
    D23d1b & $7.0\times 10^{14}$ & 1.0 & 2.445 & 0.746 & & no BH  \\
    D25d1b & $7.0\times 10^{14}$ & 1.0 & 2.450 & 0.791 & & no BH \\
    \hline
    D15d05b & $7.0\times 10^{14}$ & 0.5 & 2.424 & 0.440 & & BH \\
    D17d05b & $7.0\times 10^{14}$ & 0.5 & 2.430 & 0.541 & & BH \\
    D20d05b & $7.0\times 10^{14}$ & 0.5 & 2.437 & 0.626 & & no BH \\
    D25d05b & $7.0\times 10^{14}$ & 0.5 & 2.444 & 0.702 & & no BH 
    \\ \hline
    H5d1 & $8.0\times 10^{14}$ & 1.0 & 3.019 & 0.978 & & BH \\
    H5d05 & $8.0 \times 10^{14}$ & 0.5 & 3.020 & 0.941 & & no BH
   \end{tabular}
 \end{center}
\caption{Summary of the final outcome in the iron core collapse with
  differentially rotating initial models. The adopted equation of
  state is type 'b' for all the cases;
i.e., $(\Gamma_{1}, \Gamma_{2}, \rho_{14})$ 
  $= (1.3, 2.5, 2.0)$. }\label{Table5}
\end{table}

\begin{figure}[thb]
\vspace{-4mm}
\begin{center}
\epsfxsize=3.8in
\leavevmode
\epsffile{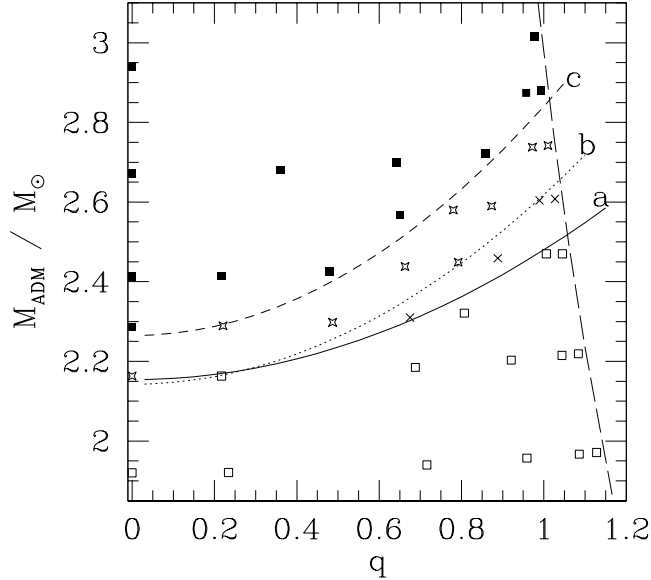} 
\end{center}
\vspace{-10mm}
\caption{The distribution map of the outcome in the iron core 
  collapse for rigidly rotating initial models. 
  The horizontal and vertical axes denote the spin parameter
  $q$ and $M_{\rm ADM}$. The
  filled squares denote the models whose final outcome is a black hole
  irrespective of the equations of state. The open stars
  denote the models whose final outcome is a black hole for the
  equations of state of 'a' and 'b', and a neutron star for the
  equation of state 'c'. The crosses denote the models whose final
  outcome is a black hole only for the equations of state 'a'. The
  open squares indicate no black hole formation irrespective of the
  equations of state. The solid, dotted,
  and dashed curves indicate the threshold mass above which a black hole is
  formed as a result of the collapse for the equations of state 
  'a', 'b', and 'c', respectively (see Sec. \ref{EffecRot} for
  locating these curves). The long dashed curve denotes the
  mass-shedding limit for rigidly rotating initial models; i.e., no equilibrium
  configurations for rigidly rotating initial condition exist in the
  right hand side of this curve.}\label{figure6}
\end{figure}
\begin{figure}[htb]
\vspace{-4mm}
\begin{center}
\epsfxsize=2.35in
\leavevmode
\epsffile{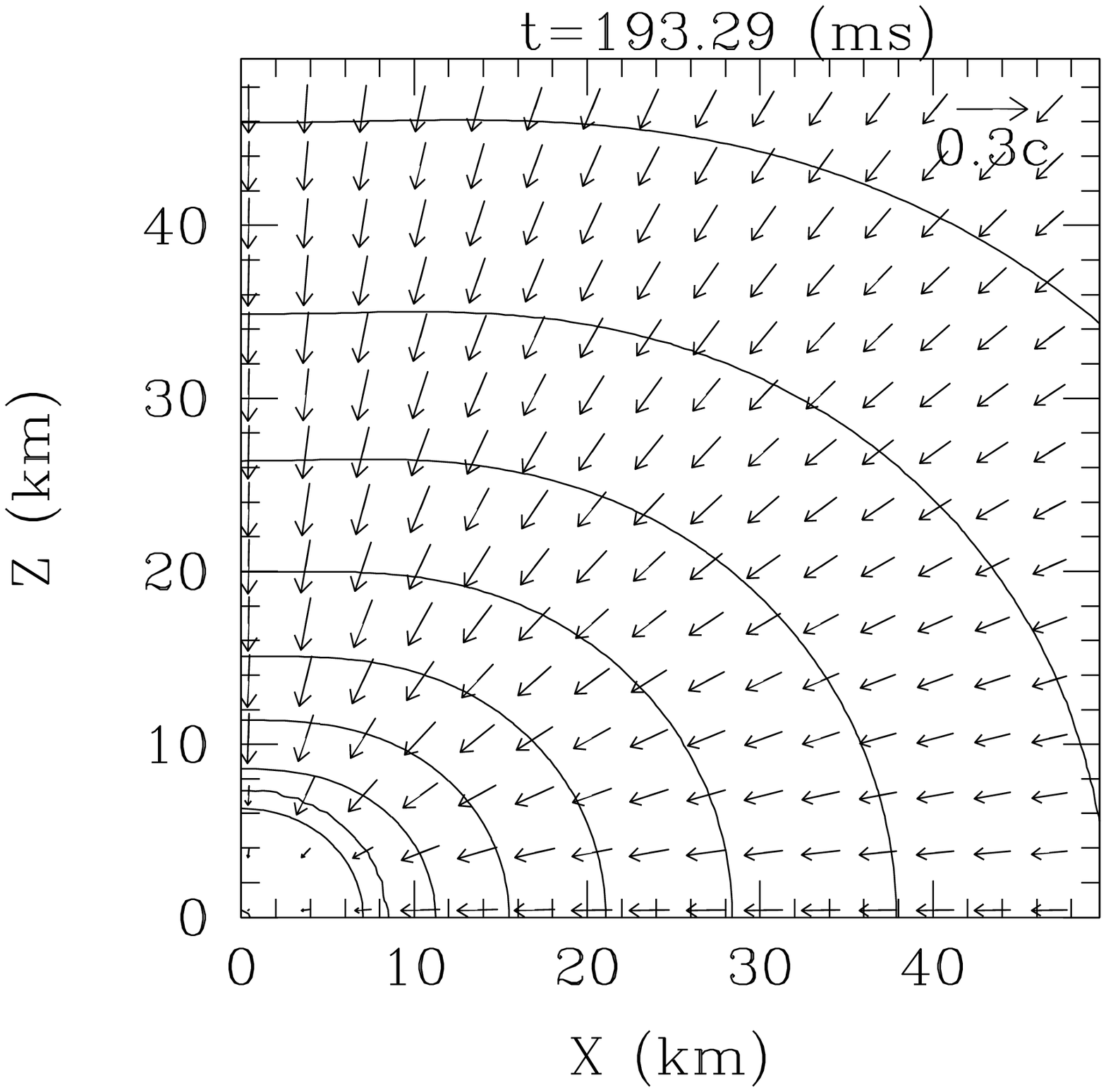} 
\epsfxsize=2.35in
\leavevmode
\epsffile{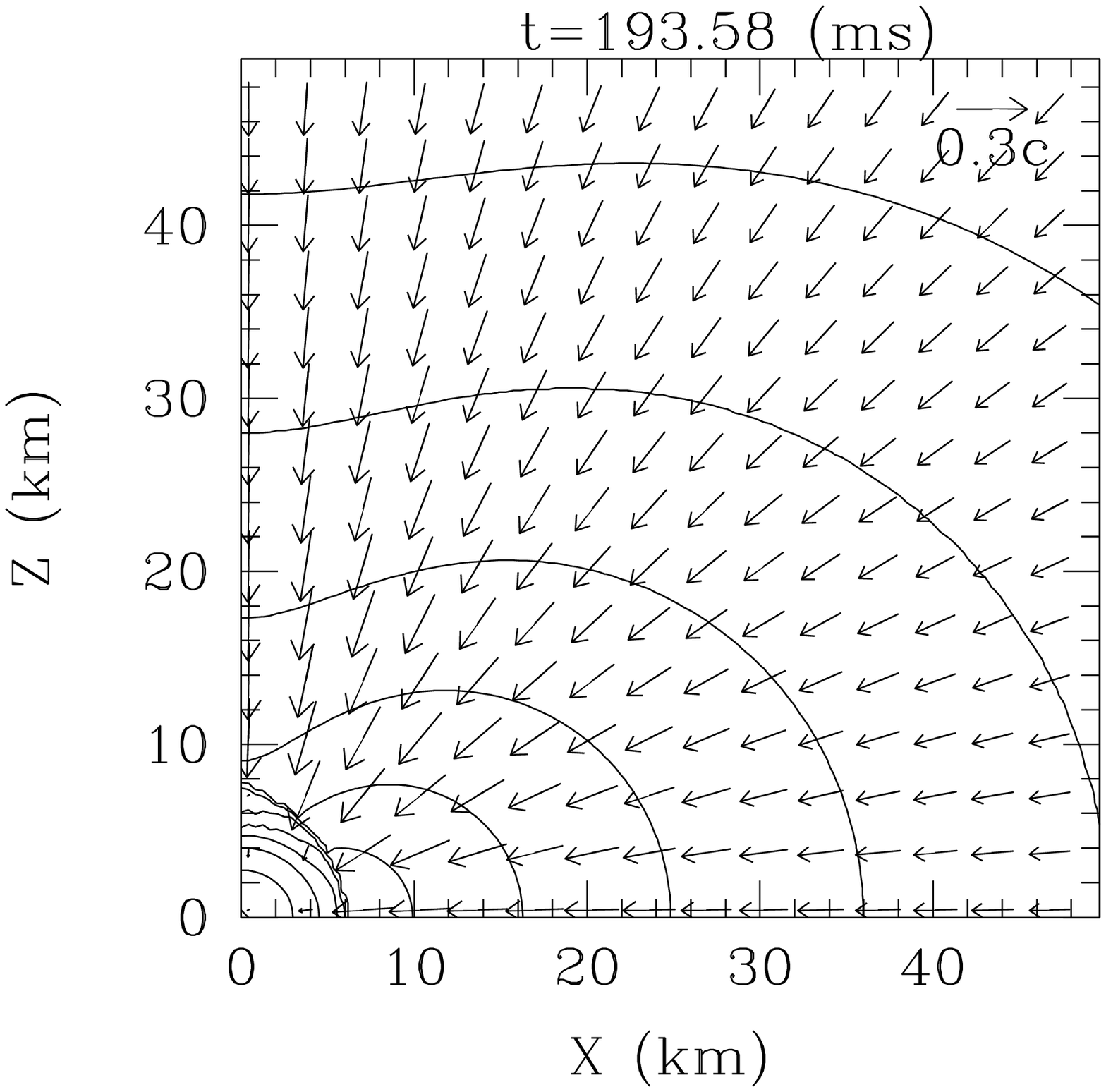}
\epsfxsize=2.35in
\leavevmode
\epsffile{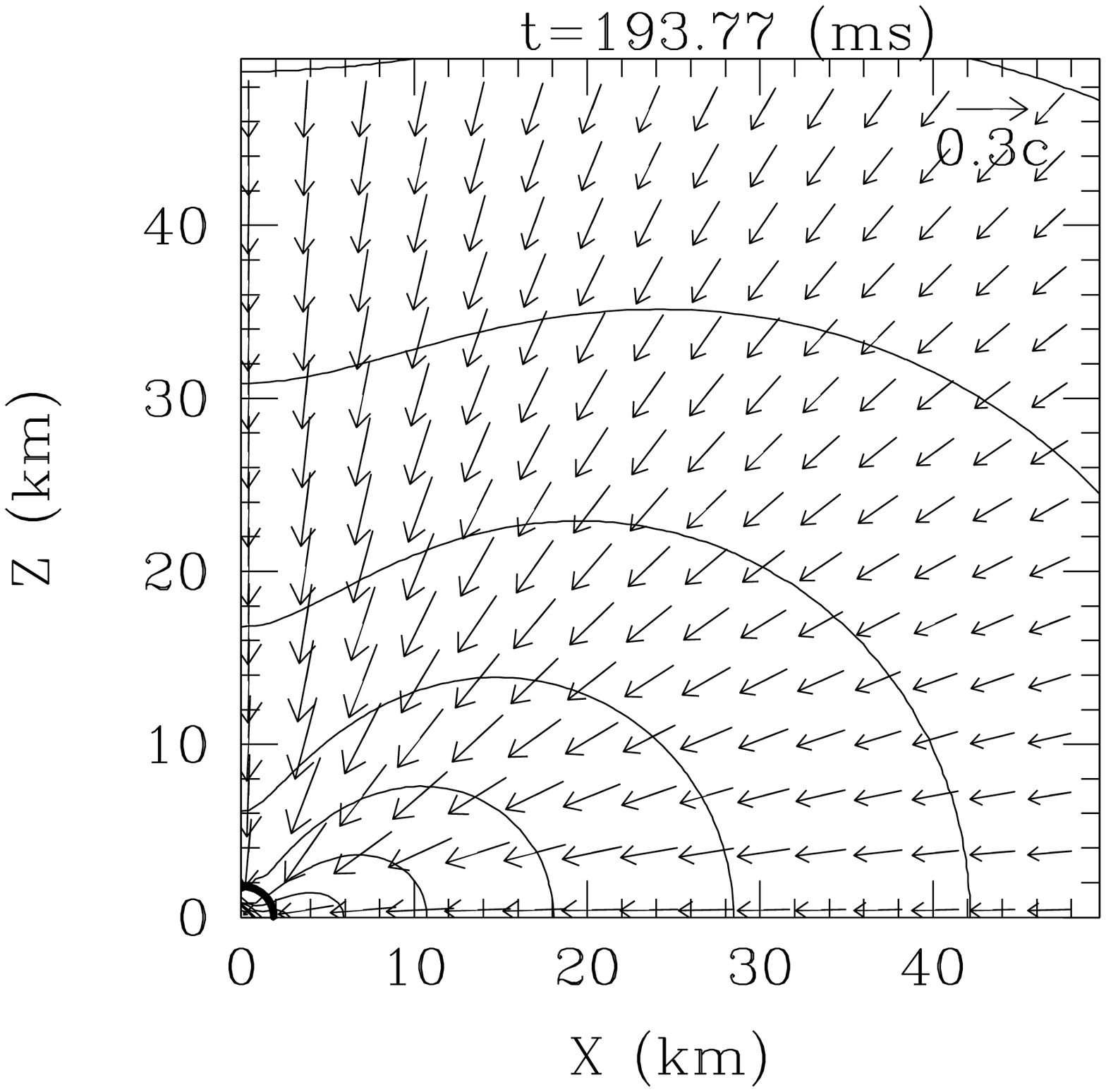}
\end{center}
\vspace{-6mm}
\caption{Snapshots of the density contour curves and velocity vectors
  in the $x$-$z$ plane for model D2
  with the equation of state 'a', at $t = 193.29$,
  193.58, and 193.77 ms. 
  The density contour
  curves are drawn for $\rho/\rho_{\rm max} = 10^{-0.4j}$, $(j=0, 1,
  2, \cdots 20)$ where $\rho_{\rm max}$ is the maximum density at the
  selected time slices.
  The thick solid curve in
  the last panel denotes the location of the apparent horizon.}\label{figure7}
\end{figure}
\begin{figure}[htb]
\vspace{-4mm}
  \begin{center}
    \epsfxsize=2.35in
    \leavevmode
    \epsffile{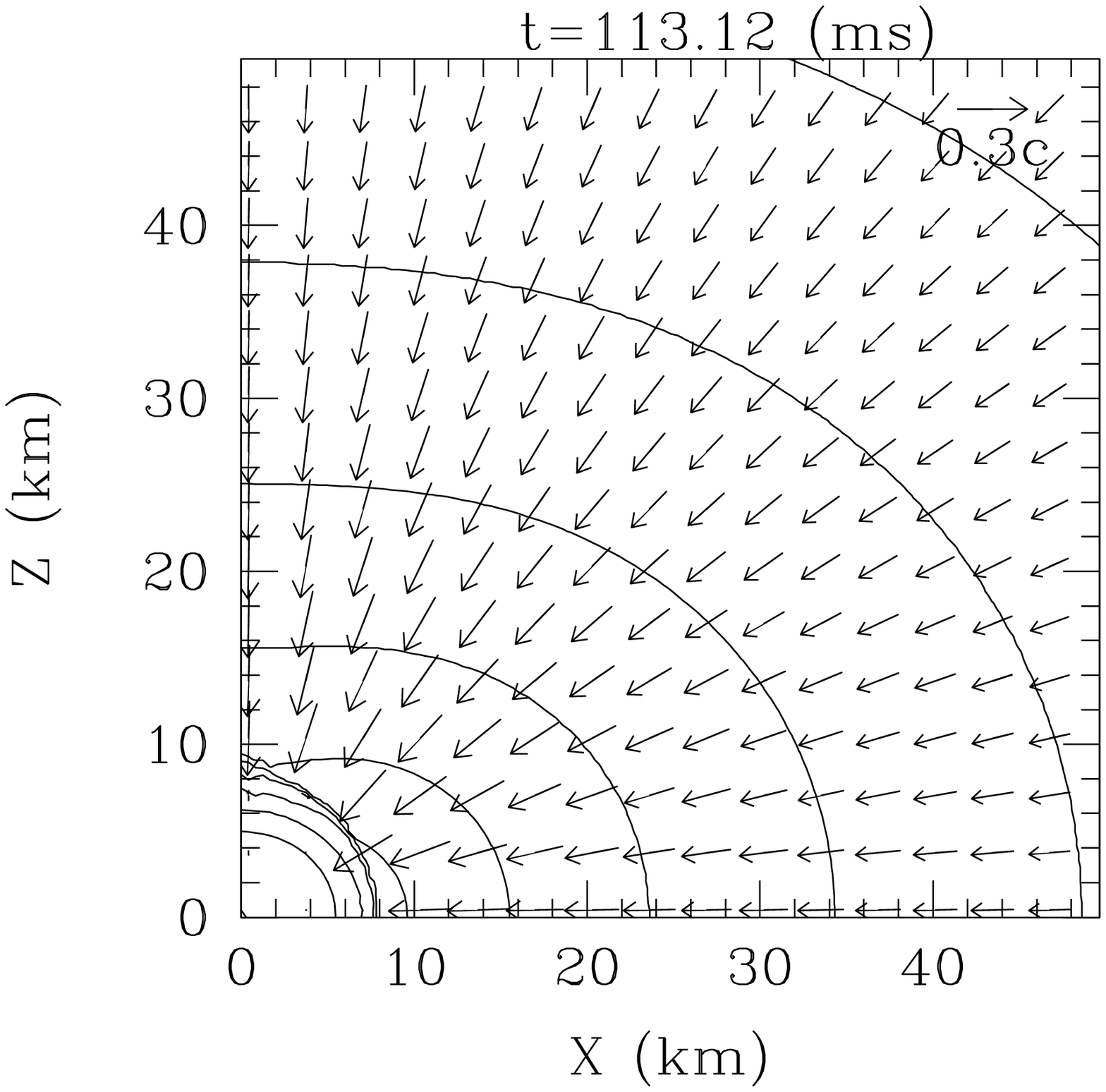} 
    \epsfxsize=2.35in
    \leavevmode
    \epsffile{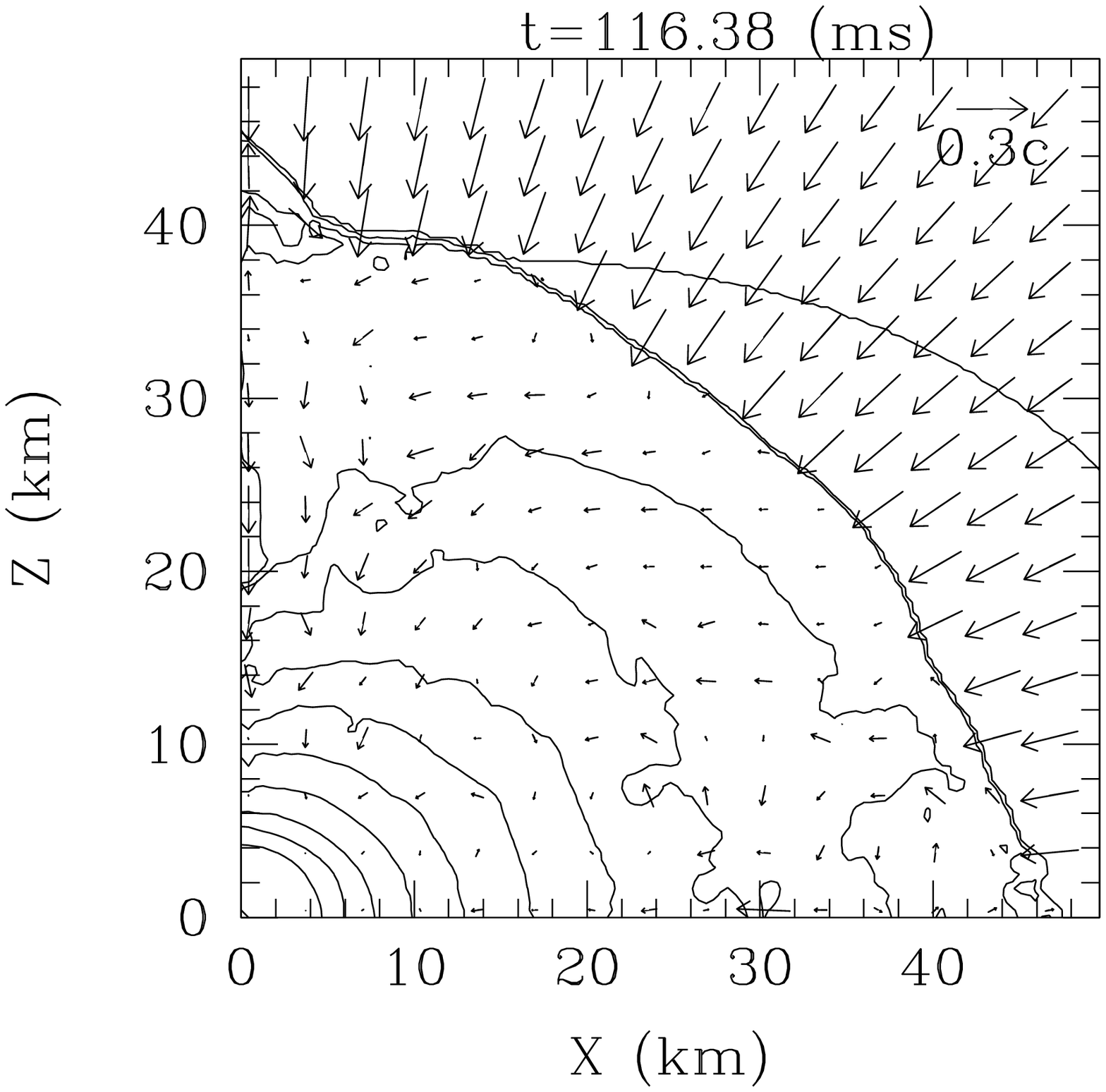}
    \epsfxsize=2.35in
    \leavevmode
    \epsffile{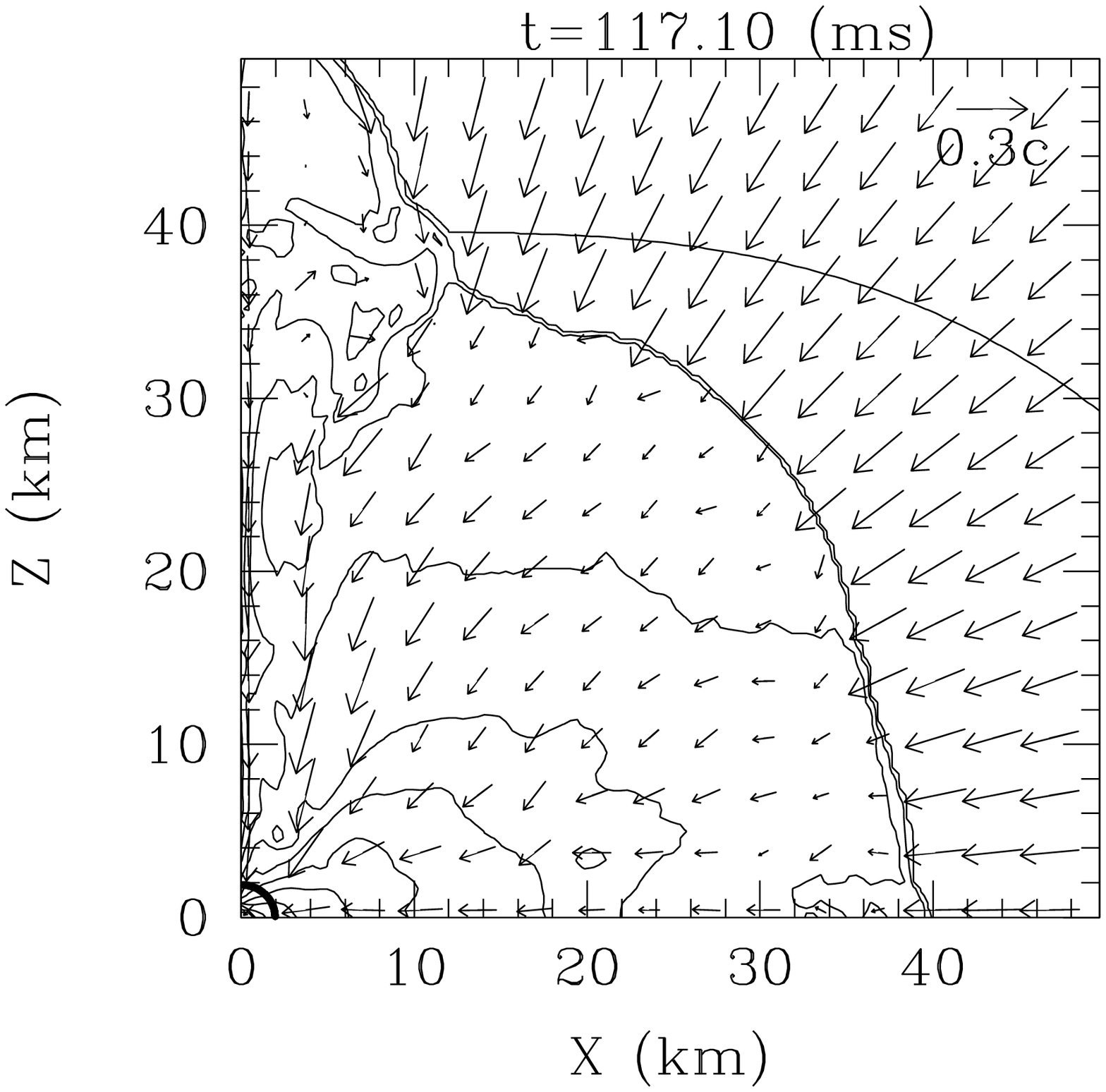}
  \end{center}
\vspace{-6mm}
\caption{The same as Fig. \ref{figure7} but for model D2
  with the equation of state 'b' at $t = 113.12$, 116.38, and 117.10 ms. 
  The thick solid curve in
  the last panel denotes the location of the apparent horizon.}\label{figure8}
\end{figure}
\begin{figure}[htb]
\vspace{-4mm}
\begin{center}
\epsfxsize=2.35in
\leavevmode
\epsffile{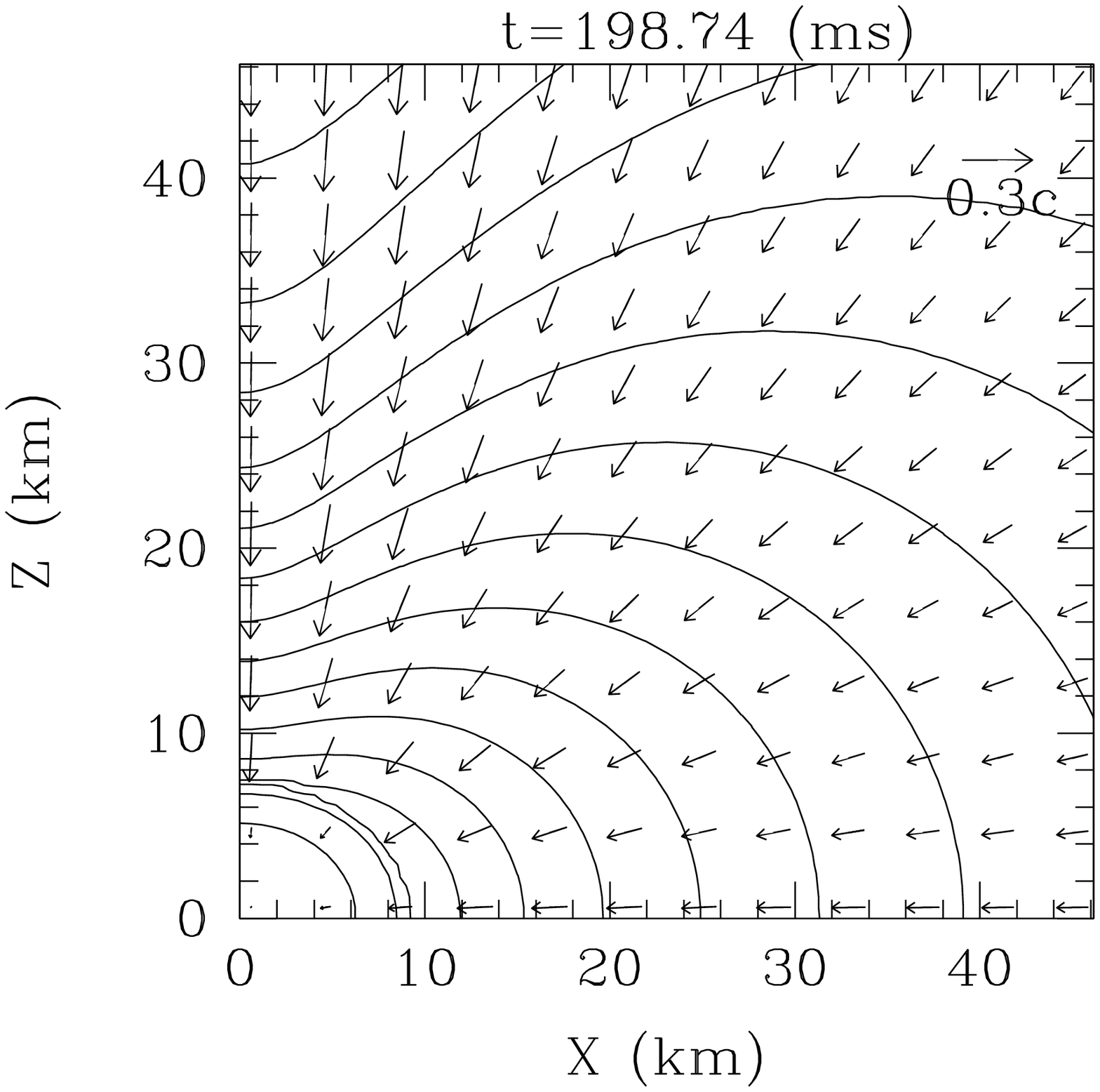}  
\epsfxsize=2.35in
\leavevmode
\epsffile{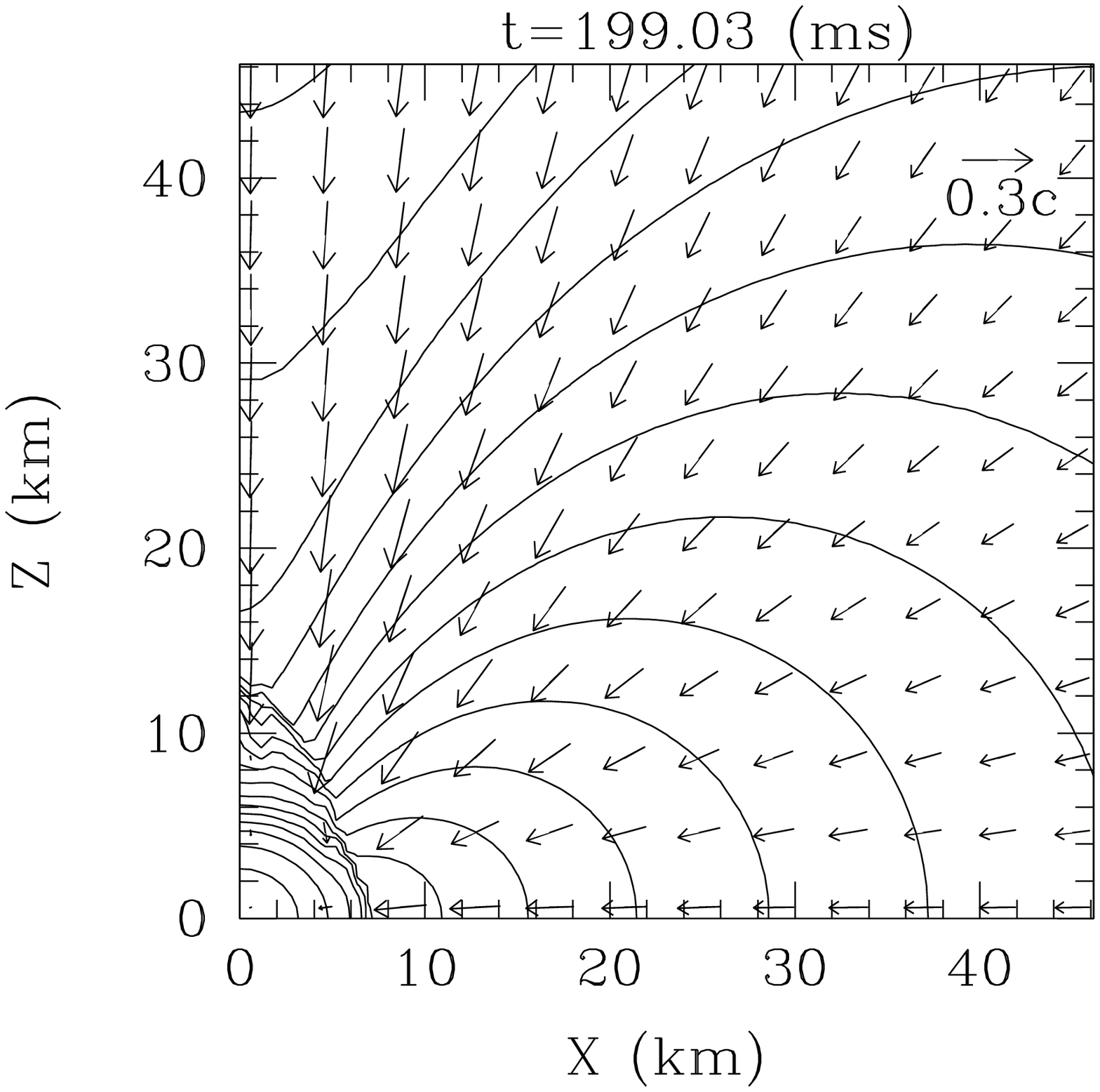}
\epsfxsize=2.35in
\leavevmode
\epsffile{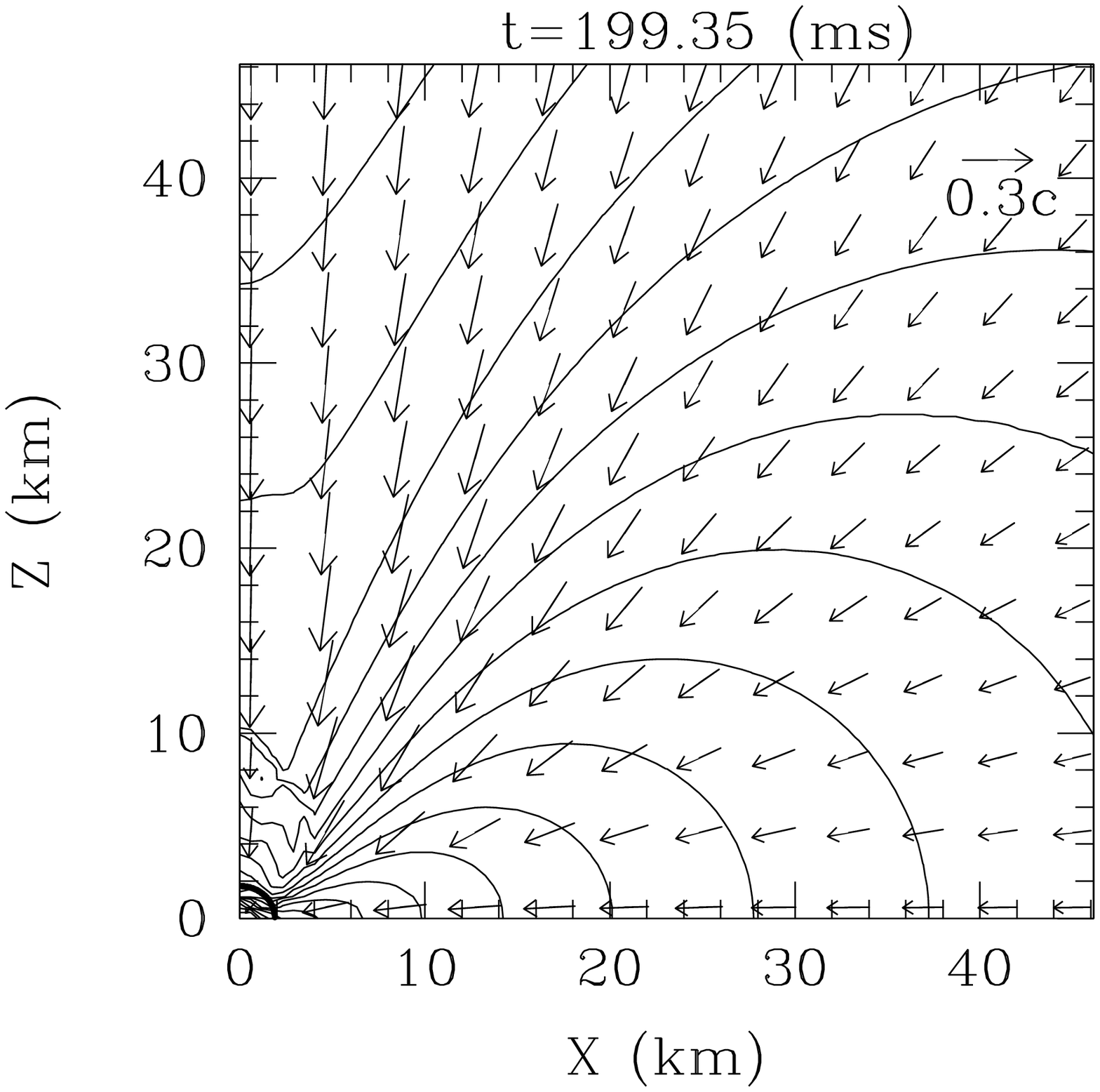}
\end{center}
\vspace{-6mm}
\caption{The same as Fig. \ref{figure7} but for model F5
  with the equation of state 'a', at $t = 198.74$, 199.03, and 199.35
  ms. The thick solid curve in the last panel denotes the location
  of the apparent horizon.}\label{figure9} 
\end{figure}
\begin{figure}[htb]
\vspace{-4mm}
\begin{center}
\epsfxsize=2.35in
\leavevmode
\epsffile{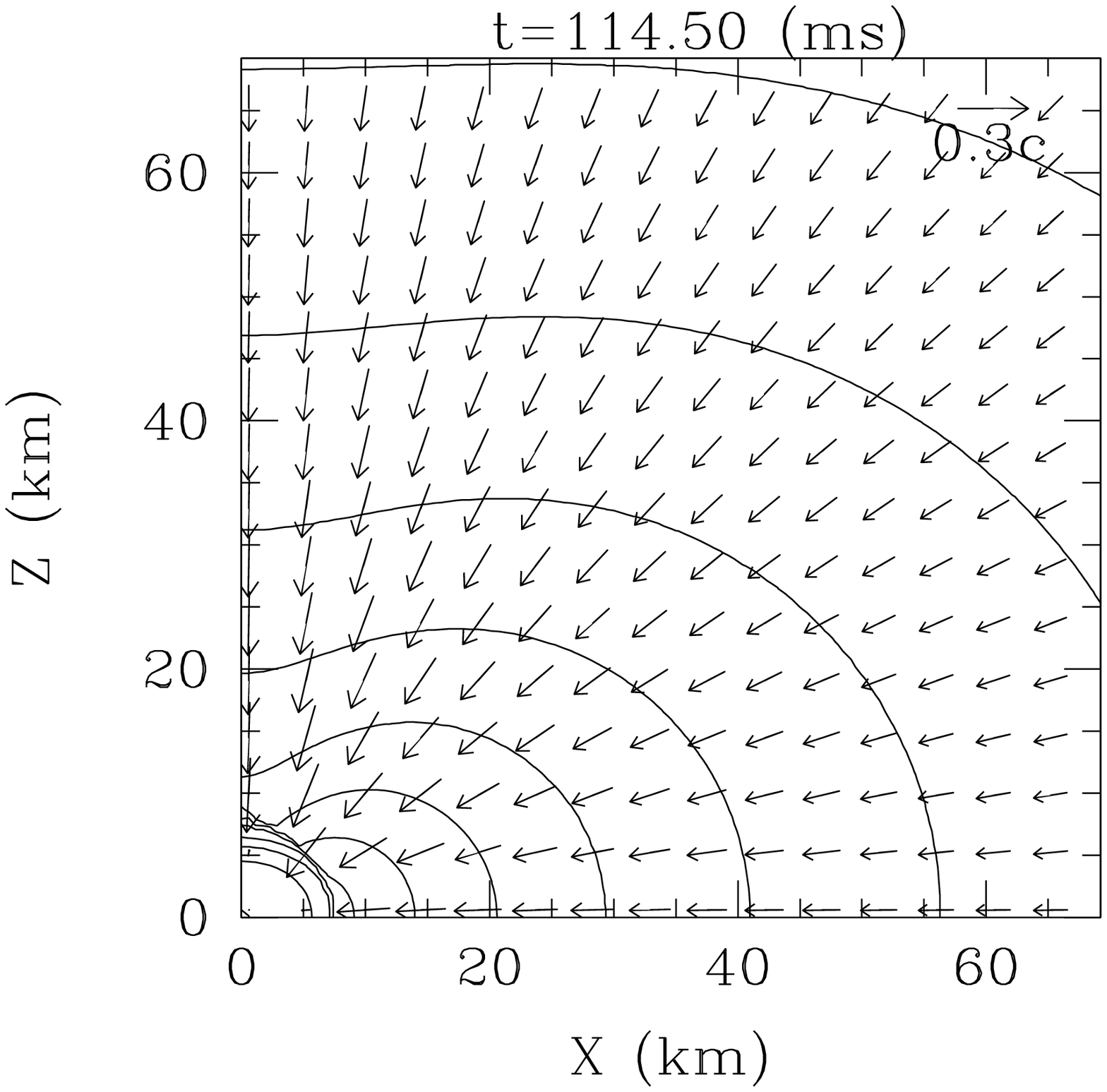} 
\epsfxsize=2.35in
\leavevmode
\epsffile{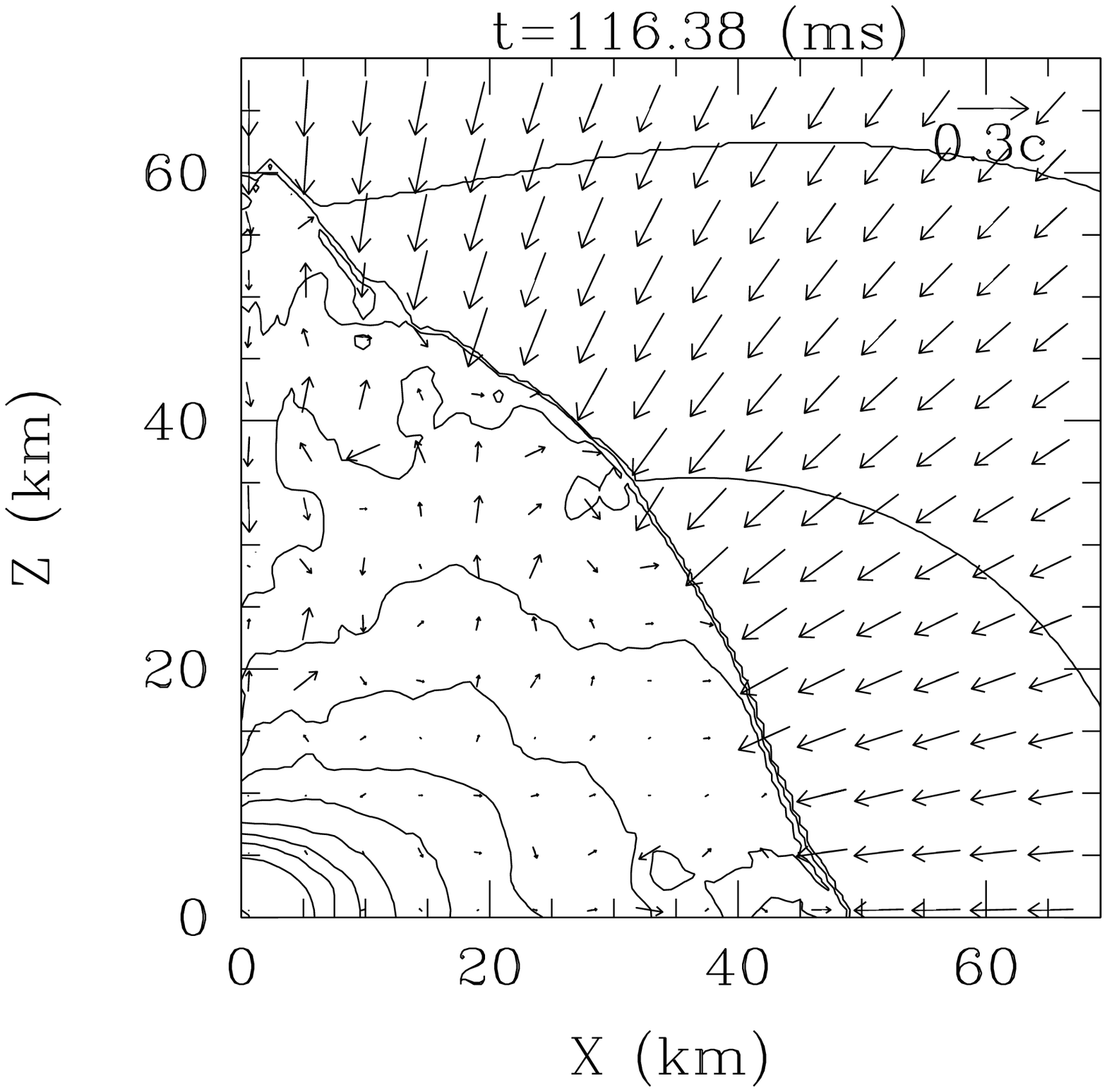} 
\epsfxsize=2.35in
\leavevmode
\epsffile{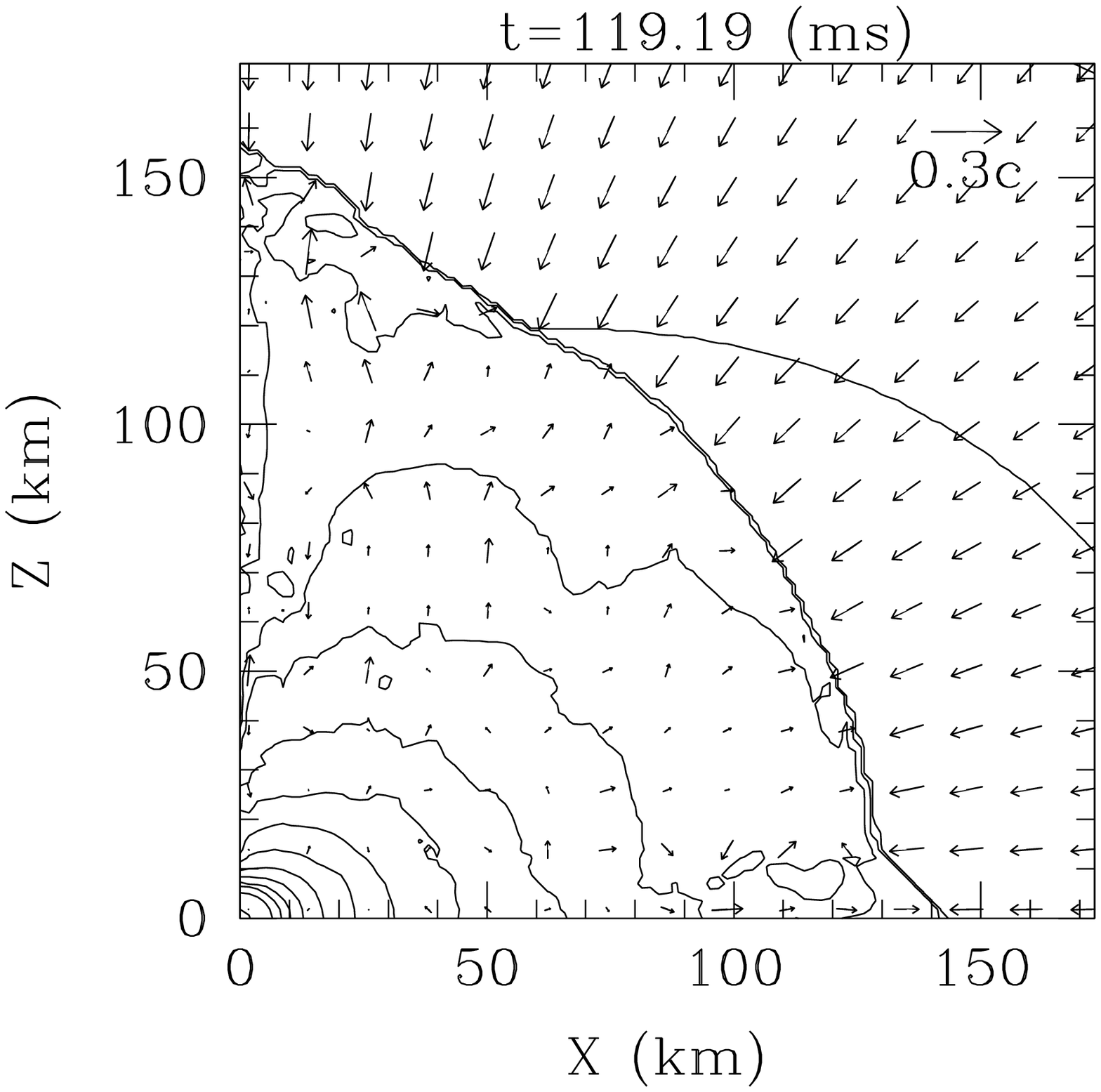} \\
\vspace{-5mm}
\epsfxsize=2.35in
\leavevmode
\epsffile{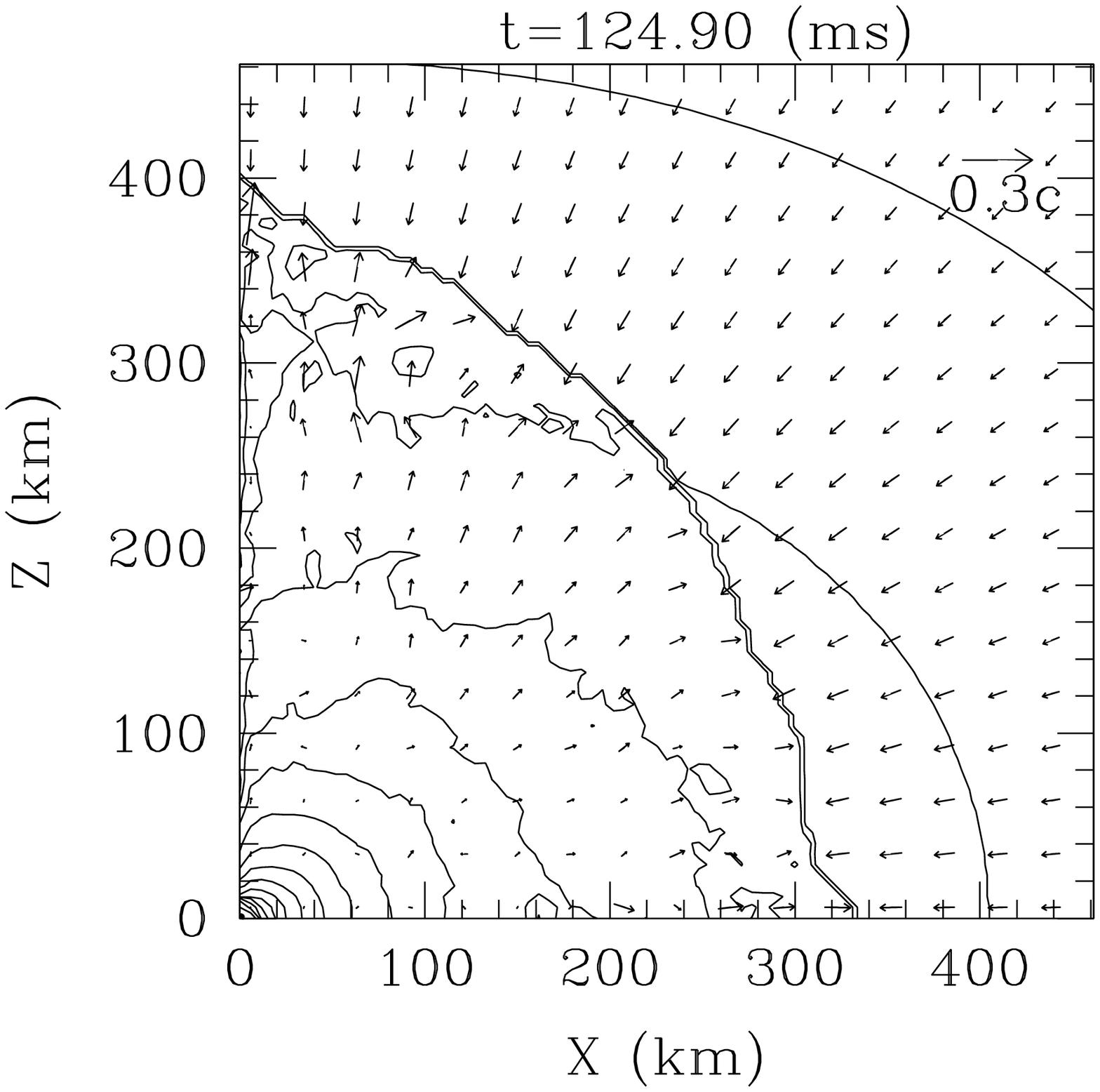} 
\epsfxsize=2.35in
\leavevmode
\epsffile{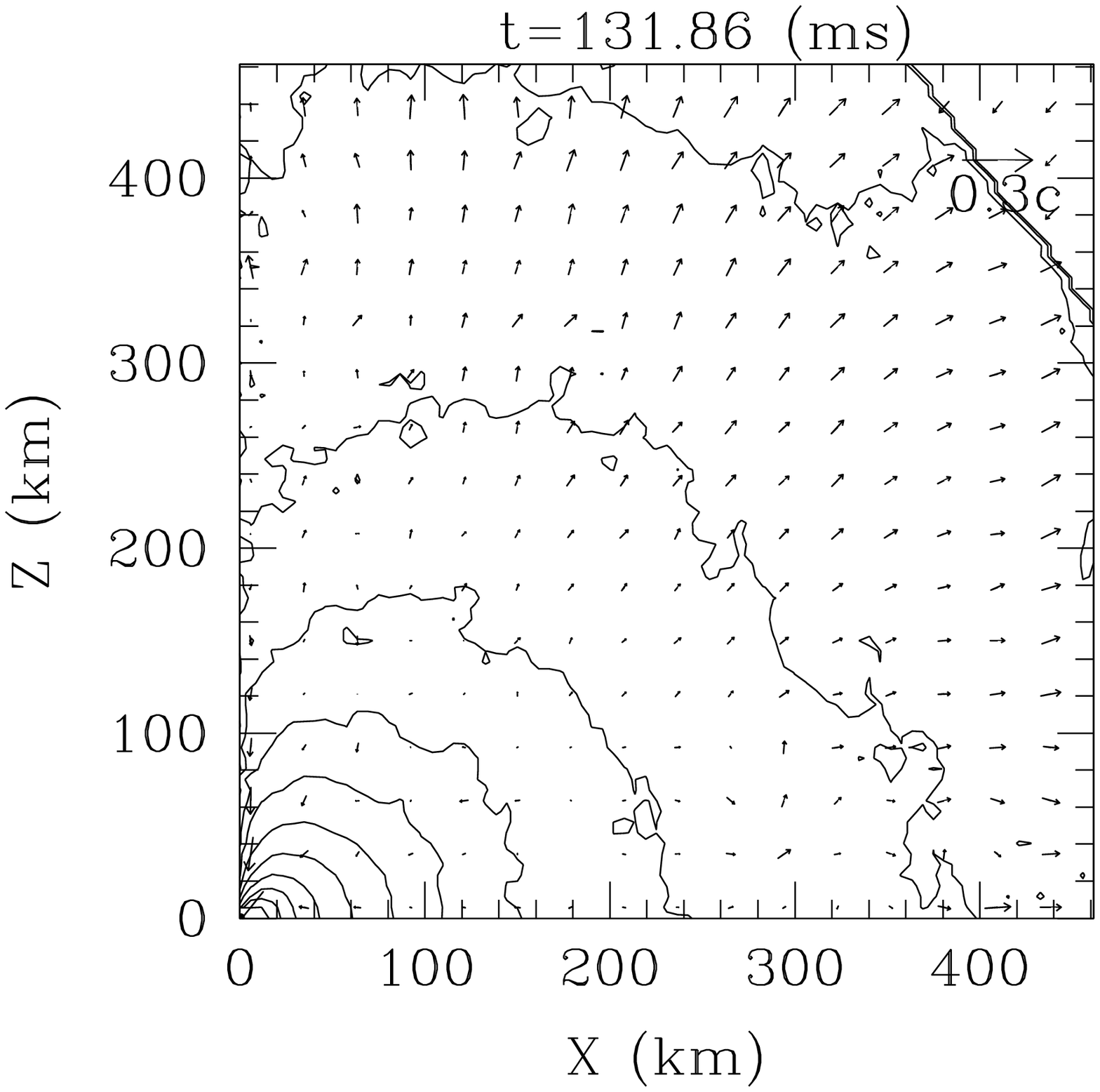}
\epsfxsize=2.35in
\leavevmode
\epsffile{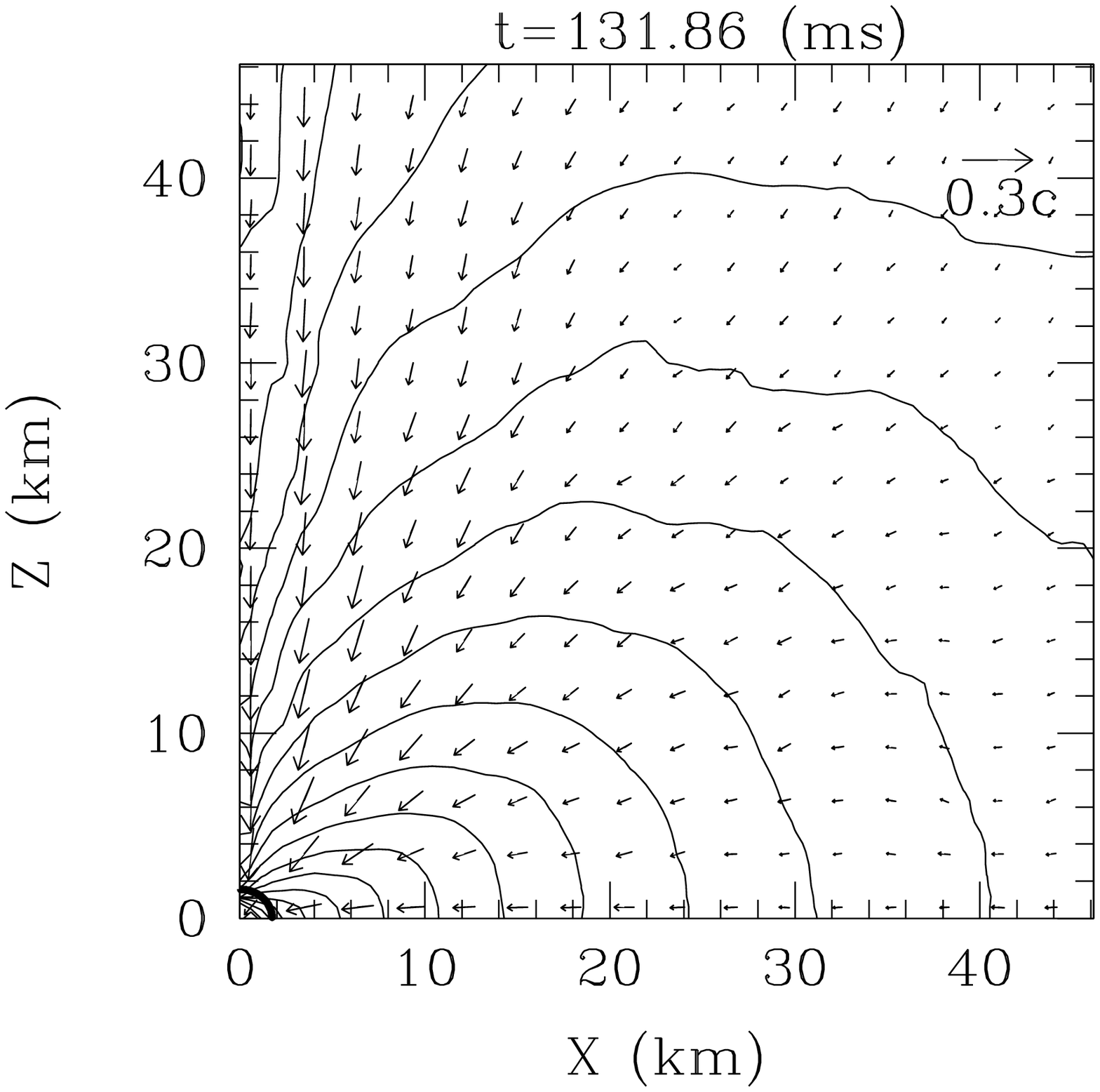}
\end{center}
\vspace{-6mm}
\caption{The same as Fig. \ref{figure7} but for model F5
  with the equation of state 'b', at $t = 114.50$, 116.38, 119.19,
  124.90, and 131.86 ms. The thick solid curve in the last panel
  denotes the location of the apparent horizon.}\label{figure10} 
\end{figure}
\begin{figure}[htb]
\vspace{-4mm}
  \begin{center}
    \epsfxsize=3.in
    \leavevmode
    (a)\epsffile{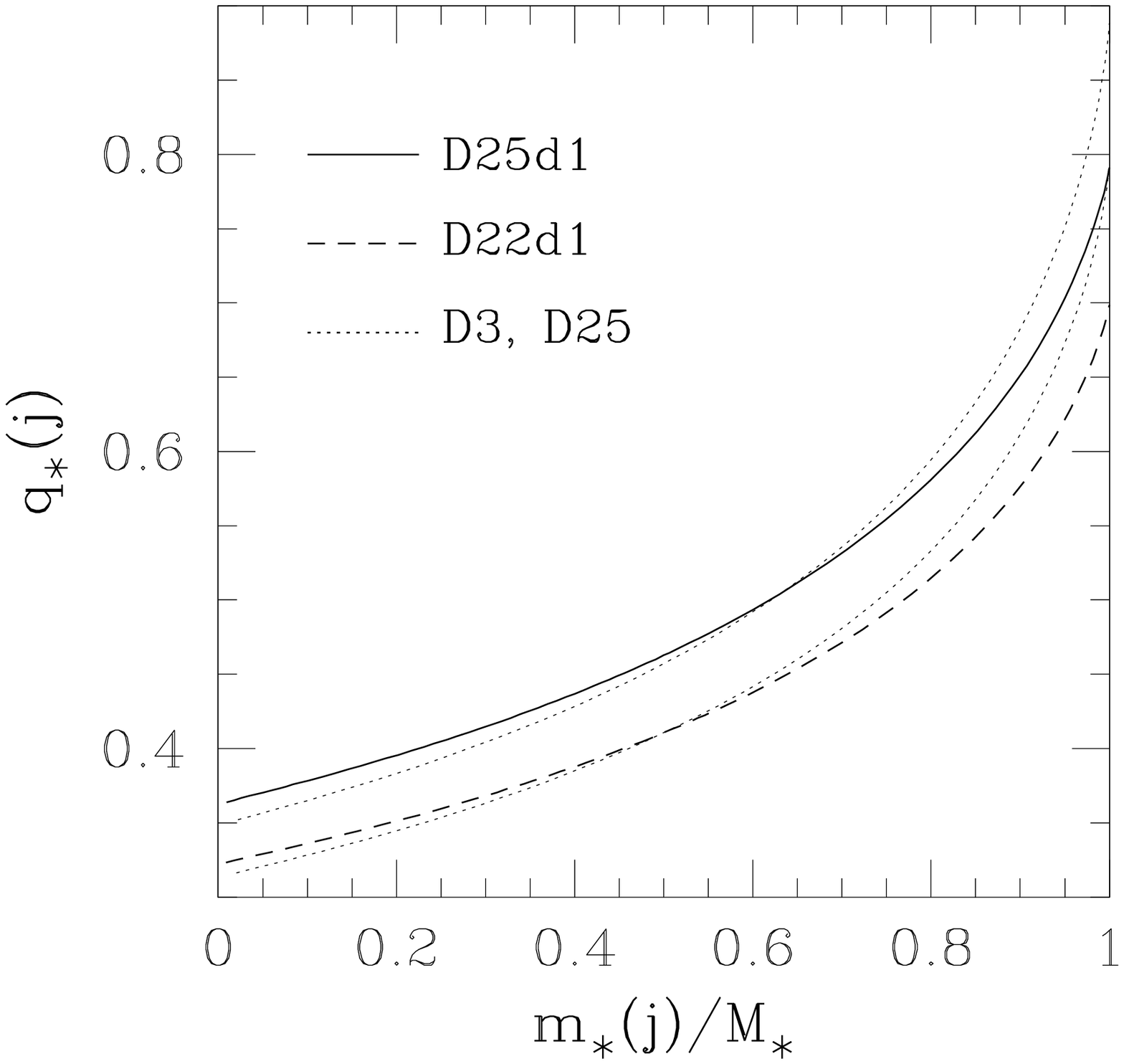} 
    \epsfxsize=3.in
    \leavevmode
    (b)\epsffile{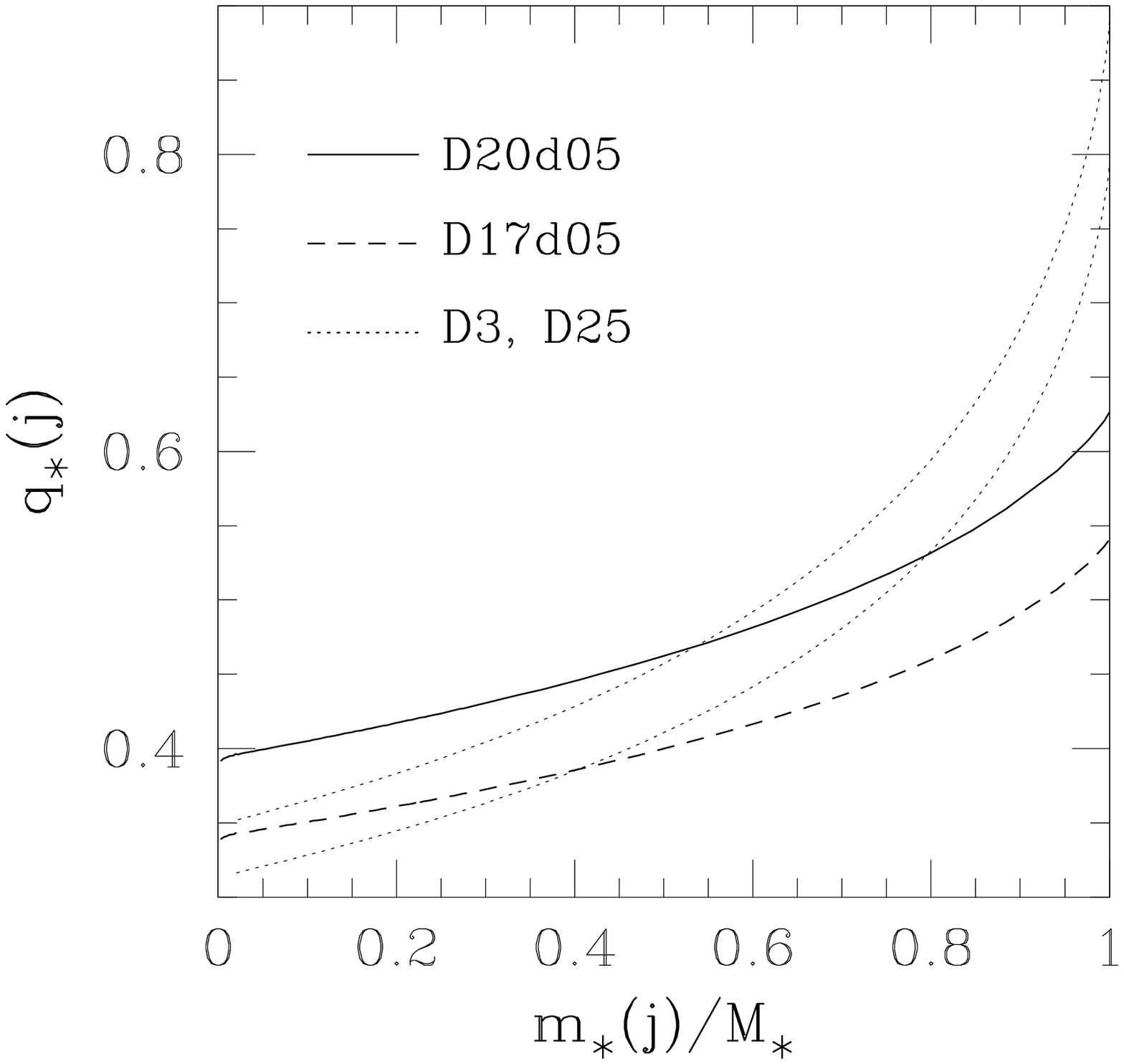} 
  \end{center}
\vspace{-2mm}
  \caption{ The spin parameter distribution inside differentially
    rotating iron cores of 
    (a) D25d1 (solid curve) and D22d1 (dashed curve); (b)
    D20d05 (solid curve) and D17d05 (dashed curve). The two dotted
    curves in both figures
    denote the spin parameter distribution for the rigidly rotating
    models D3 (upper dotted curve) and D25 (lower dotted curve). 
    Note that for model 
    D3b, a black hole is formed promptly, while for D25b, a neutron
    star is formed.
  }\label{figure11}
\end{figure}
\subsection{Criterion for prompt black hole formation}

In Table \ref{Table4}, we summarize the outcomes in the iron core collapse 
for all of rigidly rotating initial models listed in
Table \ref{Table2} and with equations of state listed in Table
\ref{Table1}.
It is important to note that even for models in which a black hole is
not formed promptly (models ``no BH''), 
a longterm fallback of matter may lead to black hole formation.
In this paper, we do not study such longterm black hole formation.
Since the simulations are performed for the iron core
(only part of the whole star), we stop
the simulations when the shock front reaches its surface
located at a radius of $\sim 1000$ km.
To follow the collapse due to the fallback,
it would be necessary to take into account not only the
iron core but the carbon-oxygen envelope. Also, 
neutrino cooling that is ignored in this paper will play a 
role for a longterm fallback with the duration $>100$ ms \cite{LAT}. 

Figure \ref{figure6} illustrates that the threshold mass for the
prompt black hole 
formation depends on the adopted equation of state as well as 
the angular momentum. 
In the following section, we describe, in more detail, the dependence
of the threshold mass on thermal pressure, rotation, differential
rotation, and adopted equations of state, separately.

\subsubsection{Contribution of thermal pressure}\label{effec_shock}

First, we focus on spherical collapse. 
Because of no rotational effect in this case, the threshold mass
depends only on equations of state.
Figure \ref{figure6} shows that 
the threshold mass for the prompt black hole
formation locates between $\approx 2.1M_{\odot}$ (model A0)
and $\approx 2.3M_{\odot}$ (model C0) depending on the adopted
equations of state. This value is by 20--40$\%$ larger than the maximum
allowed mass for the cold spherical polytrope $\approx 1.6M_{\odot}$. 

For the equations of state 'a' and 'b', the threshold mass 
is $\approx 2.1$--$2.2M_{\odot}$. On the other hand,
for the equation of state 'c' in which the value of $\Gamma_{1}$ is the 
same as the equation of state 'b' but the value of $\rho_{\rm nuc}$
is smaller, the threshold mass is $\approx 2.3M_{\odot}$.
This fact suggests that magnitude of the thermal pressure 
$P_{\rm th}$ generated by the shocks is larger for the equations of
state with the smaller value of $\rho_{\rm nuc}$. 
This difference in the strength of the shock results from the
difference in the collapse dynamics. Detailed discussions about
the dynamics of the collapse are presented in the next subsection.

\subsubsection{Dependence on $\Gamma_{1}$ and $\Gamma_{2}$}\label{secGam}

As discussed and illustrated in \cite{Zweg,HD,SS2}, the most
important parameter for the dynamics of the collapse during the infall
phase is $\Gamma_{1}$. 
For the smaller value of $\Gamma_{1}$ 
(for the larger value of $|\Gamma_{1}-4/3|$), the
depleted pressure at $t=0$ is larger. As a result, the collapse is
accelerated more and the elapsed time in the infall phase is shorter. 
Furthermore, since the depleted fraction of the 
pressure is larger in the central region than in the outer region
for the smaller value of $\Gamma_{1}$, the 
collapse in the central region proceeds more rapidly. Accordingly, the
iron core contracts less coherently and only the
collapse in the inner region of the iron core is accelerated. 
Therefore, the mass of the inner core at
the bounce phase is smaller for the smaller value of $\Gamma_{1}$. 
The smaller mass of inner core indicates that the fraction of the iron core
which undergoes the shock heating is larger since the shock wave
is generated at the outer edge of the inner core.
Therefore, the contribution of the thermal pressure to the inner core
is more important for models with the smaller value of $\Gamma_{1}$. 

Since the mass of the inner core in the infall and bounce phases is
larger for the larger value of $\Gamma_{1}$,
the degree of the overshooting of the inner core at
the bounce phase is larger, due to the stronger gravitational attraction
force. Also, for the smaller value of $\Gamma_{2}$, the degree of the
overshooting at the bounce phase is larger since a relatively higher density is
required to supply the sufficient pressure to halt the collapse. 
In the case that mass of the inner core is not so large as to collapse
to a black hole, the larger overshooting results in stronger
shock generation, since the stored energy of the inner core at the
bounce becomes larger (cf. Sec. \ref{Bipolar}).
However, the larger overshooting also increases the risk of
collapsing to a black hole since the compactness (the value of
$\alpha_{c}$) at the bounce becomes larger (smaller). 

To understand the dependence of the collapse on equations of state in more
details, we generate Figs. \ref{figure7} and \ref{figure8},
in which we display the snapshots of the density contour curves and
velocity vectors in the $x$-$z$ plane for models D2a and D2b. 
For models with the equation of state 'a' in which $\Gamma_{1} =
1.32$ and $\Gamma_{2} = 2.25$, the collapse proceeds coherently
due to the fact that $\Gamma_{1}$ is close to 4/3, and hence, 
the mass of the inner core increases to be larger than the maximum
allowed mass of the corresponding rotating neutron star at the bounce. 
Consequently, the inner core promptly collapses to a
black hole without generating shocks (cf. Fig. \ref{figure7}).
This result indicates that for the smaller value of $|\Gamma_{1}-4/3|$,
the prompt black hole formation is more liable 
and the threshold mass of the iron core for the prompt black hole
formation is smaller. 

For the equations of state 'b' and 'd', on the other hand, 
the collapse proceeds less coherently due to the smaller
values of $\Gamma_{1}$ (the larger values of $|\Gamma_{1}-4/3|$). 
As a result, the mass of the inner core at the bounce is below the maximum
allowed mass, and hence,
the inner core experiences a bounce and forms a compact protoneutron
star for a while before collapsing to a black hole.
Also, the shock is formed at the outer edge of 
the inner core and then propagates outward (cf. Fig. \ref{figure8}). 
This implies that the thermal pressure which helps supporting the
inner core is generated in contrast to the case for model D2a. 
Due to this difference, the threshold mass of the iron core for the prompt
black hole formation with the equation of state 'b' is larger than
that for 'a'. 
However, the shock formed is not strong enough to blow up the sufficient
matter outward even for model D2b: A part of 
the matter in the outer region falls into the inner core located
at the center. This fall-back leads to an accretion induced collapse 
to a black hole (see the third panel Fig. \ref{figure8}). 

On the other hand, we do not find significant difference in
the threshold mass for the prompt black hole formation between
the equations of state 'b' and 'd' (see table \ref{Table4}).
The plausible reason is described as follows. 
First, since the mass of the inner core formed at the bounce is smaller for
models with the equation of state 'd', the larger fraction of the
matter inside the  
{\it protoneutron star} will experience the shock heating. 
On the other hand, the shock wave itself is weaker for
the equation of state 'd' since the amplitude of the core bounce and
the value of $\Gamma_{\rm th}$ ($=\Gamma_{1}$) is smaller.
These two contrary effects cancel each other resulting in approximately 
the same threshold mass for the prompt black hole formation.
A more sophisticated parameter study is required to clarify the small
difference in the threshold mass.

\subsubsection{Dependence on  $\rho_{\rm nuc}$}\label{dep_rhonuc}

In Fig. \ref{figure4}, we show the evolution of $\rho_{c}$ and
$\alpha_{c}$ for models D1c and D15c, for which a black hole is formed
promptly as a result of the collapse. For comparison, the result for
model D1b is shown together.
The evolution of $\rho_{c}$ and $\alpha_{c}$ with two
equations of state 'b' and 'c' are identical during the infall phase. 
However, in the bounce and subsequent phases, increase of $\rho_{c}$
(decrease of $\alpha_{c}$) is delayed and the value of $\rho_{c}$
($\alpha_{c}$) is smaller (larger) for the equation of state 'c'. 
This is a result from the fact that 
the sudden stiffening occurs at an earlier stage of the collapse
with the equation of state 'c' due to the smaller value of $\rho_{\rm nuc}$. 

Figure \ref{figure4} also indicates that the amplitude of oscillation
of the inner core at the bounce is larger for models with the equation of
state 'c'. Several effects are responsible for this result. First,
the small value of $\rho_{\rm nuc}$ in the equation of state 'c'
results in a smaller mass of the inner core and smaller averaged density
at the bounce. As mentioned in Sec. \ref{secGam}, the smaller
mass of the inner core results in the fact that a larger amount
of the matter experiences the shock heating. 
The smaller density implies that the pressure
at the surface of the inner core is smaller, and hence, 
the work exhausted by the inner core in converting the oscillation
energy to the infalling outer envelop is smaller. 
Accordingly, the amplitudes of the bounce and the
subsequent oscillation of the inner core become larger. 
Also, the equation of state 'c' is 
``stiffer'' than 'b' in the (higher) density range between  $\rho
\approx 1.0 \times 10^{14}$ and $\approx 2 \times 10^{15}$ g/cm$^{3}$ (cf. 
Fig. \ref{figure1}). 
This also contributes to generating stronger shock waves. 
Due to these effects, the smaller value of $\rho_{\rm nuc}$ results
in the larger threshold mass for the prompt black hole formation.

\subsubsection{Effects of the rotation}\label{EffecRot}

As the value of $q$ increases, the threshold mass 
for the prompt black hole formation becomes larger
because the effect of the rotation effectively supplies additional
pressure to the iron core and reduces the amount of matter falling into the
central region. The threshold mass may be written 
approximately as a quadratic form \cite{Shiba2000}: $M_{\rm threshold} =
k~q^{2} + M_{0}$, where $M_{0}$ is the threshold mass for
the spherical case. This reflects
that the rotational kinetic energy (or the centrifugal force)
depends on $q^{2}$. Assuming this form of the threshold
mass, we approximately draw the threshold curves in Fig. \ref{figure6}. 
It is found that the rotational effect increases the
threshold mass by $\alt 15$--$25\%$ for $q \sim 1$. 
This value is comparable with the amplification factor of the
maximum mass for rigidly rotating neutron stars in equilibrium \cite{CST}.

The coefficient, $k$, depends on the strength of the shock and the
adopted equation of state. 
The threshold mass for the prompt black hole formation
at the maximum value of $q$ is larger
for the models with the smaller value of $\rho_{\rm nuc}$ as discussed in
the previous subsection 
(compare models with equations of state 'b' and 'c' in Table
\ref{Table4}). However, the dependence of the 
threshold mass at the maximum value of $q$
on $\Gamma_{1}$ is not very simple. Comparing models with
equations of state 'a' and 'b', the black hole is more liable to be formed 
with equation of state 'a', i.e., for the {\it larger} value of
$\Gamma_{1}$. On the other hand, comparing models with 
equations of state 'b' and 'd', no difference is found in the present
numerical study.
This is because the process of the black hole formation 
for models with $\Gamma_{1} = 1.32$ is significantly different from
those with $\Gamma_{1} = 1.30$ and 1.28, as described in
Sec. \ref{secGam}.

To see the effect of the rotation on the collapse dynamics in more detail, 
in Figs. \ref{figure9} and \ref{figure10}, 
we display the snapshots of the density contour curves and
velocity vectors at selected time slices
for models F5a and F5b. For these models, the value of $q$ is
nearly maximum among the rigidly rotating initial models of
a given mass. 
For model F5a, the collapse proceeds more rapidly in the direction of
rotational axis ($z$ axis) than in the equatorial plane during the infall
phase. This is because the centrifugal force is stronger in the equatorial
plane than around the $z$ axis (see the first panel of Fig. \ref{figure9}).
Accordingly, the collapsing inner region is deformed to be an oblate
shape. As the collapse proceeds, the density contour in the outer part
of the inner core is deformed to be a concave structure.
At the bounce, this leads to formation of a steep density gradient 
along the rotational axis (see the second panel of Fig. \ref{figure9}).
This would result in stronger shock waves along the
rotational axis than in the equatorial plane.
For model F5a, however, the mass of the inner core at the bounce
is so large that the strong gravitational attraction force prevents
the shock from being propagated. Consequently, the inner core promptly 
collapses to a black hole.
On the other hand, for less massive cases, shock waves propagate
outward. In the shock propagation, this asymmetry generates
anisotropic shocks (see also Sec. \ref{NSform}).

For model F5b, shock waves propagate outward since 
the mass of the inner core at the bounce is smaller than the maximum
allowed mass due to the smaller value of $\Gamma_{1}$. 
In this case, however, the density gradient along the
$z$ axis is not so steep as that for model F5a (see the first panel of Fig.
\ref{figure10}), although the mass and the angular momentum of the
progenitor are identical between F5a and F5b. As a result, the
asymmetry in the shock front is small (see the second to fourth panels
of figure \ref{figure10}). The reason is that the mass of the inner core at
the bounce is smaller for model F5b. For rigidly rotating initial
models, the centrifugal force is stronger for larger cylindrical
radius, and hence, the smaller mass of the inner core 
implies that the effect of rotation is less important.

For model F5b, the shock is strong enough to reach the
surface of the iron core (see the fifth panel of Fig. \ref{figure10}).
However, in a region behind the shock, the fluid elements (in
particular) around the rotational axis fall back into a protoneutron
star formed at the center, and eventually, the protoneutron star
collapses to a black hole (see the last panel in Fig. \ref{figure10}). 
This illustrates that a black hole may be
formed in a rather long time scale even for models in which the black
hole is not formed promptly, if the progenitor is rotating. 

\subsubsection{Effects of differential rotation}

Table \ref{Table5} shows the outcomes of the collapse for all of
differentially rotating initial models listed in Table \ref{Table3}.
In the simulation for differentially rotating models, we only
adopt the equation of state 'b'. Model D25d1 with $A = 1.0$  
does not form a black hole, while its rigidly rotating
counterpart D25 for which the mass and the angular momentum are
approximately the same as those of D25d1 collapses to a black hole.
For differentially rotating model D20d05
for which $A=0.5$, and the mass and the angular momentum
are slightly smaller than those for rigidly rotating model D2, 
black hole is not formed.
For $A=0.5$, black hole is not formed promptly even from a very massive
initial model H in which $M_{\rm ADM} \approx 3.0M_{\odot}$
(cf. Table \ref{Table5}). These indicate that the black hole is
less liable to be formed promptly for the higher degree of
differential rotation. 

The results found here are quite reasonable since with the decrease
of $A$, the angular velocity in the inner region increases. 
To clarify this effect in a more quantitative manner, we generate
Fig. \ref{figure11}, which 
shows the spin parameter distribution 
$q_{\ast}(j)$ defined by Eq. (\ref{spin_distri}).
First, note that a black hole is formed for models D25 and D22d1
but not for D3 and D25d1 with the equation of state 'b'. 
The two dotted curves in Figs. \ref{figure11}(a) and \ref{figure11}(b) 
denote $q_{\ast}(j)$ of these two models D3b (upper dotted curve)
and D25b (lower dotted curve), respectively. 
The solid and dashed curves in Fig. \ref{figure11}(a) denote 
$q_{\ast}(j)$ for models D25d1 and D22d1 for which $A=1.0$. 
Although the total value of the spin parameter of the 
differentially rotating iron cores of D25d1 ($q=0.791$) is smaller than that 
for rigidly rotating model D3 ($q=0.888$), the value of $q_*$ 
around the central region ($m_{\ast}(j) \alt M_{\ast}/2$)
exceeds that of D3. As shown in \cite{SS3} in detail, 
the value of $q_*$ around the center plays a crucial role in 
determining the criterion for prompt black hole formation; i.e,
the iron core with a larger value of $q_*$ around the center
is less liable to collapse to a black hole. 
Therefore, it is reasonable that black hole is not formed for model D25d1. 
On the other hand, the value of $q_*$ around the center for 
model D22d1 ($q=0.698$) is as large as that 
for model D25 around the central region. This implies the rotational
centrifugal force is not strong enough to prevent the iron core from
collapsing to a black hole.

Figure \ref{figure11}(b) shows $q_{\ast}(j)$ for models D20d05 (solid 
curve) and D17d05 (dashed curve) with $A=0.5$. A black hole
is not formed for D20d025 while it is formed for D17d05. 
Figure \ref{figure11}(b) clarifies 
that the spin parameter distribution is flatter with
the smaller value of $A$. Therefore, the values of the 
global spin parameter $q$ for $A=0.5$ are much smaller than those of
the rigidly rotating counterparts. On the other hand, 
the value of $q_*$ around the central region becomes larger. 
As a result, black hole formation is more effectively prevented. 
This quantitatively indicates that the threshold mass for the
prompt black hole formation is larger for models with larger degree of
differential rotation.

\subsection{Prediction of the final system}\label{Prediction}

\begin{figure}[htb]
  \vspace{-4mm}
  \begin{center} 
    \epsfxsize=3.in
    \leavevmode
    \epsffile{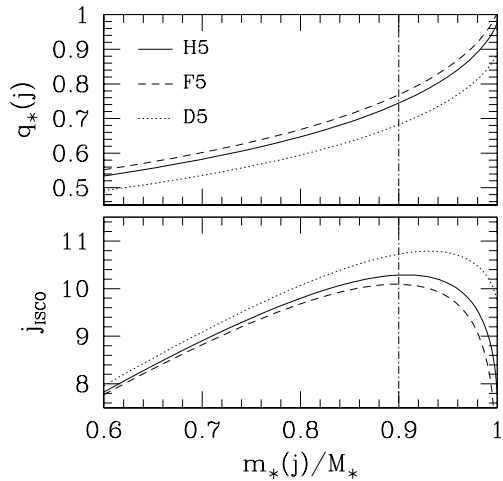}
  \end{center}
  \vspace{-7mm}
  \caption{The distribution of $q(j)$ (top panel) and the value of
  $j_{\rm ISCO}$ as a function of $m_{\ast}(j)/M_{\ast}$ for models H5
  (solid curve), F5 (dashed curve), and D3 (dotted curve). The dotted
  dashed vertical line denotes $m_{\ast}(j)/M_{\ast} = 0.9$ where
  $j_{\rm ISCO}$ for F5 takes the maximum value
  approximately.}\label{figure12}
\end{figure}

Because of the assumption that the viscous effect is negligible during the
collapse, the specific angular momentum $j=\hat{u}_{\varphi}$ of each fluid
element is conserved in the axisymmetric system and so do the
functions $m_*(j)$ and $J_*(j)$. 
Using this fact, the final outcome after the black hole formation
can be predicted \cite{ShiSha,ShibaSA,SS3}. 

Let us consider innermost stable circular orbit (ISCO) around
the growing black hole located at the center.
Assuming that a fluid element around a black hole is approximately
in a circular orbit, it will fall into the seed black hole eventually
if its value of $j$ is smaller than that at the ISCO ($j_{\rm ISCO}$). 
The value of  $j_{\rm ISCO}$ depends on the mass and the angular momentum
of the black hole and changes as 
the ambient fluid elements accrete onto the black hole. 
If $j_{\rm ISCO}$ increases as a result of the accretion, the more ambient
fluid elements will fall into the black hole.
On the other hand, if $j_{\rm ISCO}$ decreases during the accretion,
no more fluid element will fall into the black
hole, and as a result, the dynamical growth of the black hole will
terminate. Therefore, if $j_{\rm ISCO}$ has a maximum (hereafter denoted as 
$\jisom$), the black hole will grow until $j$ reaches $\jisom$. 
Namely, it is reasonable to expect that 
the final values of the mass and the spin parameter of the black hole
will be 
\beqn
&&M_{\rm BH} \approx m_{\ast}(\jisom), 
\label{approx-MBH} \\
&&q_{\rm BH} \approx q_{\ast}(\jisom)
\label{approx-JBH}.
\eeqn
with the disk mass $M_{\rm ADM}-m_*(\jisom)$.

To estimate the value of $j_{\rm ISCO}$, we assume that the spacetime
metric can be instantaneously approximated by that of a Kerr spacetime
of the mass $m_*(j)$ and the spin $q_*(j)$. 
On these approximations, we can compute $j_{\rm ISCO}$ of a 
growing black hole as \cite{BPT,ST}, 
\beq \label{def-jISCO}
 j_{\rm ISCO} = 
\frac{ \sqrt{m_{\ast}(j) \riso}
\left( \riso^{2} - 2 q_{\ast}(j)m_{\ast}(j) \sqrt{m_{\ast}(j) \riso} 
+ (q_{\ast}(j)m_{\ast}(j))^{2}\right)}
{\riso \left(\riso^{2} -3m_{\ast}(j) \riso 
+ 2 q_{\ast}(j)m_{\ast}(j) \sqrt{m_{\ast}(j) \riso}\right)^{1/2}} , 
\eeq
where
\beqn
&& \riso = m_{\ast}(j) \left[ 3 + Z_{2} -
	 \left\{(3-Z_{1})(3+Z_{1}+2Z_{2})\right\}^{1/2} \right],
\nonumber \\
&& \ \ Z_{1} = 1 
+ \left[1- q_{\ast}(j)^{2} \right]^{1/3}
\left[ \{1+ q_{\ast}(j)\}^{1/3} + \{1-q_{\ast}(j)\}^{1/3}\right], \nonumber \\
&& \ \ Z_{2} = \left[ 3q_{\ast}(j)^{2} + Z_{1}^{2} \right]^{1/2} . \nonumber 
\eeqn

In Fig. \ref{figure12}, we show the quasi-local spin parameter
distributions inside the iron core, defined by
Eq. (\ref{spin_distri}), and the value of $j_{\rm ISCO}$
evaluated by Eq. (\ref{def-jISCO}) for models H5, F5, and D3. 
As described above, the dynamical evolution of the formed black
hole will terminate when $j_{\rm ISCO}$ reaches a local 
maximum. Then, the mass, $M_{\rm BH}$, and spin, $q_{\rm BH}$, of the
the dynamically evolved black hole are given by
Eqs. (\ref{approx-MBH}) and (\ref{approx-JBH}).

Figure \ref{figure12} indicates that for model F5 and H5, the inner
region collapse to form a black hole of  
$M_{\rm BH}/M_{\ast} \approx 0.90$ and $q_{\rm BH} \approx 0.76$. 
On the other hand, the matter in the outer region of high
specific angular momentum will form a massive disk of 
$M_{\rm disk}/M_{\ast} \approx 0.10$ around the black hole. 
Similarly, for model D3, a black hole of $M_{\rm BH}/M_{\ast} \approx
0.93$ and $q_{\rm BH} \approx 0.73$ will be formed. 
Such a rapidly rotating black hole surrounded by a massive 
disk is one of the promising candidates for the central engine of the 
long-duration gamma-ray bursts \cite{Woosley,Paczynski,MW}.

Although the above prediction is quite reasonable \cite{SS3},
confirmation of this prediction requires 
to carry out a simulation until the black hole plus disk
system is formed. To accomplish this, the so-called black hole excision
techniques are required. For the case of the collapse with stiff
($\Gamma = 2.0$) equations of state, we have confirmed it
possible to continue the simulation
more than $\sim 100M$ after the formation of black hole by using
techniques based on  the so-called simple excision developed by
Alcubierre and Br\"ugmann \cite{AB}. We plan to wrestle with the
confirmation of the above prediction by extending the techniques to
the collapse with soft equations of state.

\subsection{Protoneutron star formation}\label{NSform}
%
\begin{figure}[p]
\vspace{-4mm}
\begin{center}
\epsfxsize=2.35in
\leavevmode
\epsffile{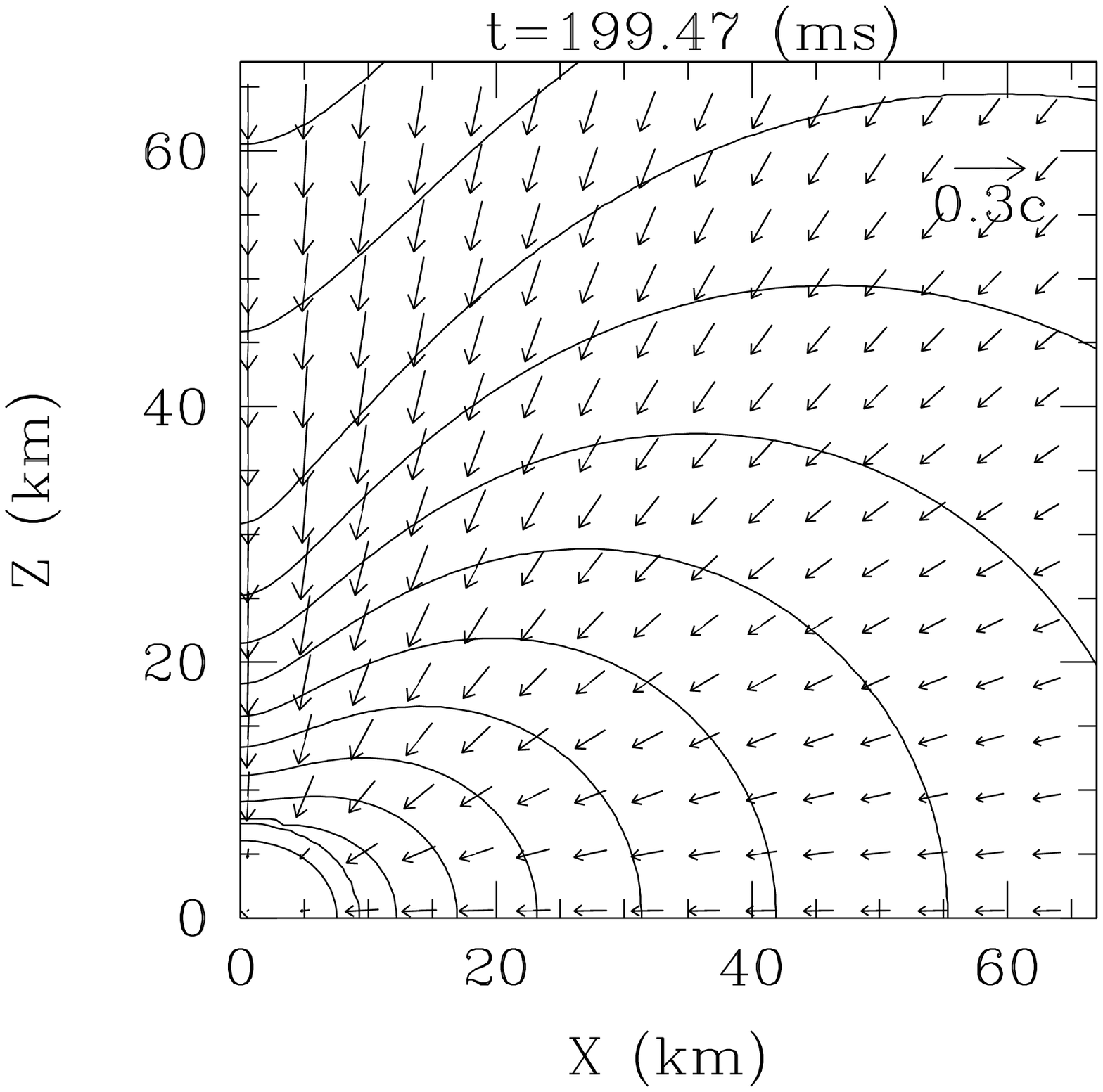} 
\epsfxsize=2.35in
\leavevmode
\epsffile{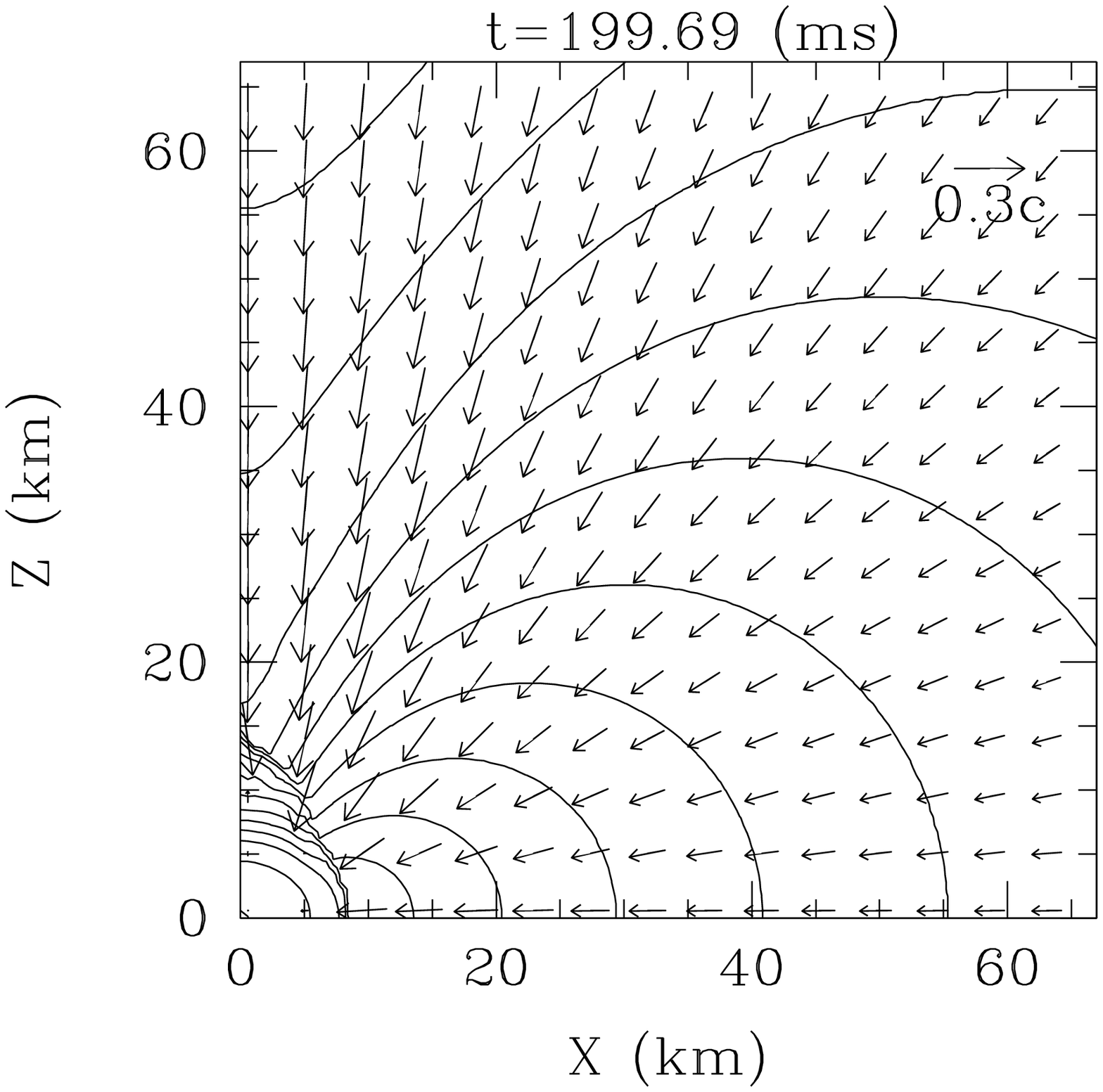} 
\epsfxsize=2.35in
\leavevmode
\epsffile{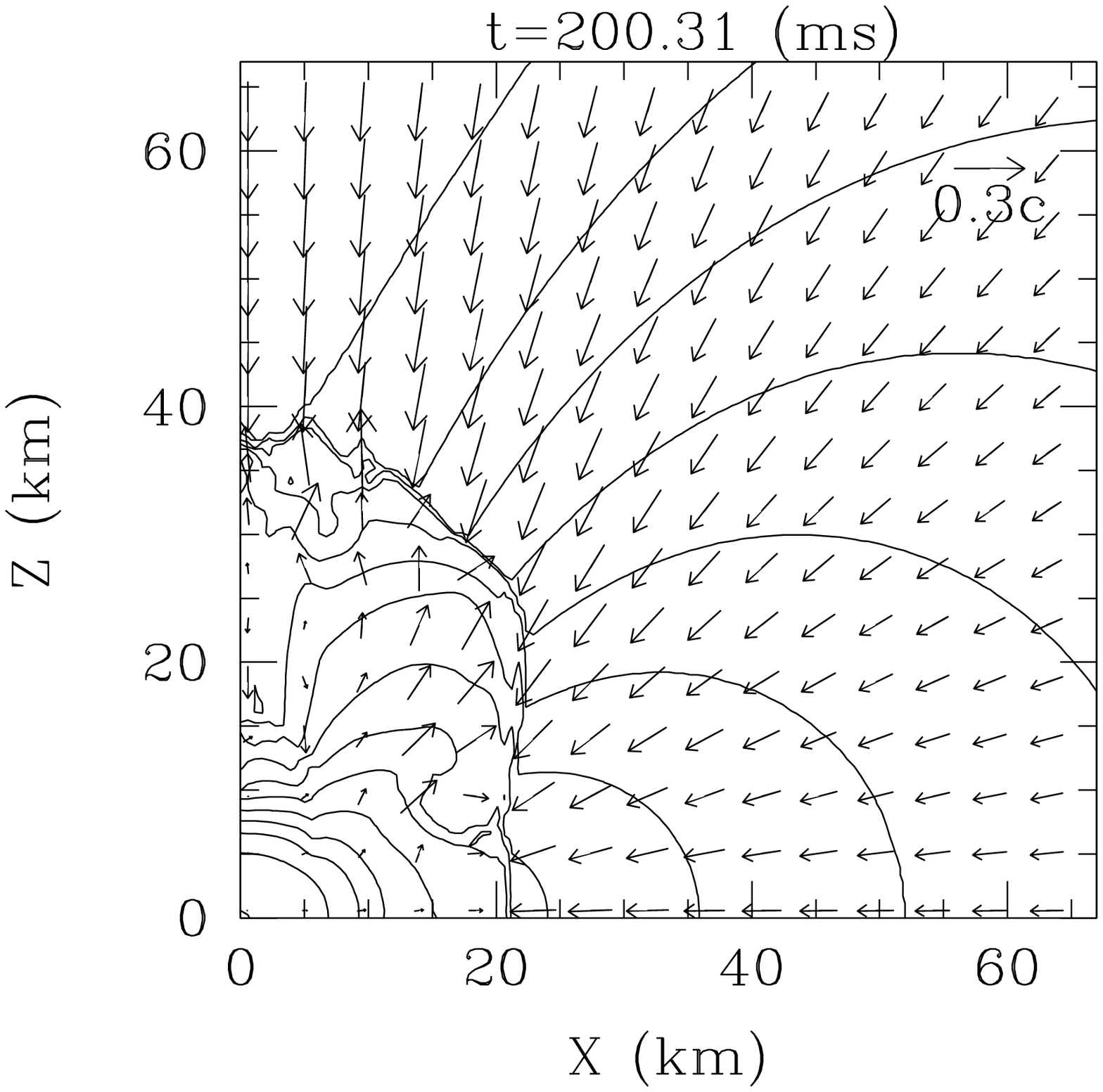} \\
\vspace{-5mm}
\epsfxsize=2.35in
\leavevmode
\epsffile{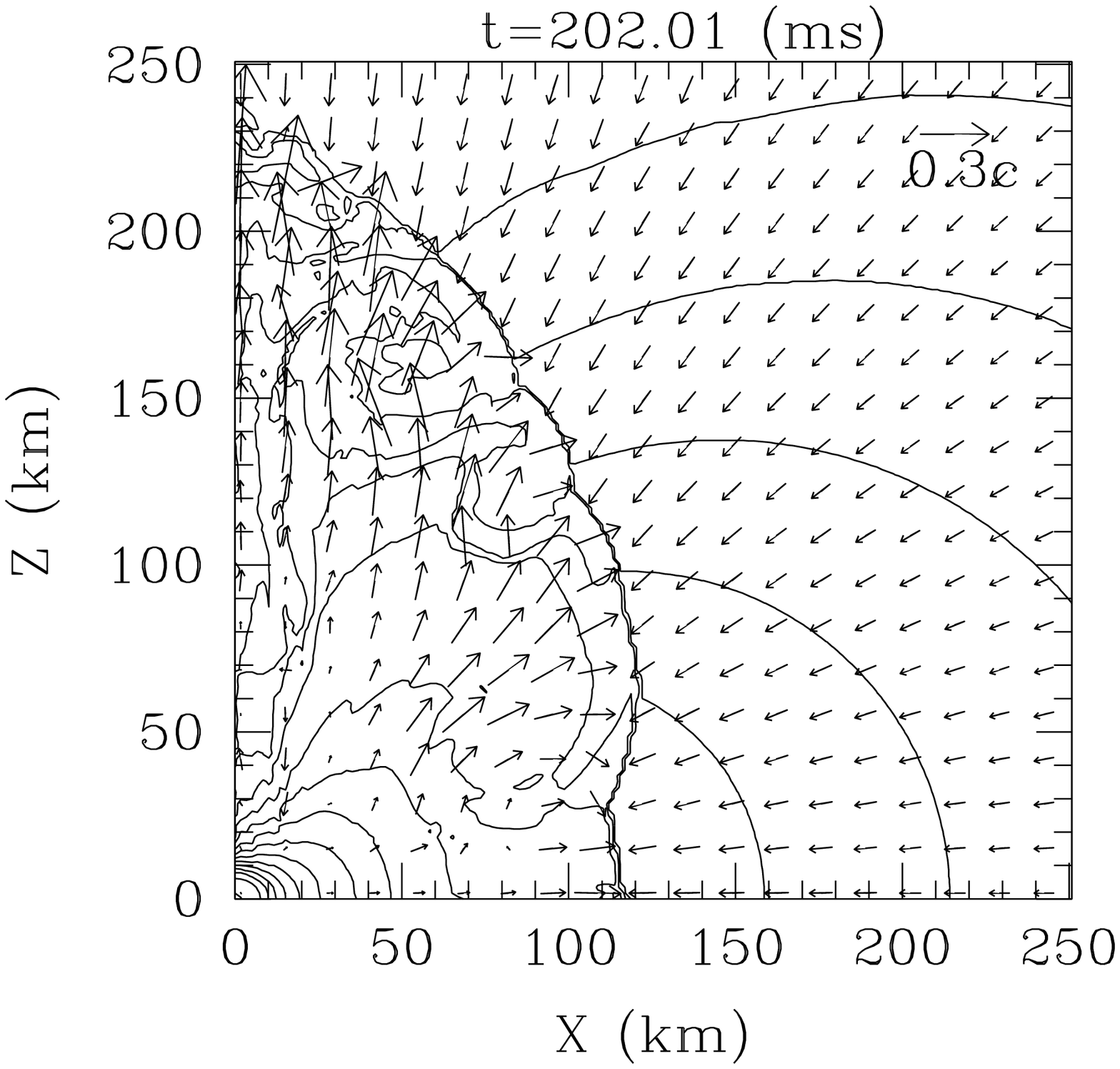} 
\epsfxsize=2.35in
\leavevmode
\epsffile{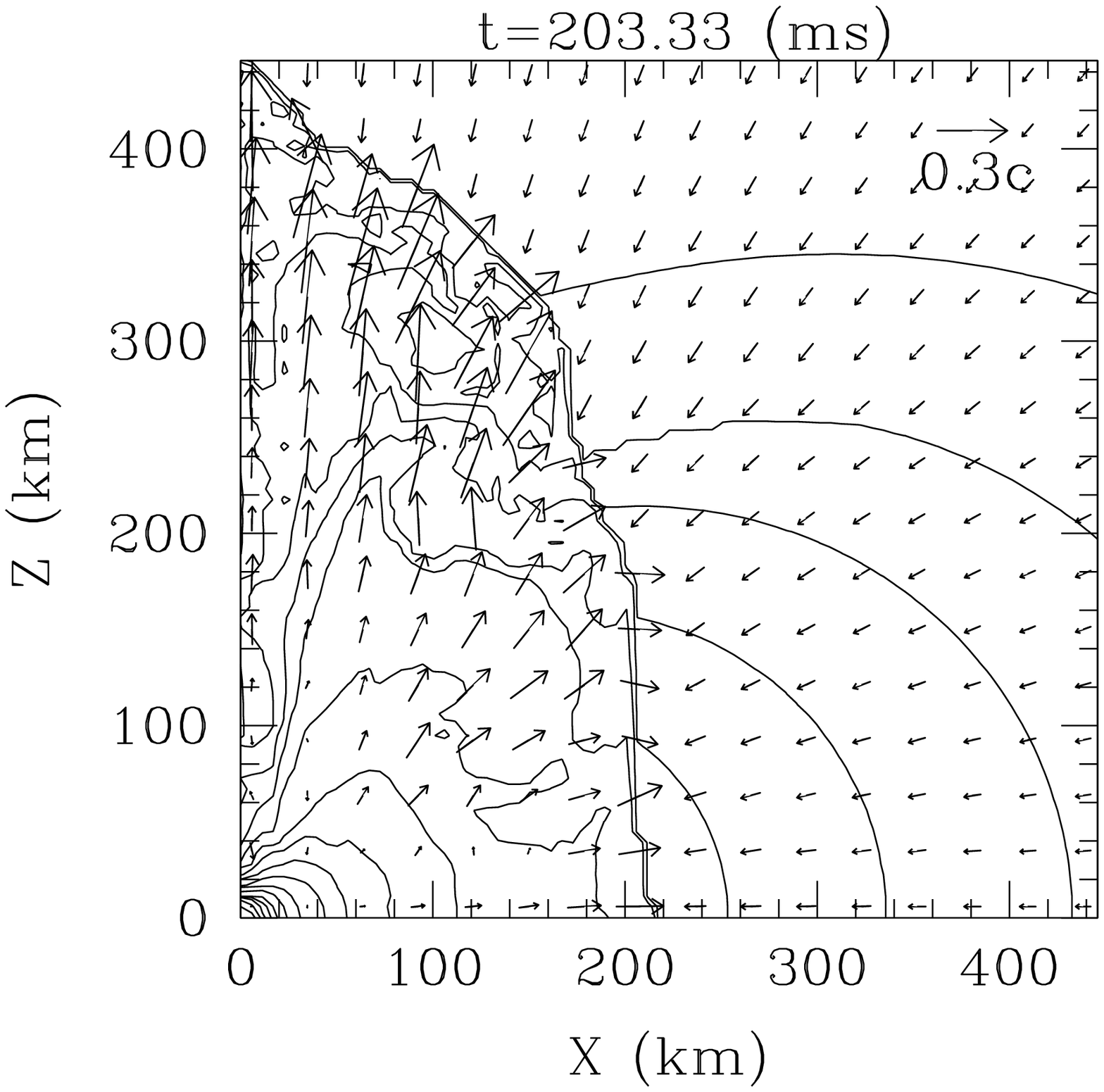}
\epsfxsize=2.35in
\leavevmode
\epsffile{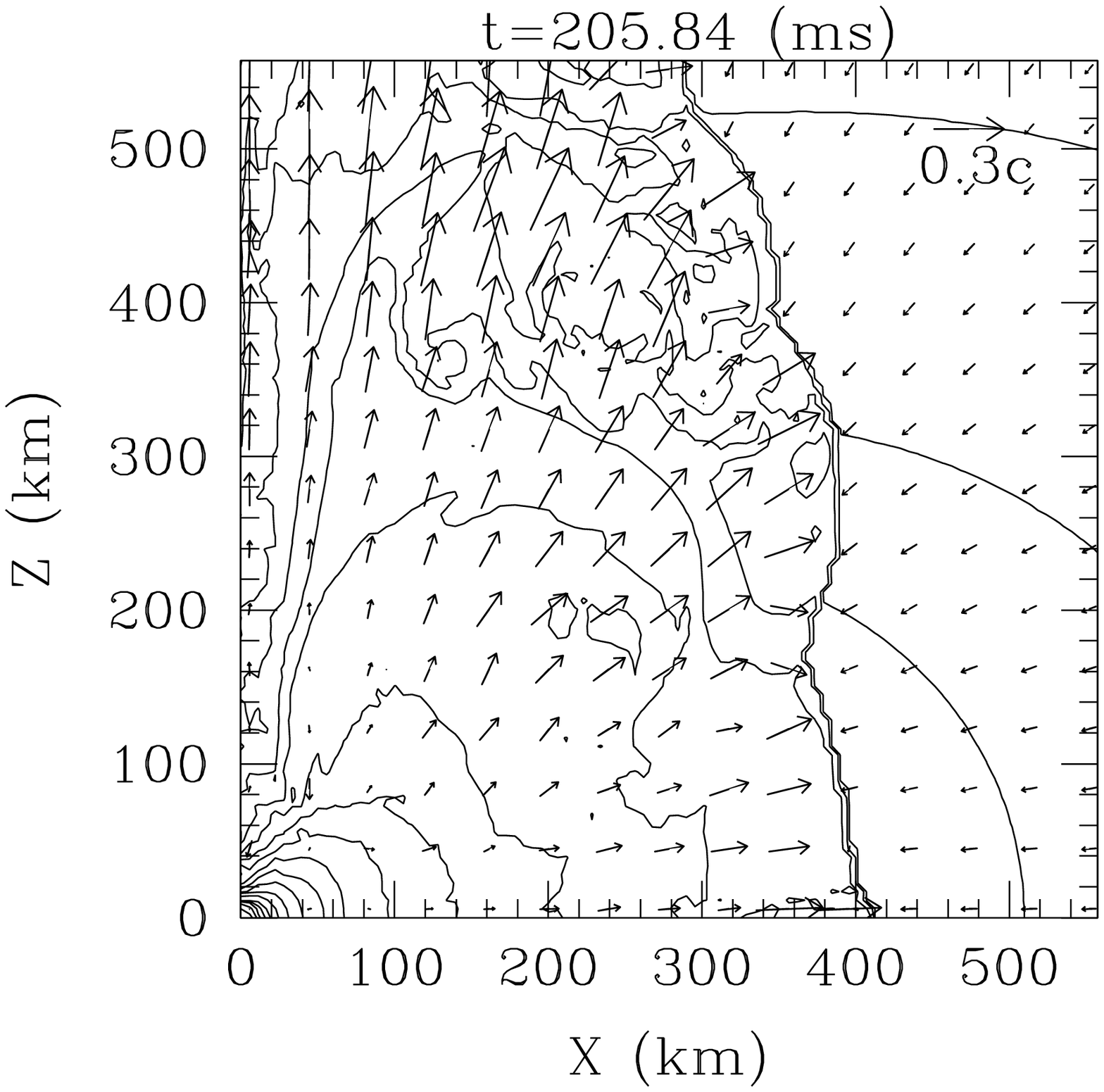}
\end{center}
\vspace{-4mm}
\caption{The same as Fig. \ref{figure7} but for model D5
  with the equation of state 'a', at $t = 199.47$,
  199.69, 200.31, 202.01, 203.33, and 205.84 ms.}\label{figure13}
\end{figure}
%
%
\begin{figure}[p]
\vspace{-4mm}
\begin{center}
\epsfxsize=2.35in
\leavevmode
\epsffile{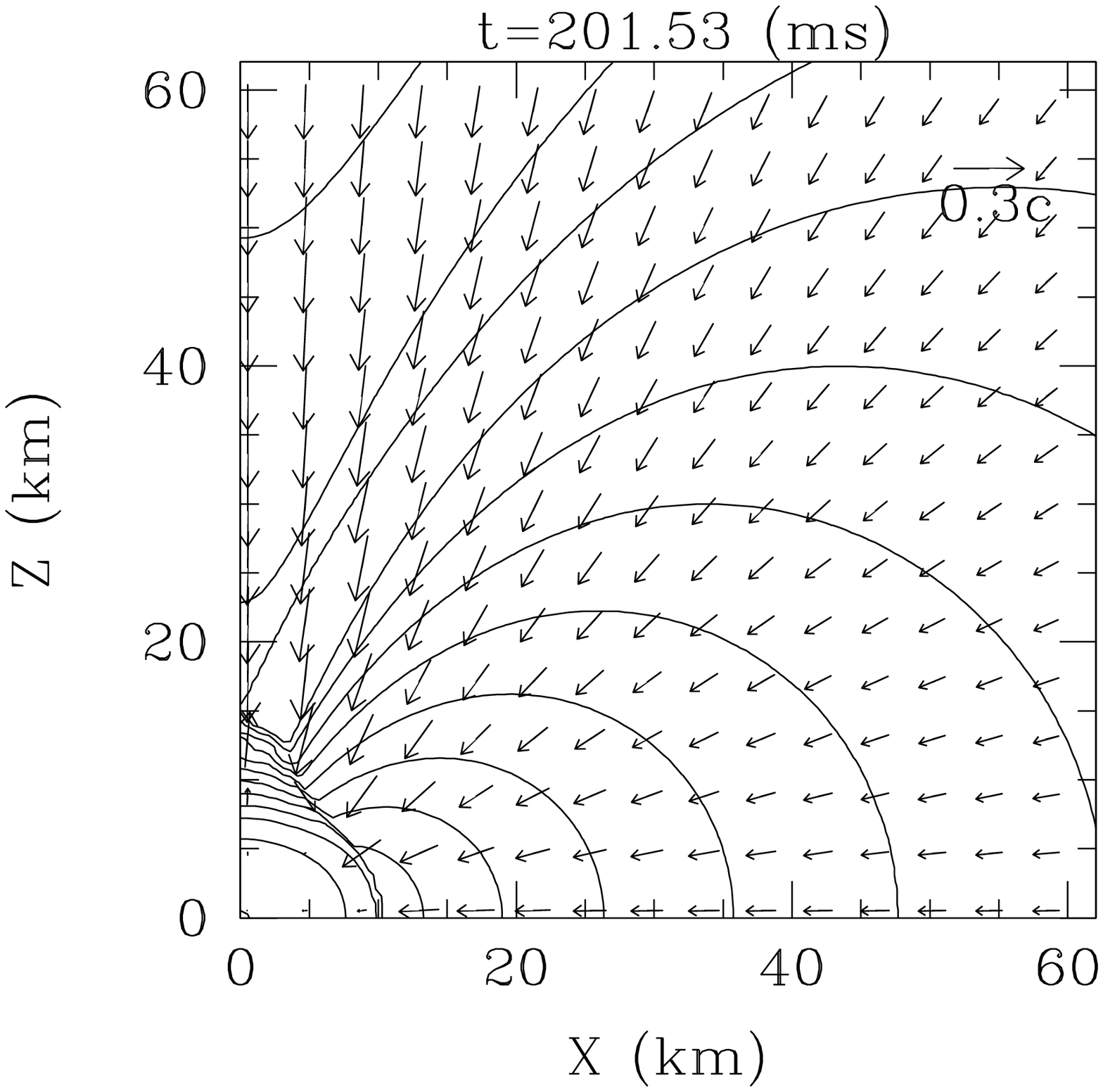} 
\epsfxsize=2.35in
\leavevmode
\epsffile{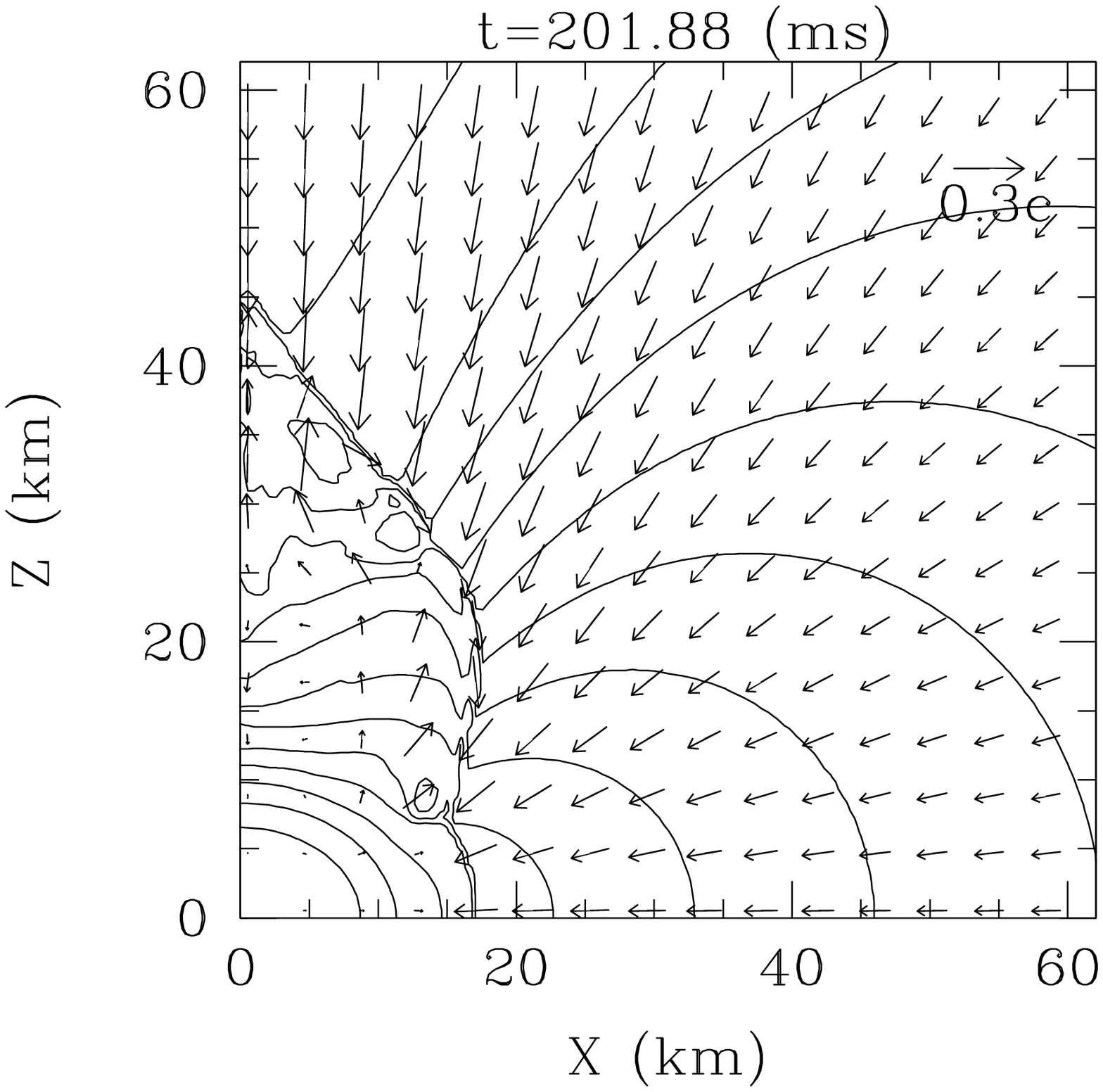}  
\epsfxsize=2.35in
\leavevmode
\epsffile{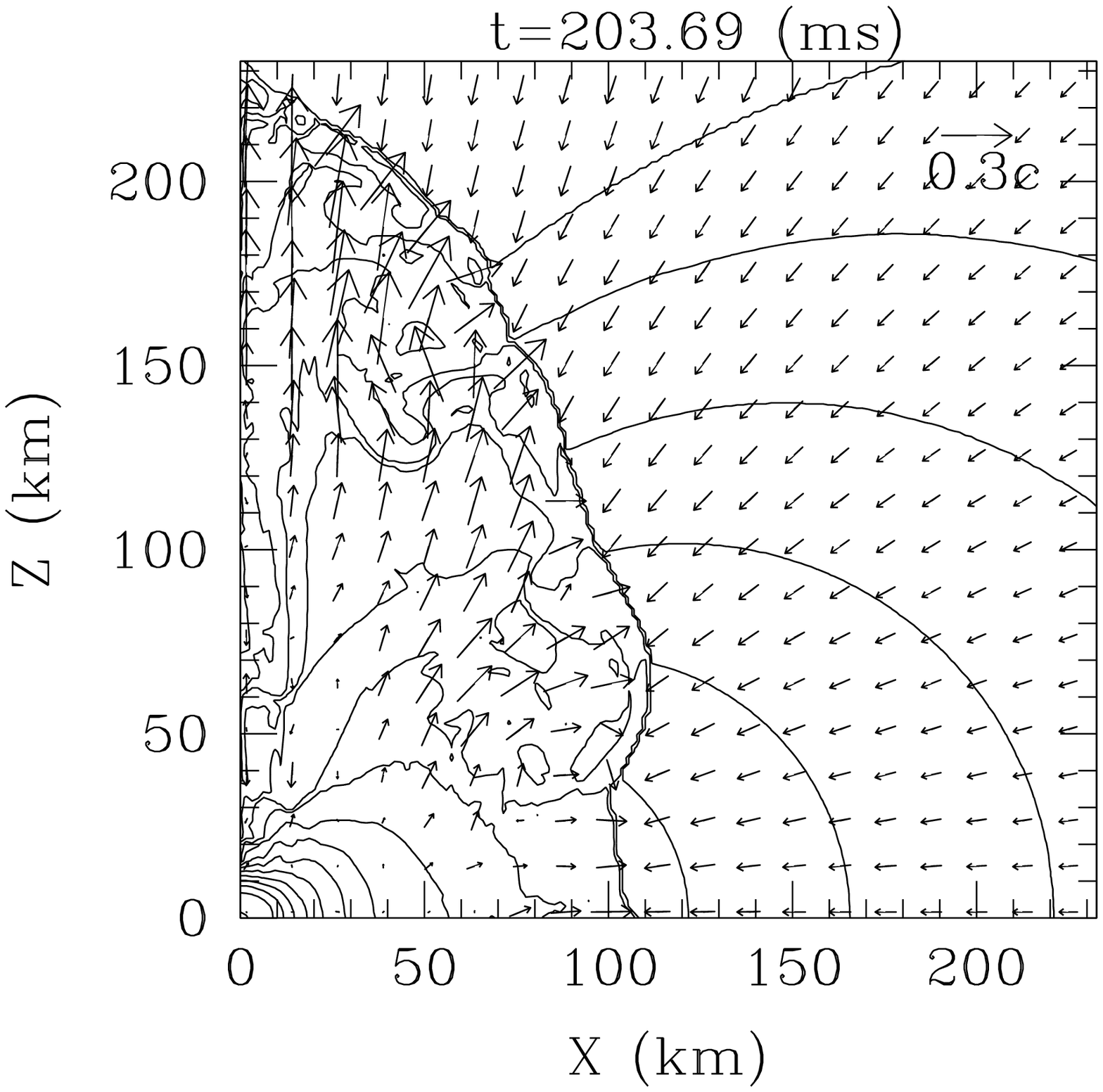}  
\end{center}
\vspace{-7mm}
\caption{The same as Fig. \ref{figure7} but for model A5
  with the equation of state 'a', at $t = 201.53$, 201.88, and 203.69 
  ms.}\label{figure14} 
\end{figure}
\begin{figure}[htb]
\vspace{-4mm}
  \begin{center}
\epsfxsize=2.35in
\leavevmode
\epsffile{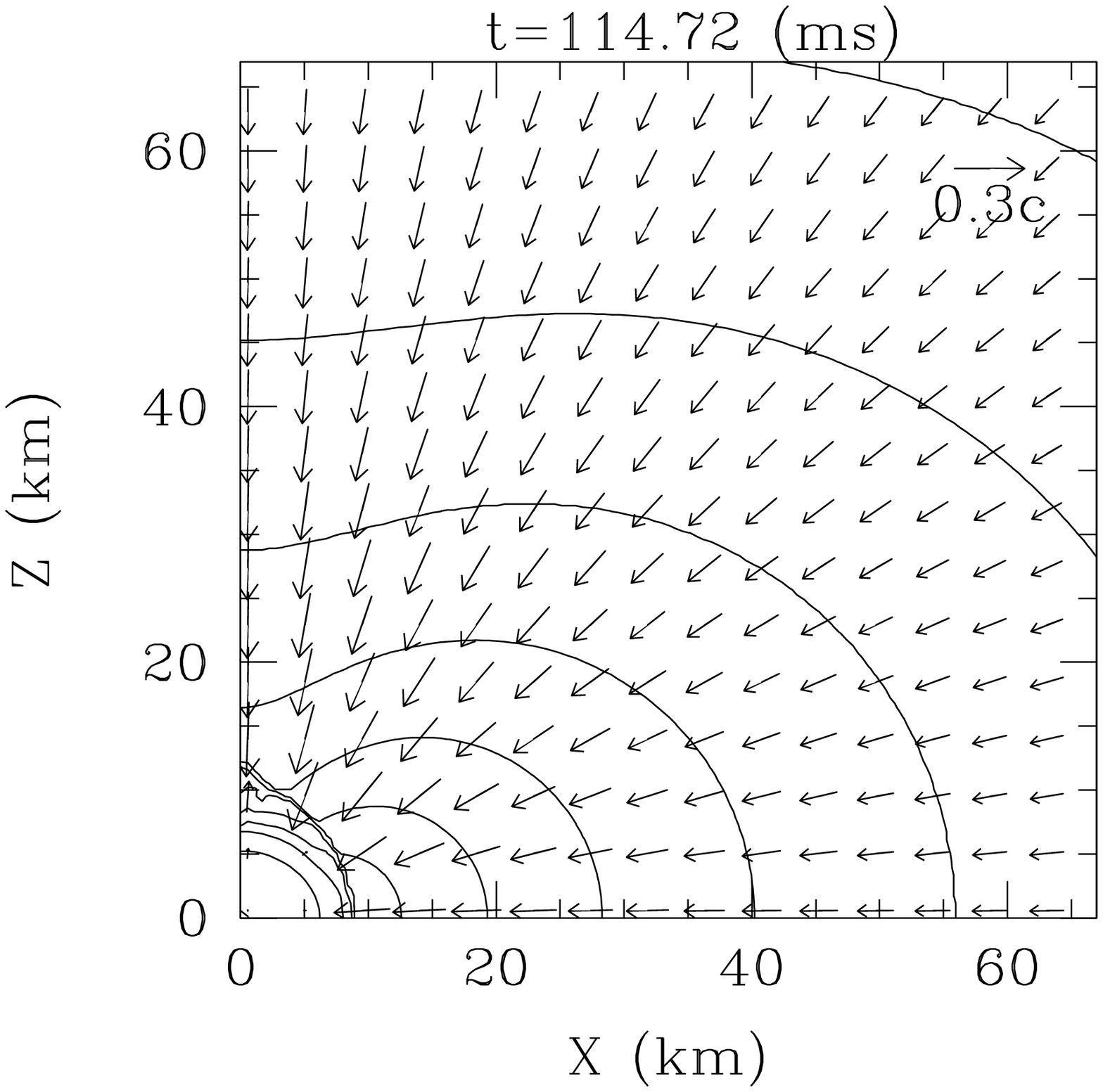} 
\epsfxsize=2.35in
\leavevmode
\epsffile{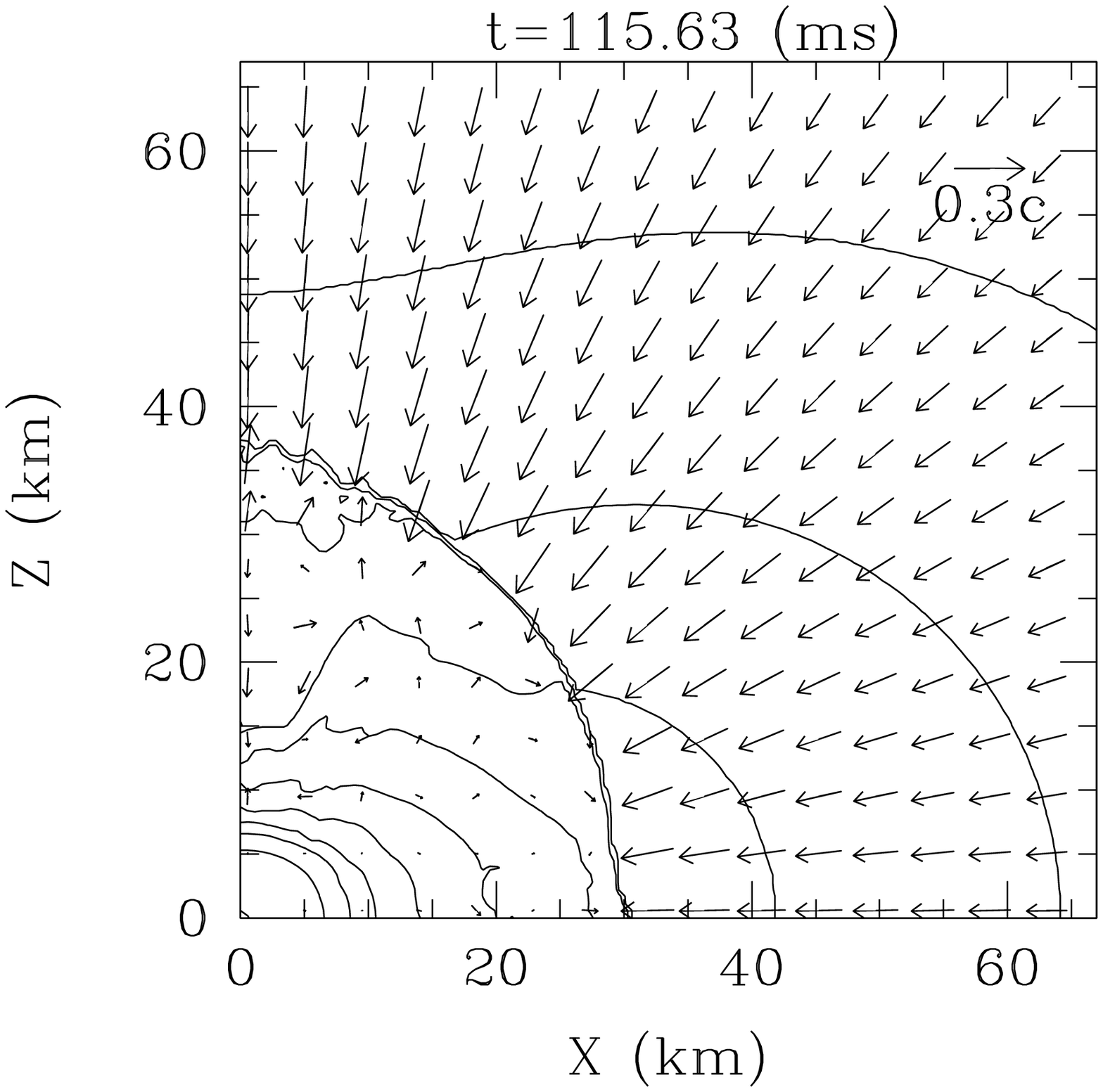} 
\epsfxsize=2.35in
\leavevmode
\epsffile{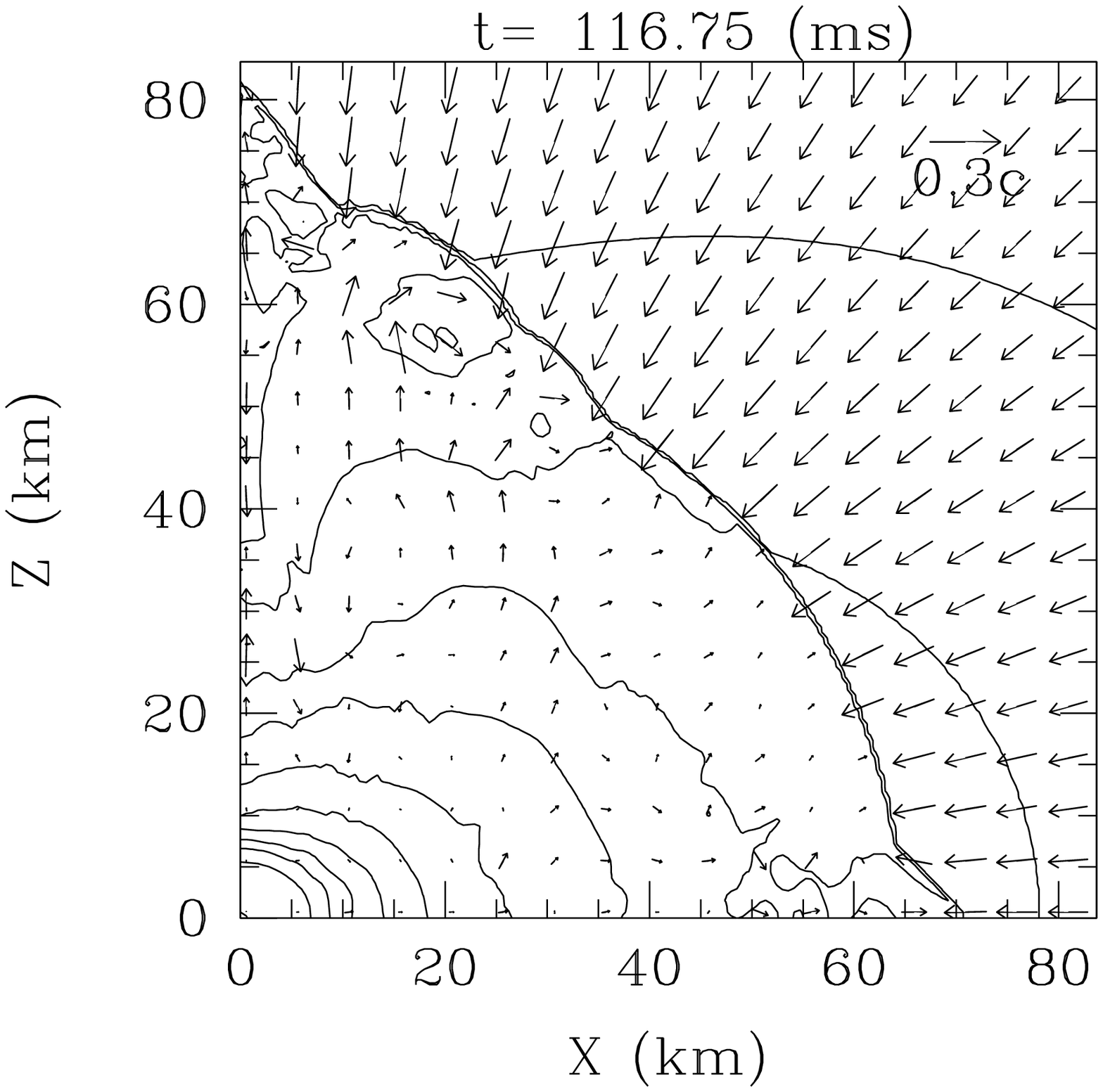} \\
\vspace{-5mm}
\epsfxsize=2.35in
\leavevmode
\epsffile{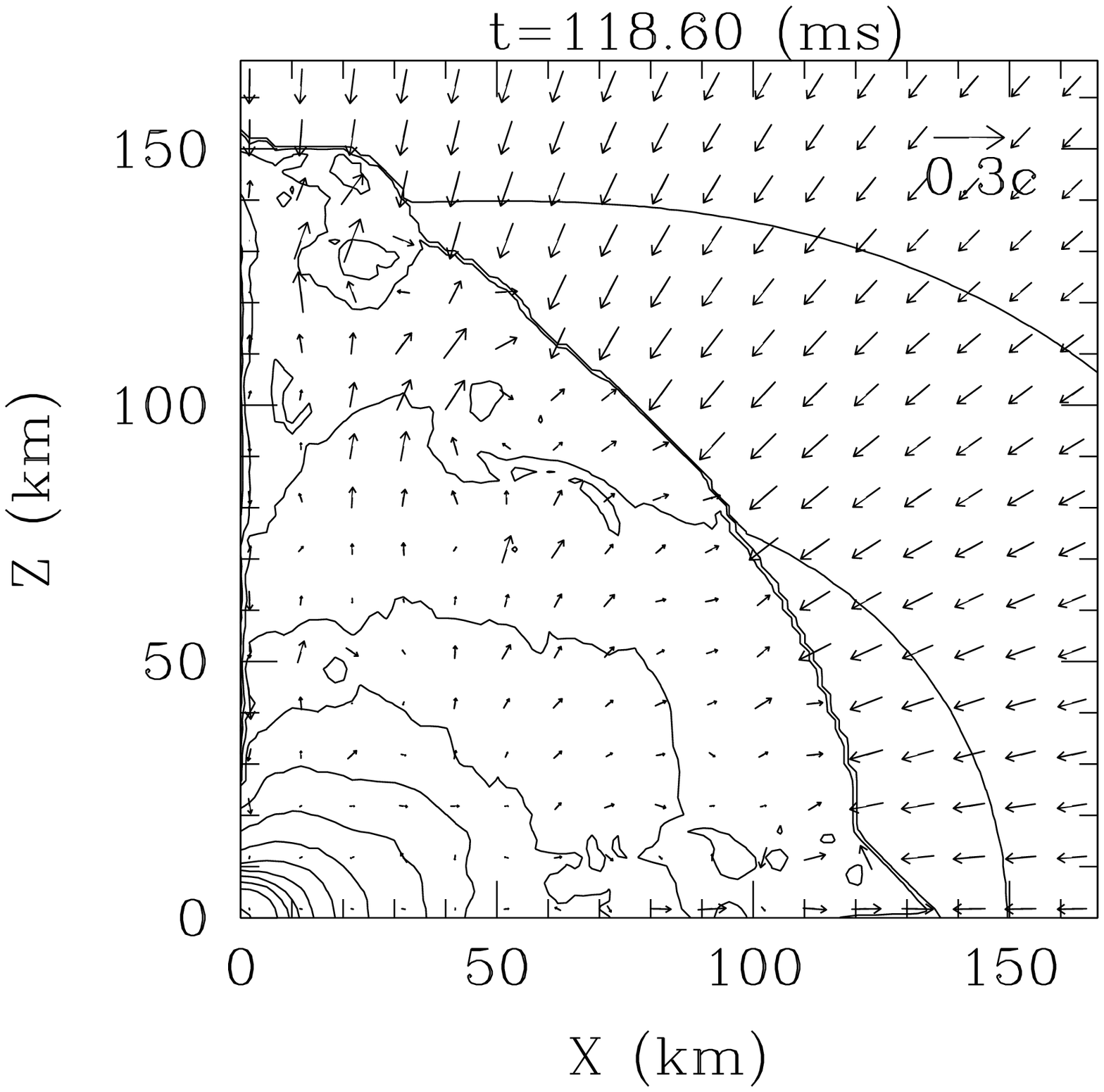} 
\epsfxsize=2.35in
\leavevmode
\epsffile{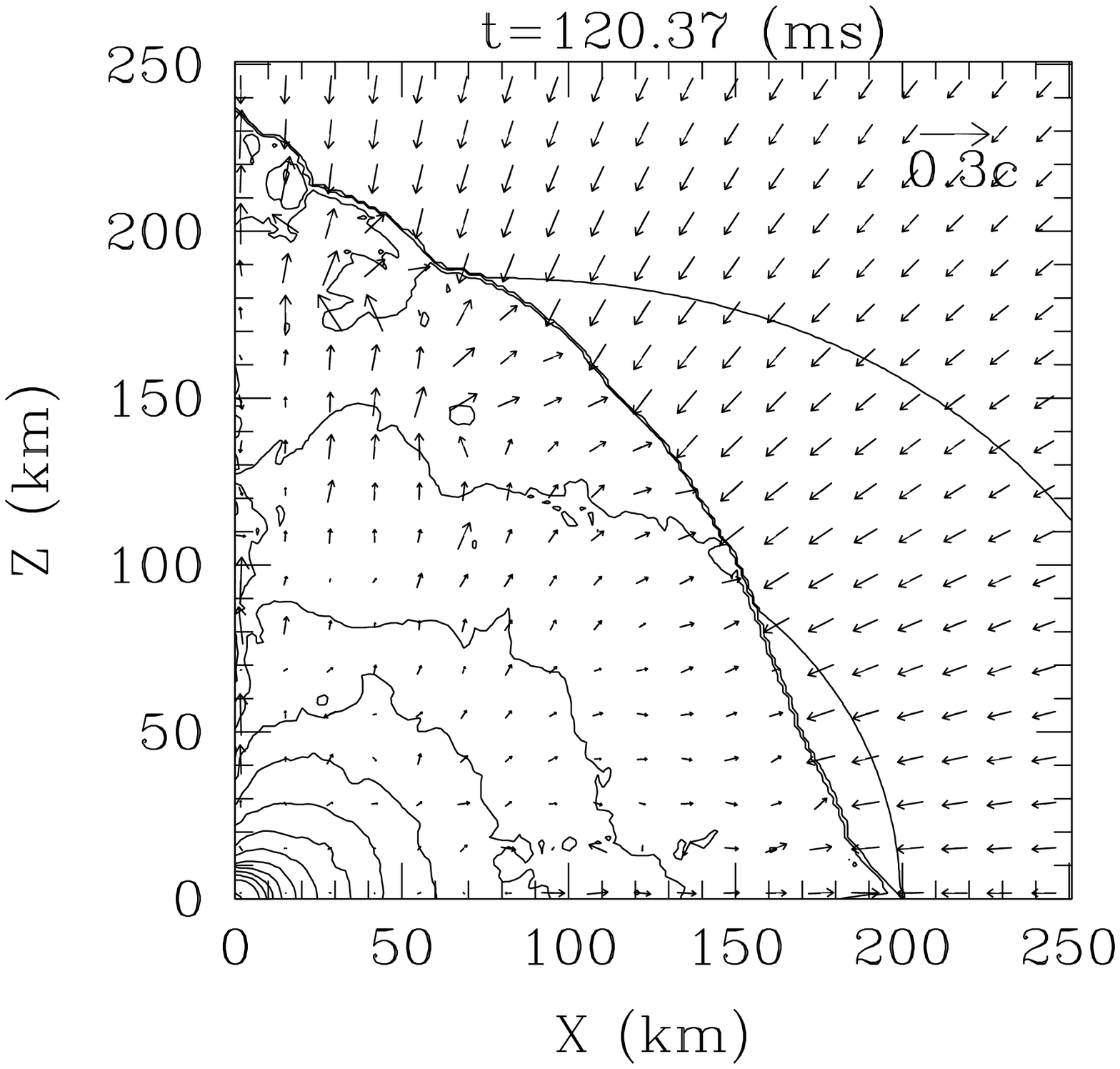}
\epsfxsize=2.35in
\leavevmode
\epsffile{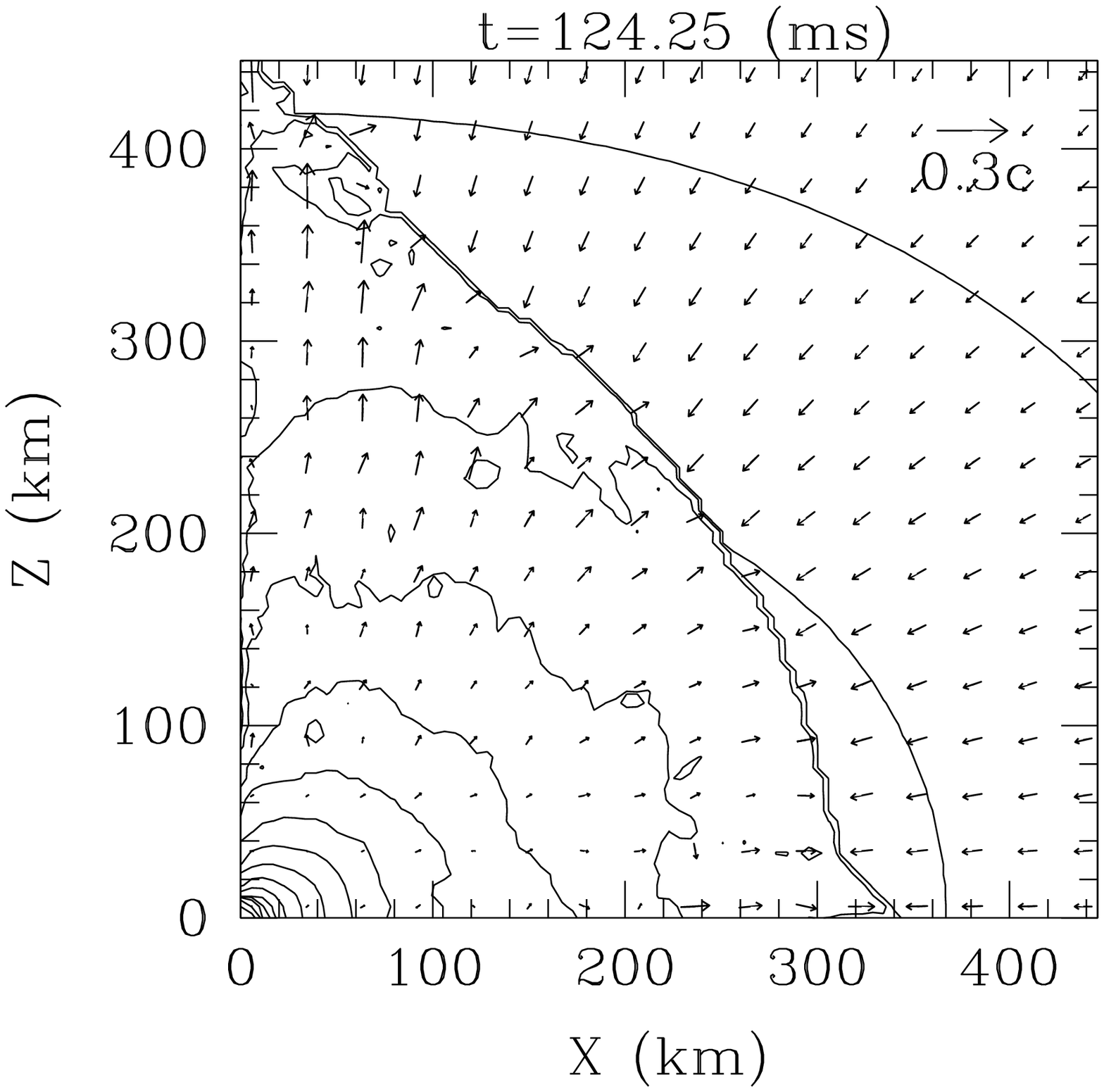}
\end{center}
\vspace{-4mm}
\caption{The same figure as Fig. \ref{figure7} but for model D5
  with the equation of state 'b', at $t = 114.72$,
  115.63, 116.75, 118.60, 120.37, and 124.25 ms.}\label{figure15}
\end{figure}
\begin{figure}[htb]
\vspace{-4mm}
  \begin{center}
\epsfxsize=2.35in
\leavevmode
\epsffile{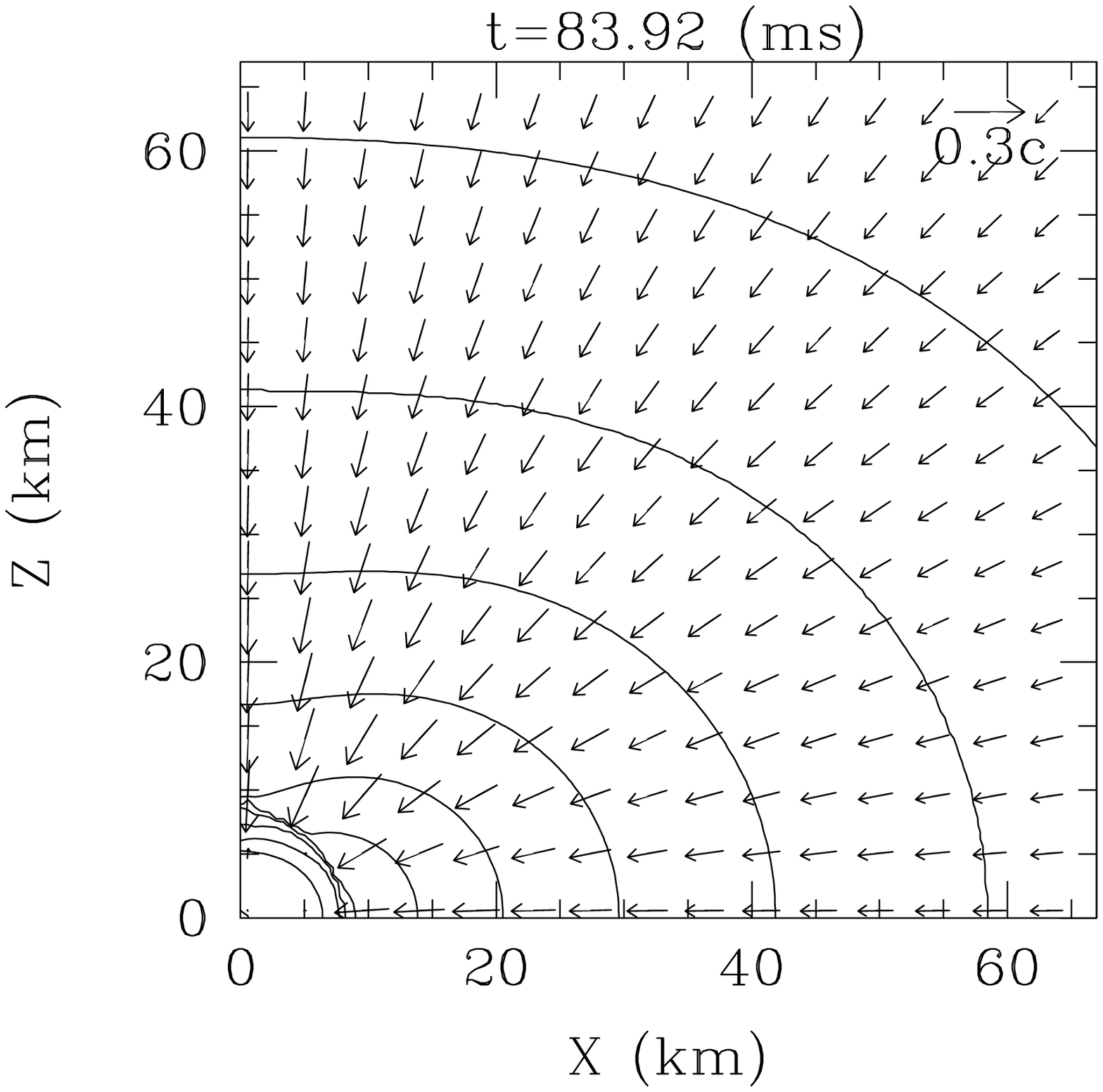} 
\epsfxsize=2.35in
\leavevmode
\epsffile{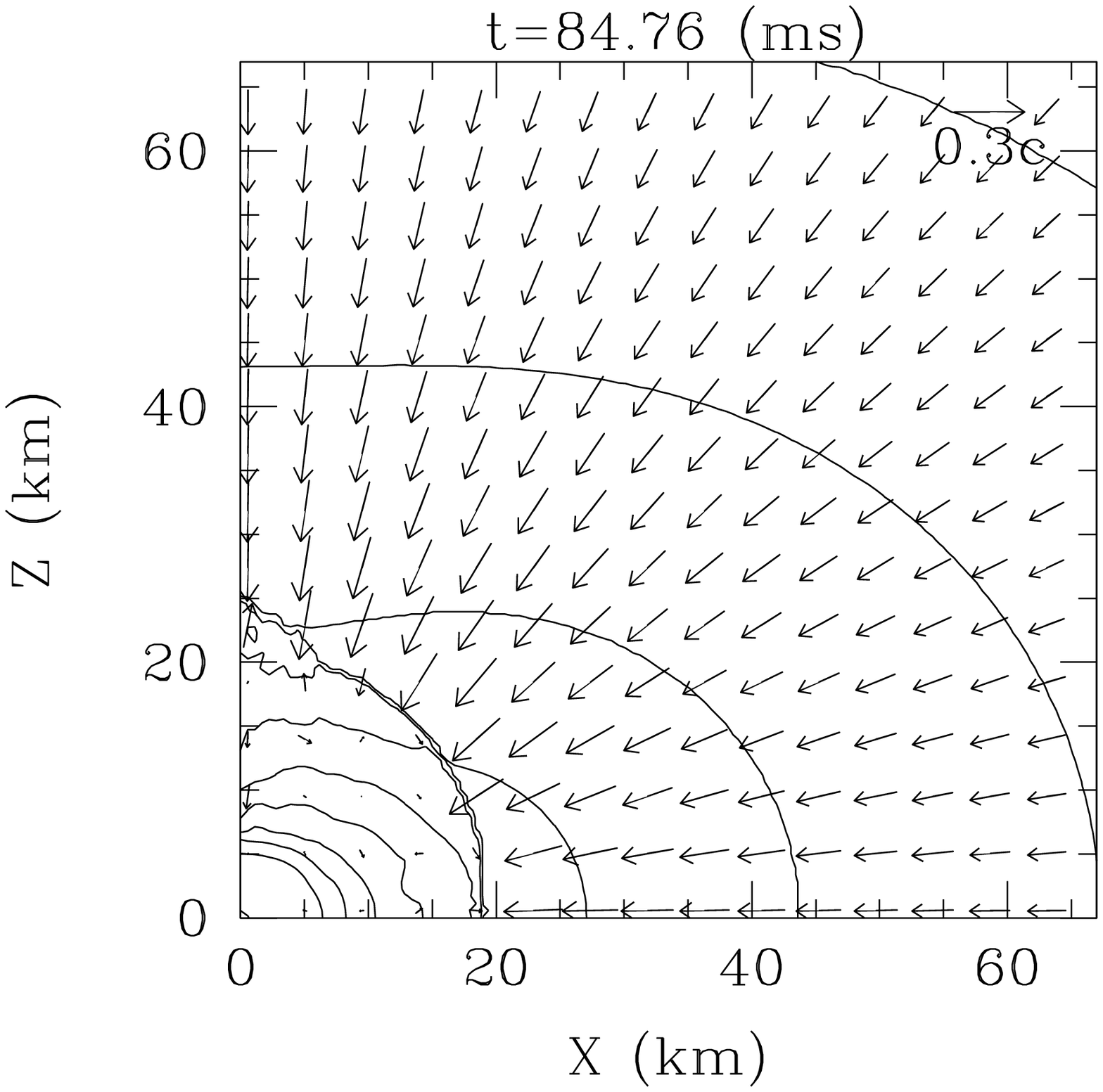} 
\epsfxsize=2.35in
\leavevmode
\epsffile{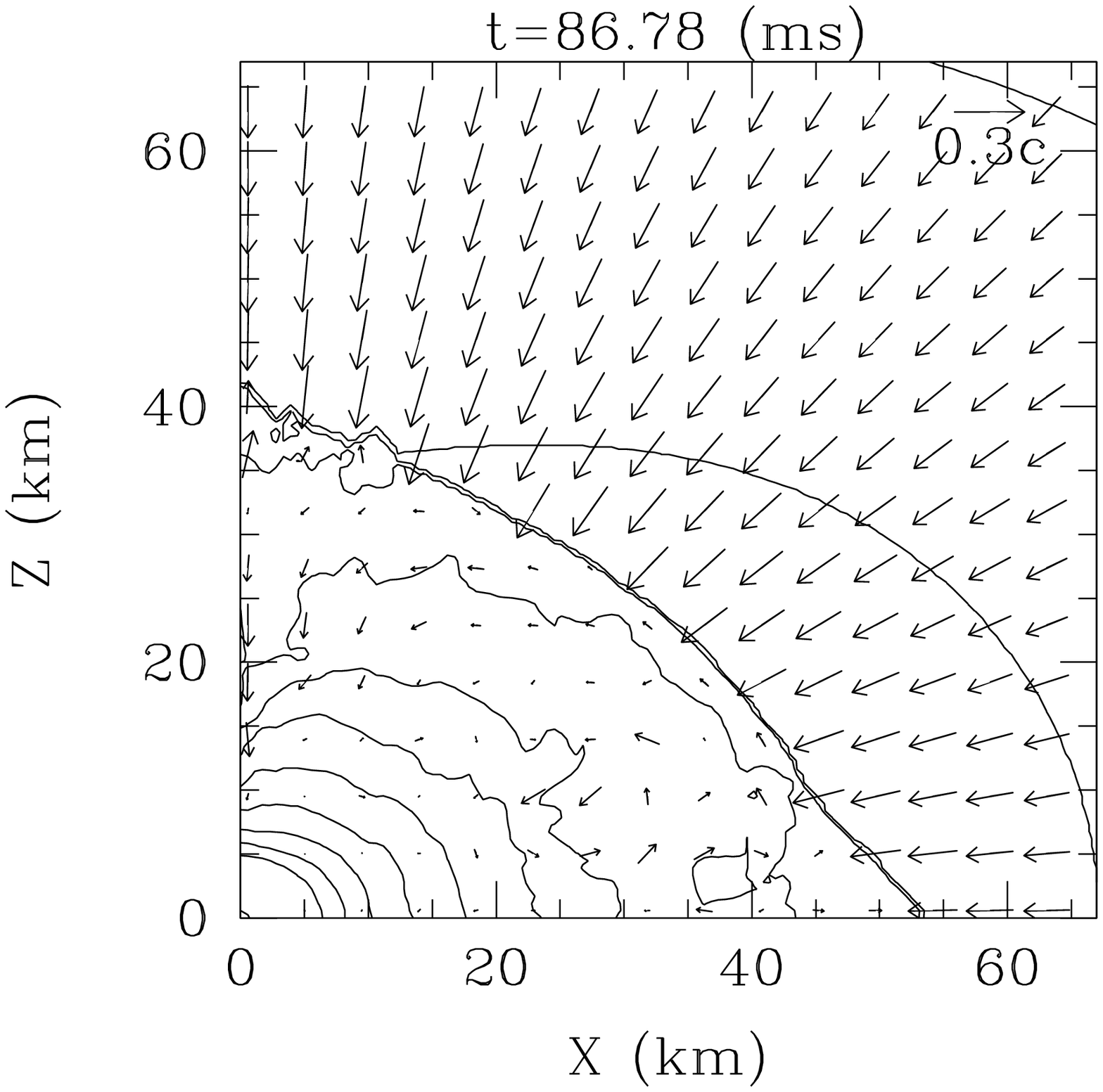} \\
\vspace{-5mm}
\epsfxsize=2.35in
\leavevmode
\epsffile{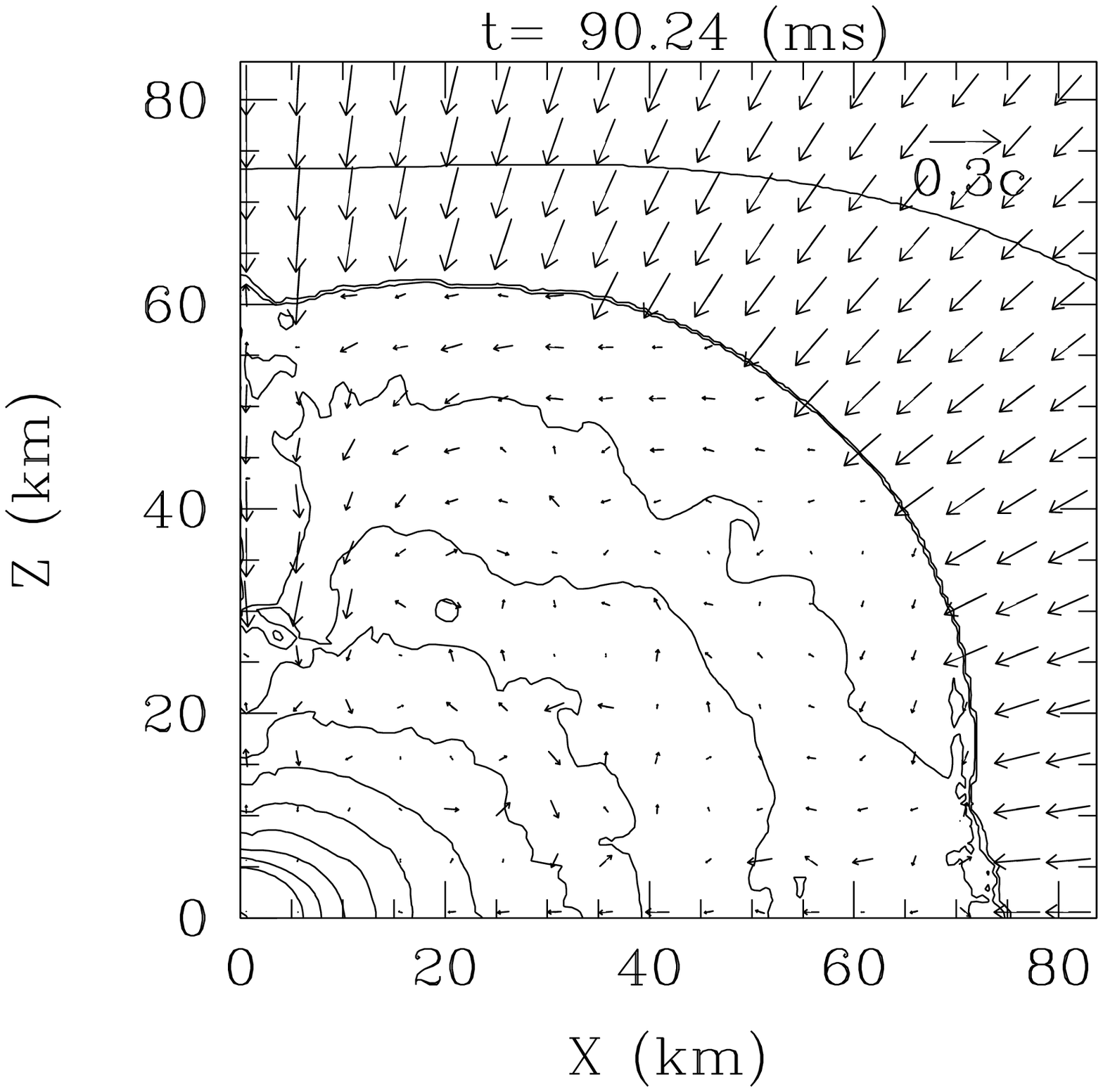} 
\epsfxsize=2.35in
\leavevmode
\epsffile{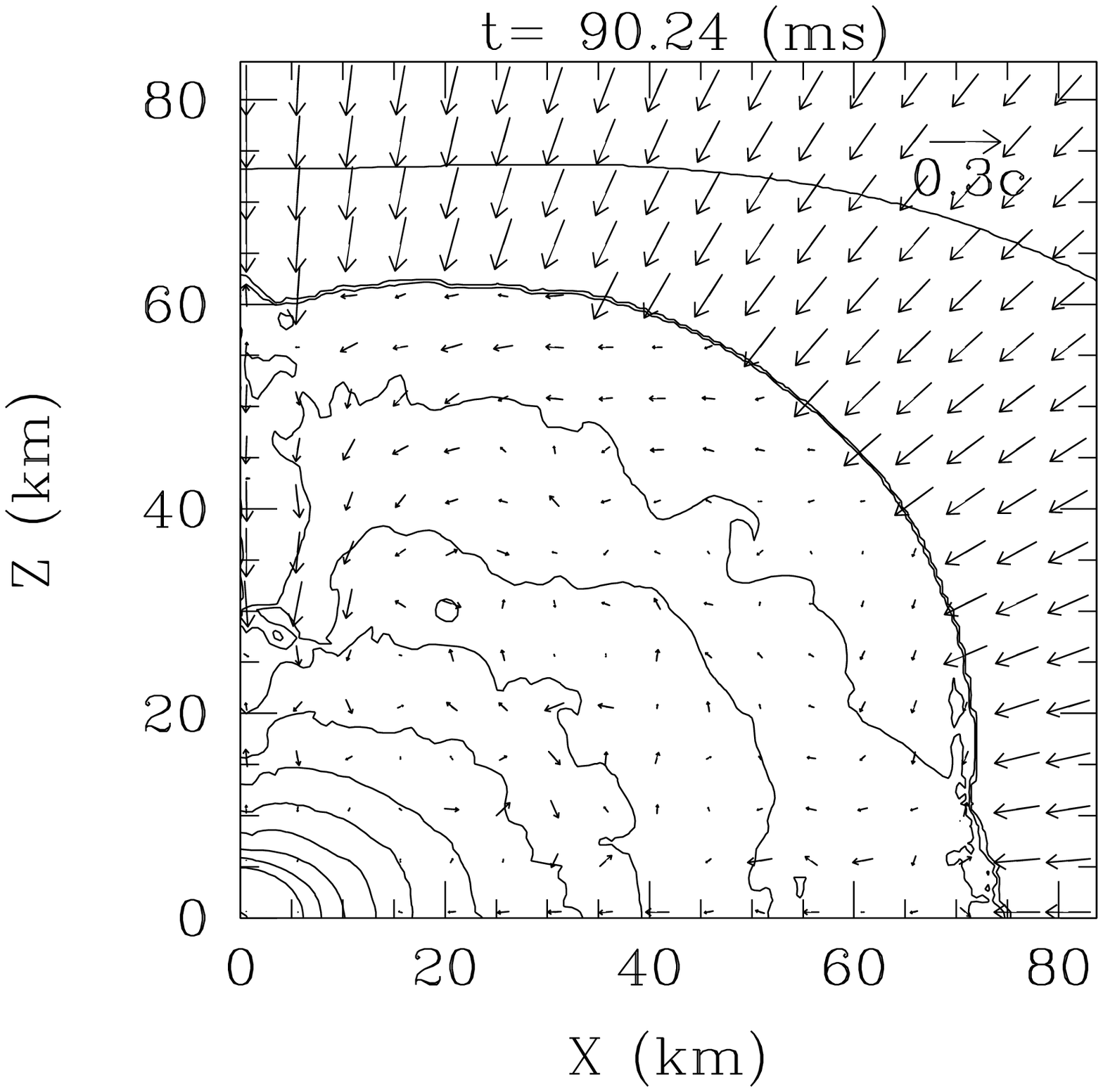}
\epsfxsize=2.35in
\leavevmode
\epsffile{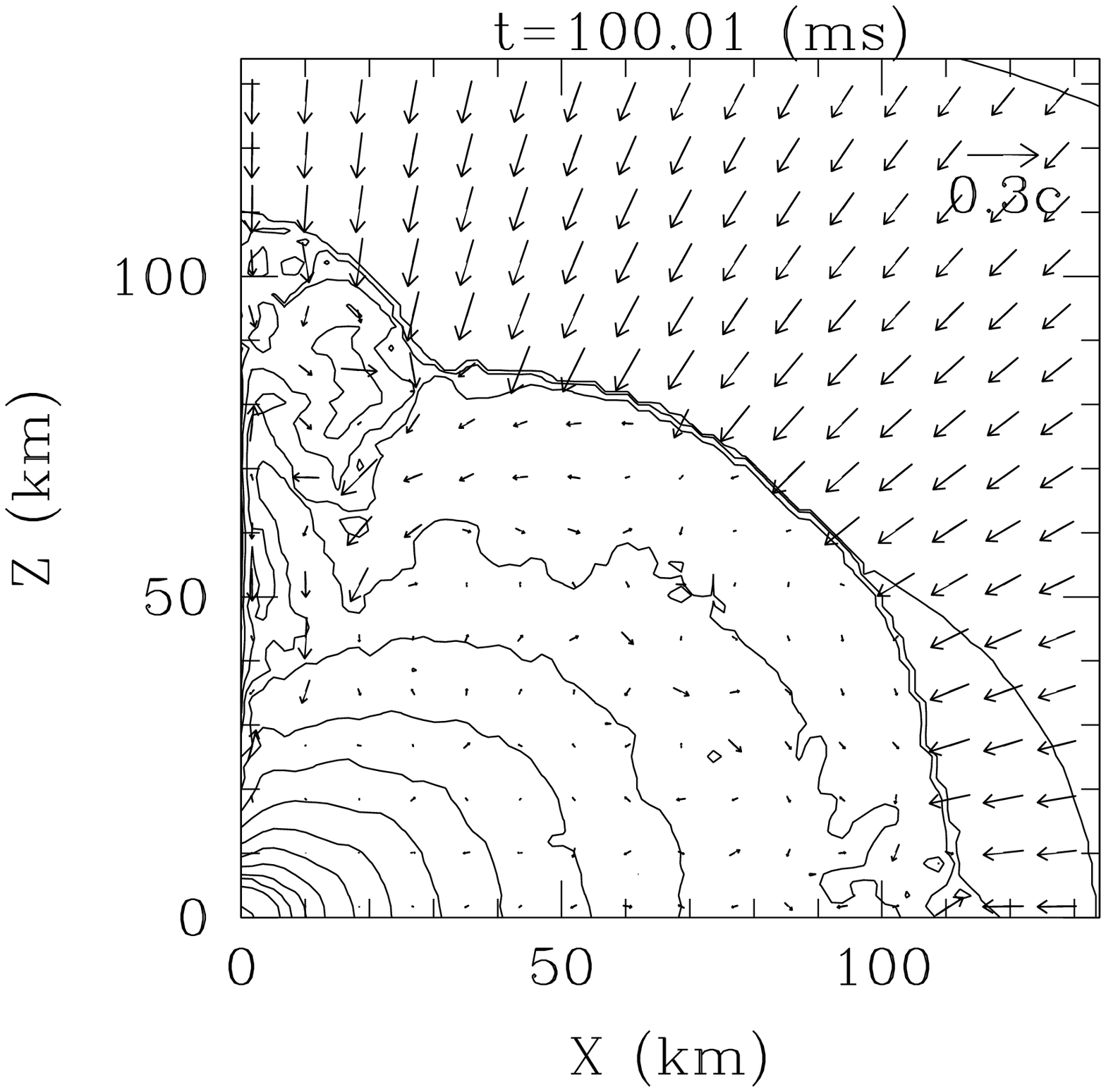}
\end{center}
\vspace{-4mm}
\caption{The same as Fig. \ref{figure7} but for model D5
  with the equation of state 'd', at $t = 83.92$,
  84.76, 86.78, 88.92, 90.24, and 100.01 ms.}\label{figure16}
\end{figure}
\subsubsection{Dependence of explosion on the equations of state}
\label{Bipolar}

As discussed in the previous subsections, the dynamics of the collapse 
and the threshold mass for the prompt black hole formation depend sensitively
on the equations of state. In the case of neutron star formation,
we also find that features of the explosion depend sensitively on
the equations of state.
We illustrate this fact focusing on rapidly rotating model D5 
in which black hole is not formed promptly irrespective of the
equations of state. 

In Fig. \ref{figure13}, we display the contour plots for model D5a.
For the collapse with the equation of state 'a' which is stiff in the
subnuclear density (with larger value of $\Gamma_{1}$) while soft in the
supranuclear density (with smaller value of $\Gamma_{2}$), the steep density
gradient is formed around the rotational axis of the inner core. 
On the other hand, such steep gradient is not formed around the
equatorial plane (see the second panel of Fig. \ref{figure13}). 
Due to the steep gradient, strong  
shock waves are generated along the rotational axis.
Consequently, a jet-like explosion is seen 
(see the third panel of Fig. \ref{figure13}).
The explosion velocity in the direction of the rotational axis
is much larger than that in the equatorial plane
(see the fourth and fifth panels of Fig. \ref{figure13}): 
The maximum speed becomes $\approx 0.8c$ around the rotational axis
near the shock.
Thus, the shock wave reaches the surface of the iron core
much more quickly (see the last panel of Fig. \ref{figure13}).
At this time, a funnel structure is formed around the rotational axis. 

The structure of bipolar shock waves depends on mass of progenitor.
For model A5a in which $M_{\rm ADM} \approx 1.97M_{\odot}$, 
a steep density gradient is also formed along the rotational axis 
(see Fig. \ref{figure14}), but it is formed only 
in narrower region than that for model D5a
(compare the first panel of Fig. \ref{figure14} and
the second panel of Fig. \ref{figure13}). 
Reflecting this result, the structure of the shock
front is sharper than that for model D5a and is deformed to be 
spear-like (compare the second panel of  Fig. \ref{figure14}
and the third panel of Fig. \ref{figure13}).

Such bipolar and energetic supernova explosion is required
to explain the observations of several hypernovae, which are suggested to 
be associated with a gamma-ray burst
\cite{Wang1,Iwamoto,Maeda1,Wang2,Nomoto}.
Indeed, the energetic explosion along the rotational axis is
preferable to avoid the baryon-loading problem 
in the fire-ball model of the gamma-ray bursts \cite{MW}. 
The present results suggest that the collapse of rapidly rotating
iron cores of $M \agt 2M_{\odot}$ with a class of equations of state
similar to 'a' is preferable for realizing such a state. 

To see the dependence of the features of the explosion on
the equations of state, we also display the contour plots 
for models D5b and D5d in Figs. \ref{figure15} and \ref{figure16}. 
As the first panel of these figures show, 
the magnitude of the density gradient along the rotational axis
is not very different from that in the equatorial plane, and hence, 
structure of the shock is not very aspherical
(see the second panel of the Figs. \ref{figure15} and \ref{figure16}).

The shock for model D5b is strong enough to quickly propagate
through the outer core, sweeping the matter falling into the shock.
For model D5d, on the other hand, the shock is not as strong as
that for D5b, and hence, the propagation speed is slow. 
Also, the density gradient formed around the shock is not very 
different between along the rotational axis and in the equatorial
plane, and hence, the structure of the shock becomes eventually oblate
(not prolate) due to the effect of the rotation
(see the third panel of Fig. \ref{figure16}). 
As the matter behind the shock falls and beats the protoneutron
star at the center, it oscillates and the shocks are formed continuously.
Thus, the shocks propagates outward and helps
sweeping the infalling matter outward (see the fourth and fifth panels of Fig.
\ref{figure16}). The features found for models D5b and D5d
are not found in model D5a. This fact implies that the
explosion mechanism depends sensitively on the equations of state
for rapidly rotating iron core collapse. 

It should be also mentioned that the bipolar explosion has not been
found in the previous study \cite{SS2} in which the same parametric
equation of state with $(\Gamma_{1}, \Gamma_{2}) = (1.32, 2.5)$ 
and smaller core mass of $1.5M_{\odot}$ is adopted. 
(Note that this equation of state is stiffer in supranuclear
density than the equation of state 'a' of the present paper). 
This fact implies that an equation of state with not 
only a large value of $\Gamma_{1}(=1.32)$ but also a small value of
$\Gamma_{2} \leq 2.25$ may be required for producing the bipolar
explosion.
\subsubsection{Rotational profile of the protoneutron star} \label{Sec-rot}

%
\begin{figure}[p]
  \vspace{-6mm}
  \begin{center}
    \epsfxsize=2.3in
    \leavevmode
    \epsffile{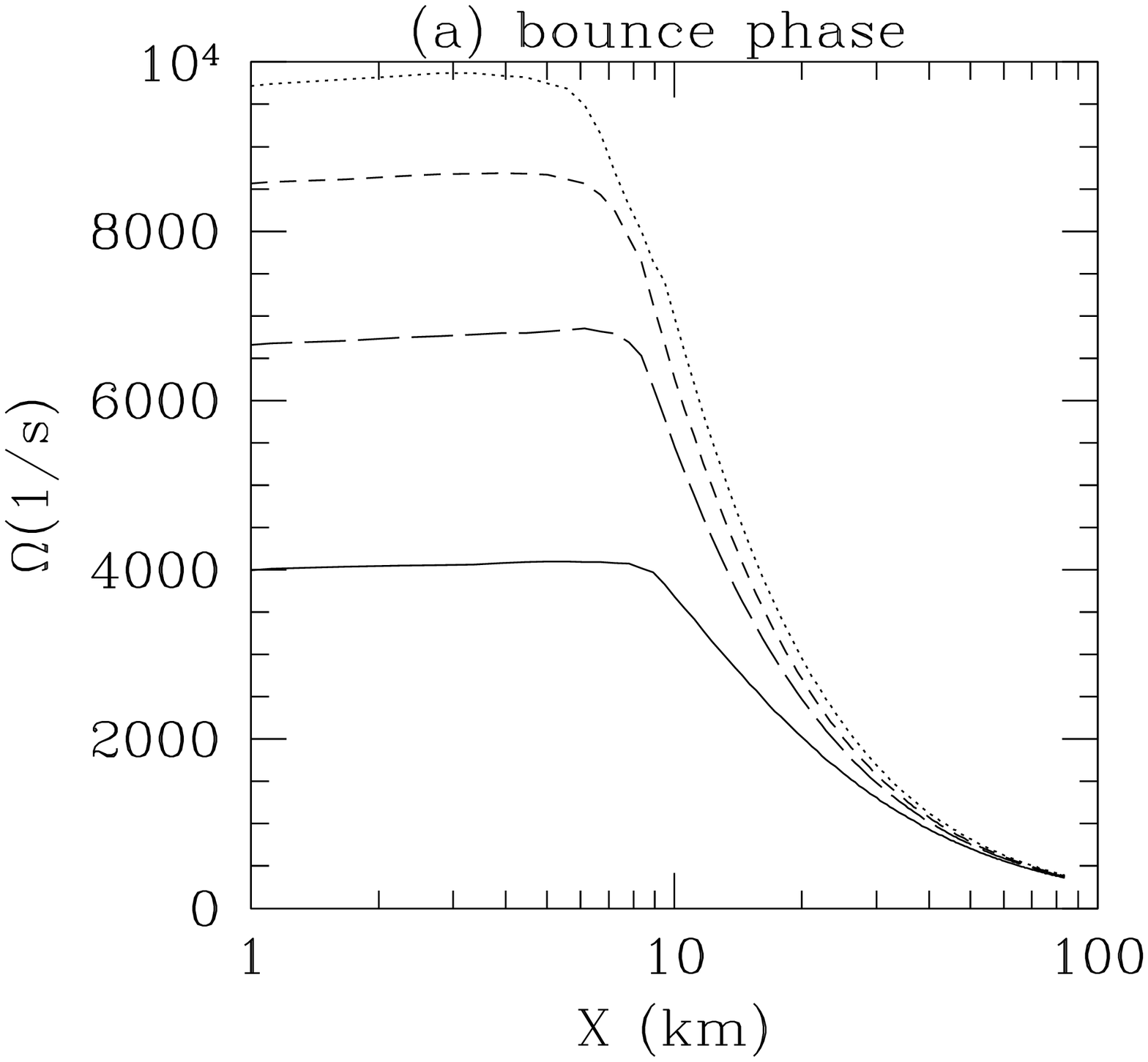} 
    \epsfxsize=2.3in
    \leavevmode
    \epsffile{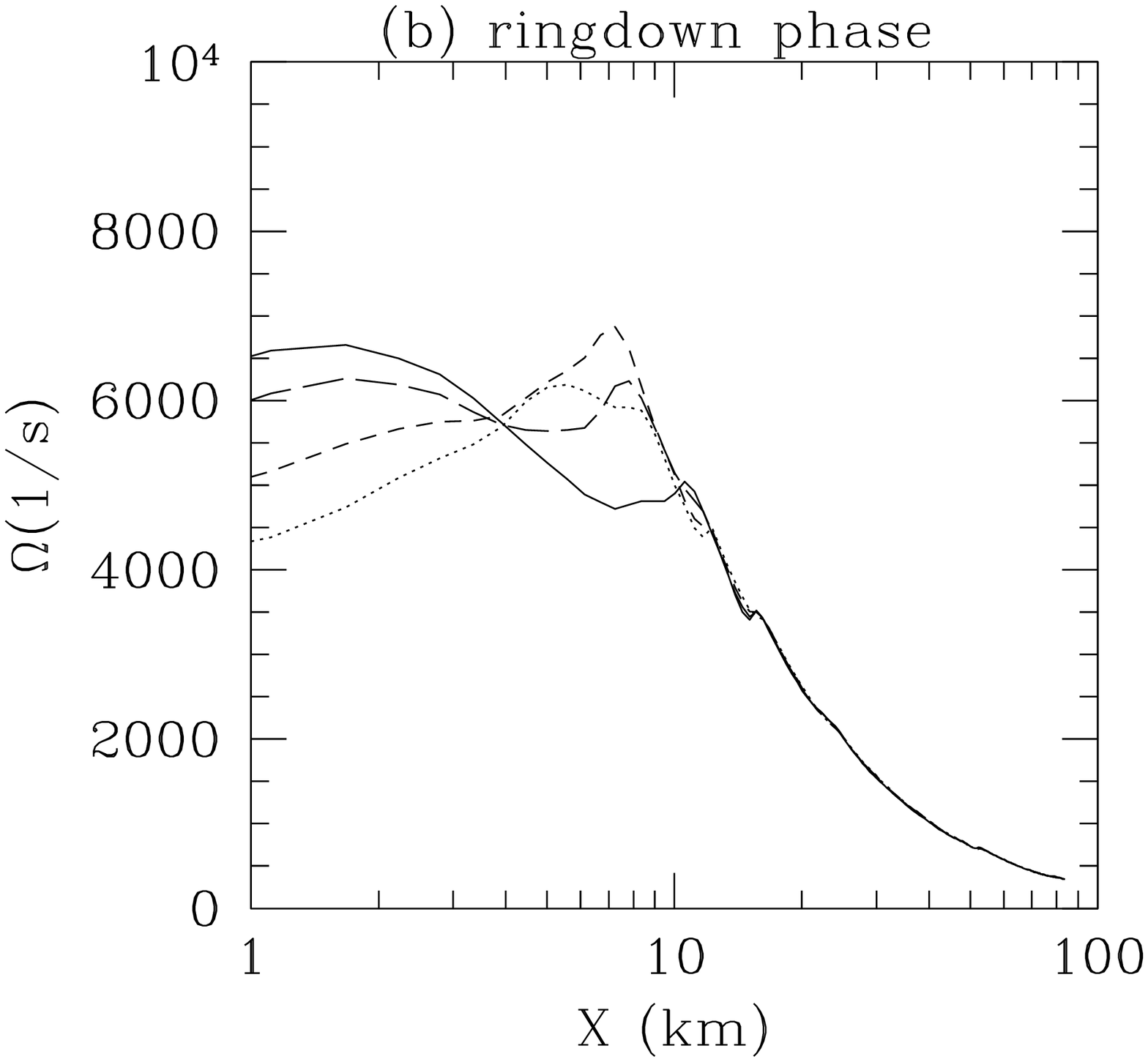}  
  \end{center}
  \vspace{-6mm}
  \caption{ The angular velocity profile inside the iron core
    along the $x$-axis for model D5a at selected time slices:
    (a) at the bounce phase of $t=199.43 $ (solid curve), 199.62 (long dashed
    curve), 199.73 (dashed curve), and 199.86 ms (dotted curve); (b) at
    the ringdown phase of $t=206.49$ (solid curve), 206.75 (long dashed
    curve), 206.94 (dashed curve), and 207.12 ms (dotted curve). 
  }\label{figure17}
\end{figure}
%
\begin{figure}[p]
  \vspace{-6mm}
  \begin{center}
    \epsfxsize=2.3in
    \leavevmode
    \epsffile{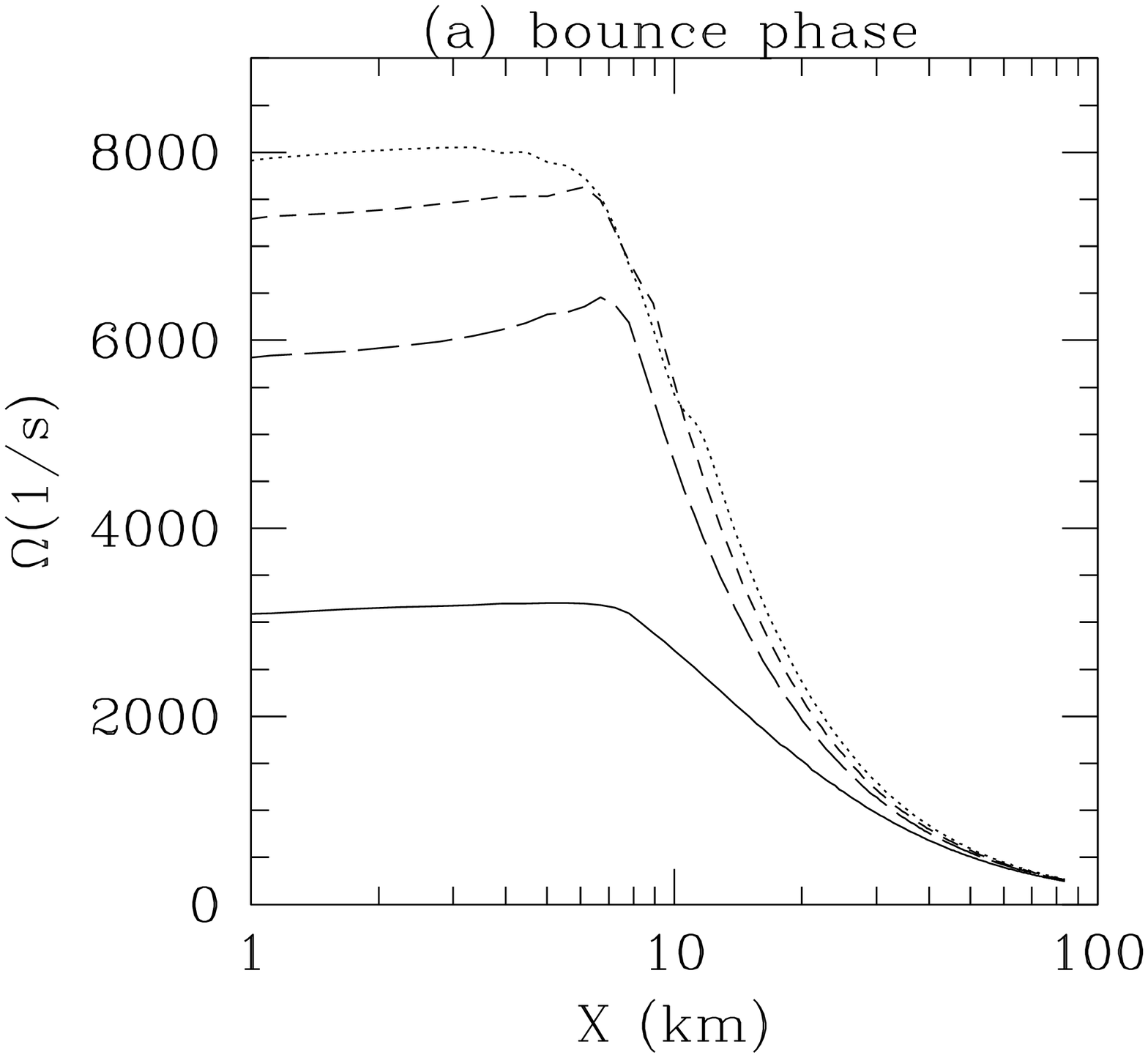} 
    \epsfxsize=2.3in
    \leavevmode
    \epsffile{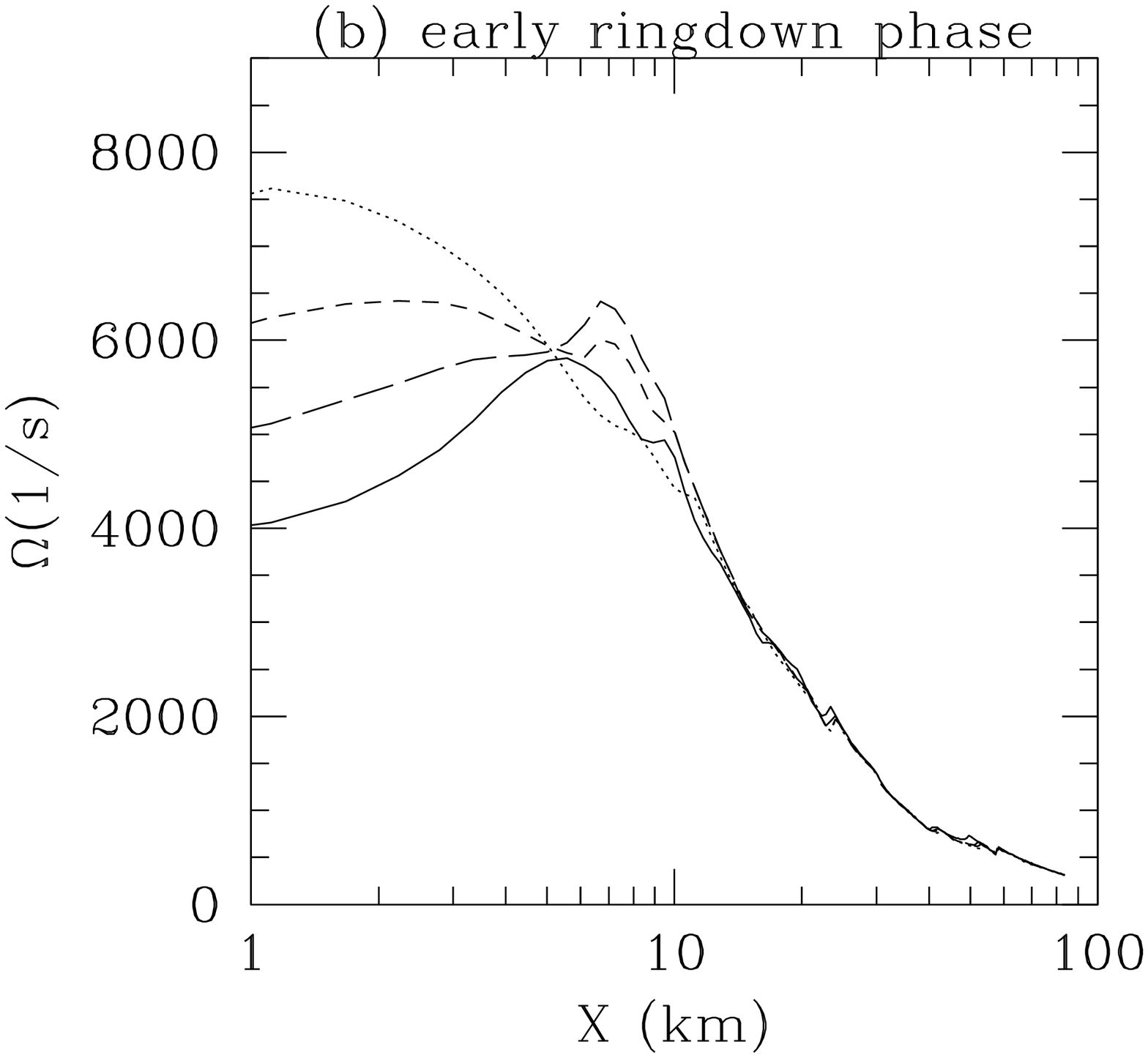} 
    \epsfxsize=2.3in
    \leavevmode
    \epsffile{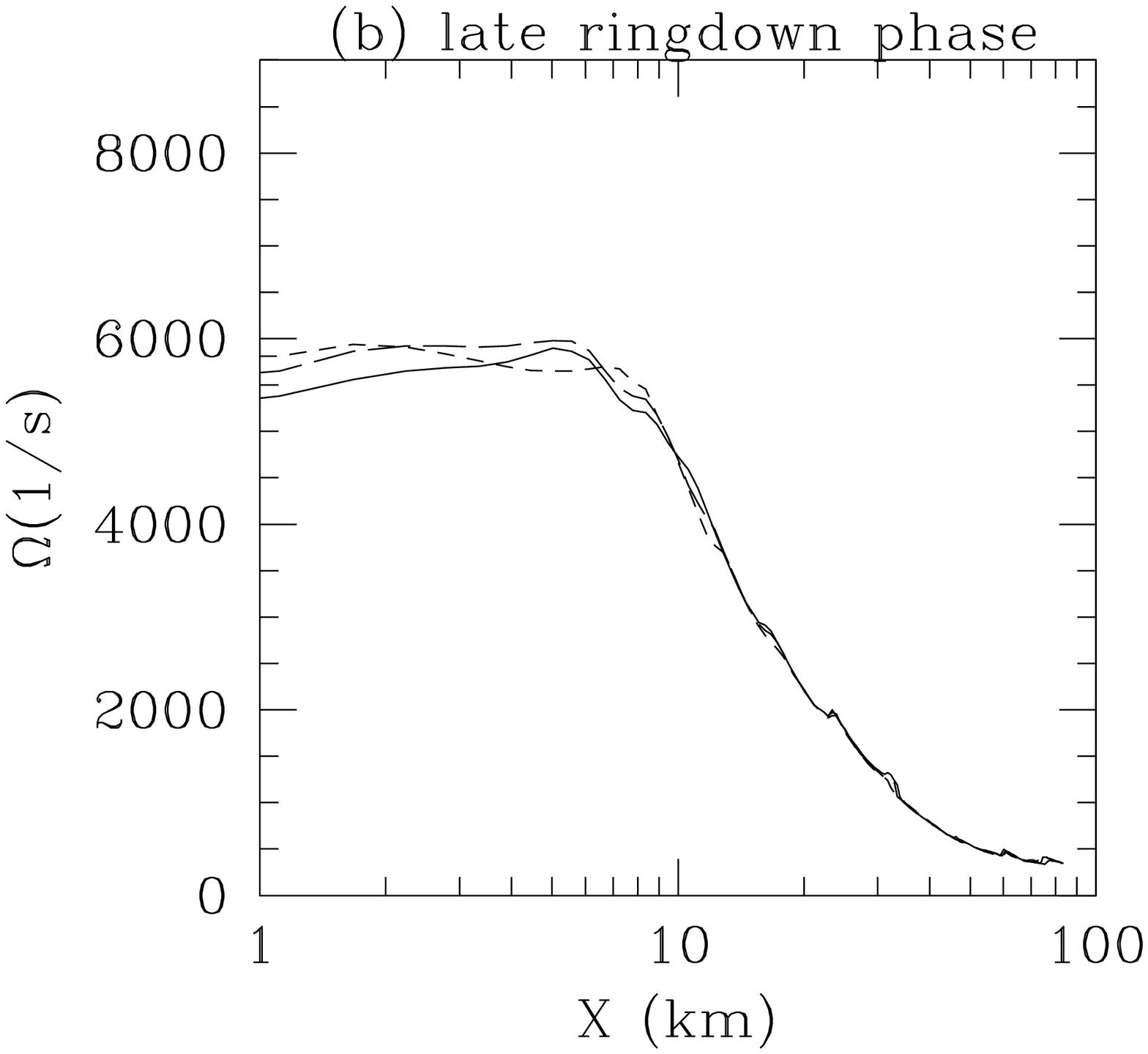} 
  \end{center}
  \vspace{-6mm}
  \caption{ The same as Fig. \ref{figure17} but for model D5b at
    selected time slices: (a)
    at the bounce phase of $t=114.39$ (solid curve), 114.61 (long dashed
    curve), 114.74 (dashed curve), and 114.87 ms (dotted curve); (b) at
    the early ringdown phase of $t=118.86$ (solid curve), 119.16 (long dashed
    curve), 119.29 (dashed curve), and 119.53 ms (dotted curve); (c) at
    the late ringdown phase of $t=122.06$ (solid curve), 122.36 (long dashed
    curve), 122.67 ms (dashed curve). 
  }\label{figure18}
\end{figure}
%
\begin{figure}[p]
  \vspace{-6mm}
  \begin{center}
    \epsfxsize=2.3in
    \leavevmode
    \epsffile{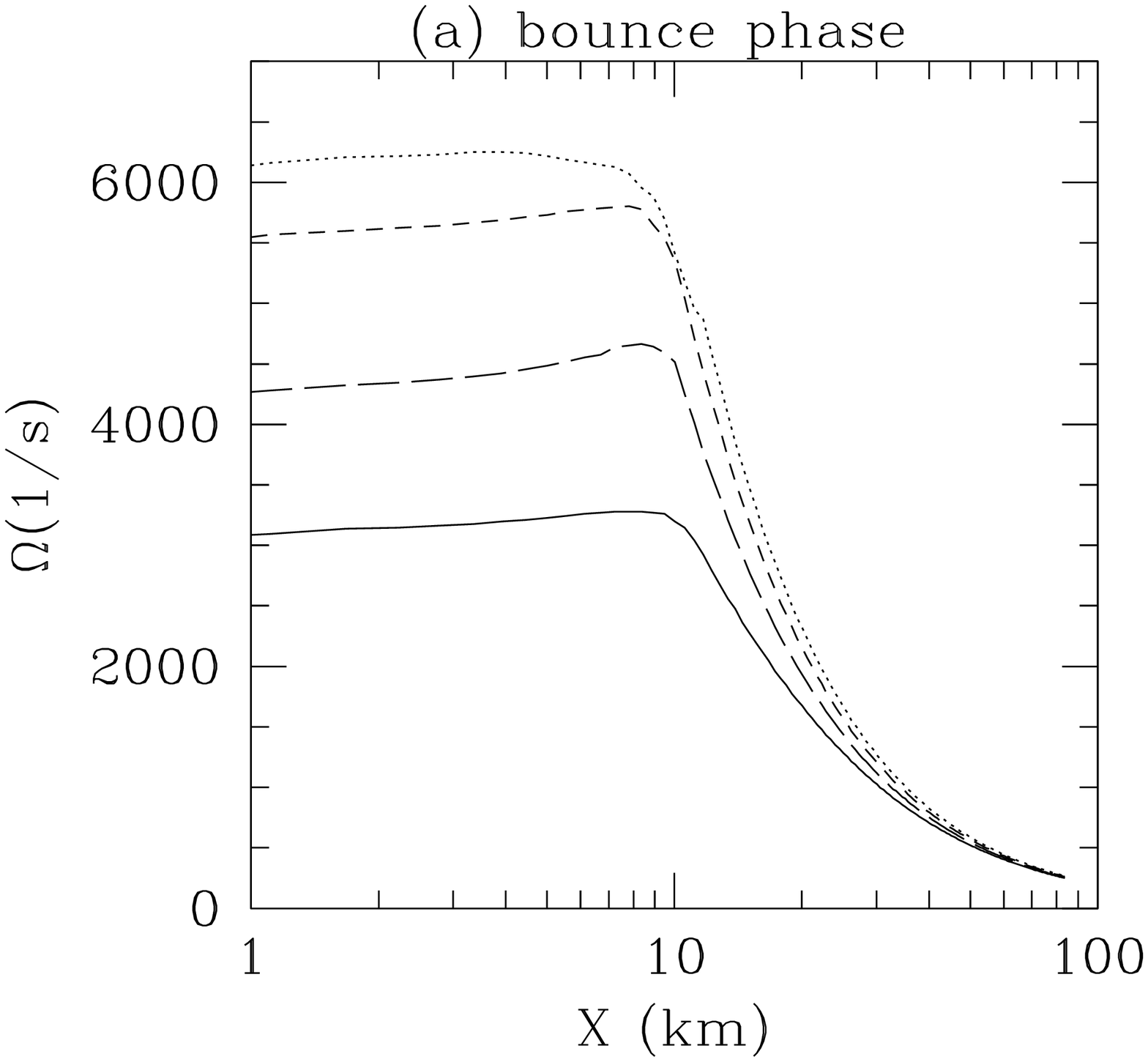} 
    \epsfxsize=2.3in
    \leavevmode
    \epsffile{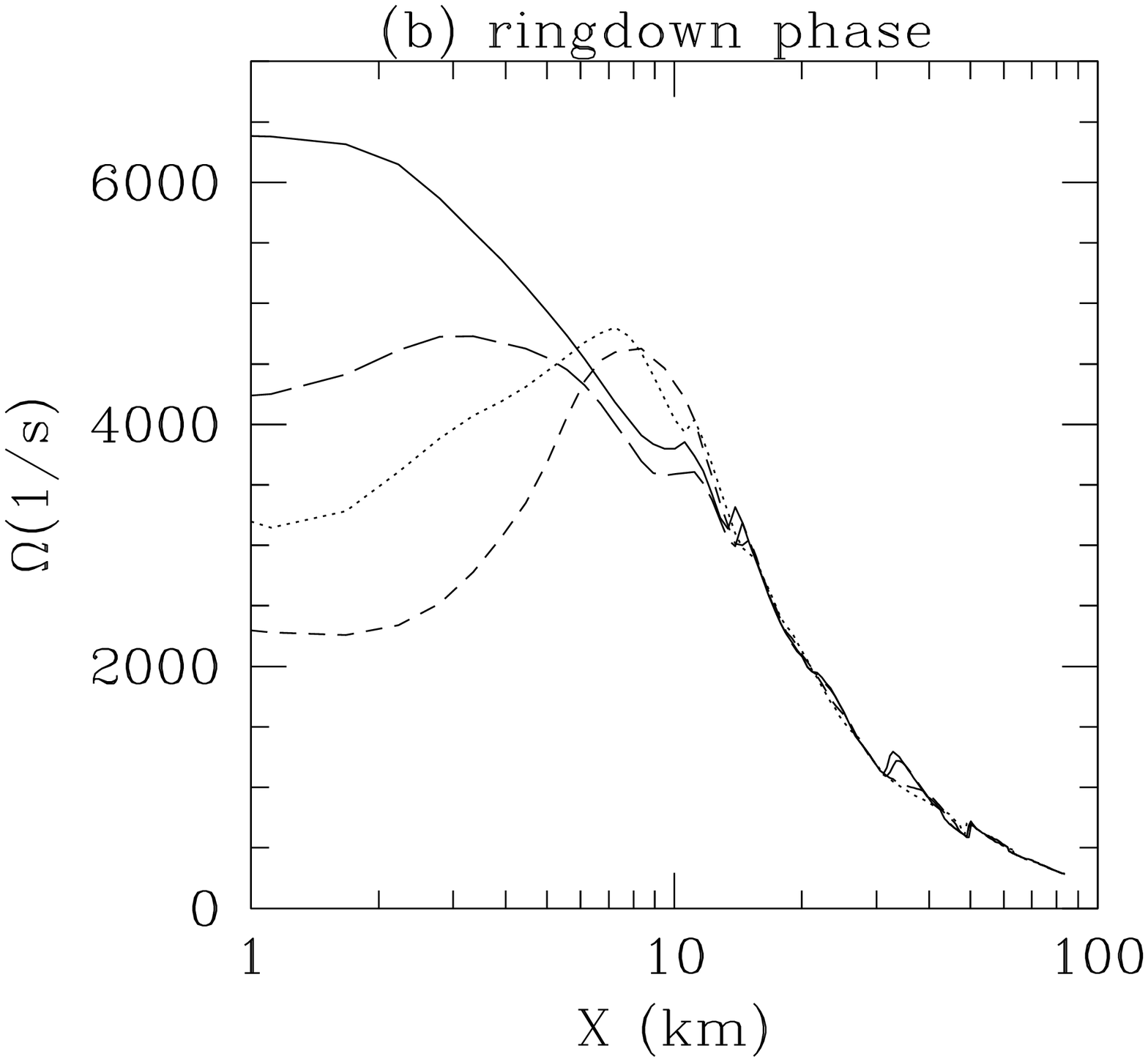}  
  \end{center}
  \vspace{-6mm}
  \caption{ The same as Fig. \ref{figure17} but for model D5c at
    selected time slices: (a)
    at the bounce phase of $t=114.46 $ (solid curve), 114.59 (long dashed
    curve), 114.72 (dashed curve), and 114.83 ms (dotted curve); (b) at
    the ringdown phase of $t=120.96$ (solid curve), 121.11 (long dashed
    curve), 121.67 (dashed curve), and 121.97 ms (dotted curve). 
  }\label{figure19}
\end{figure}
\begin{figure}[htb]
  \vspace{-6mm}
  \begin{center}
    \epsfxsize=2.3in
    \leavevmode
    \epsffile{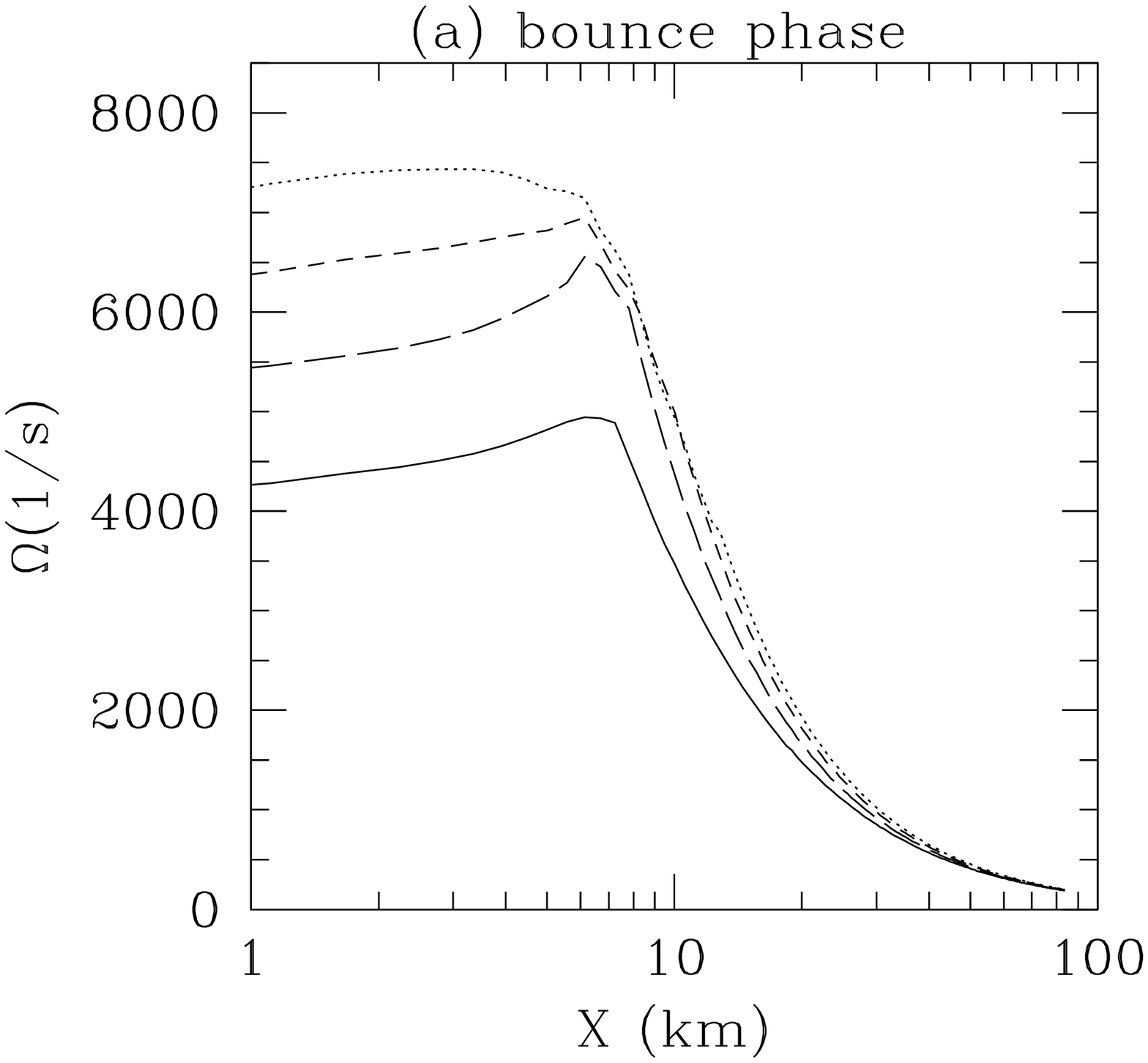} 
    \epsfxsize=2.3in
    \leavevmode
    \epsffile{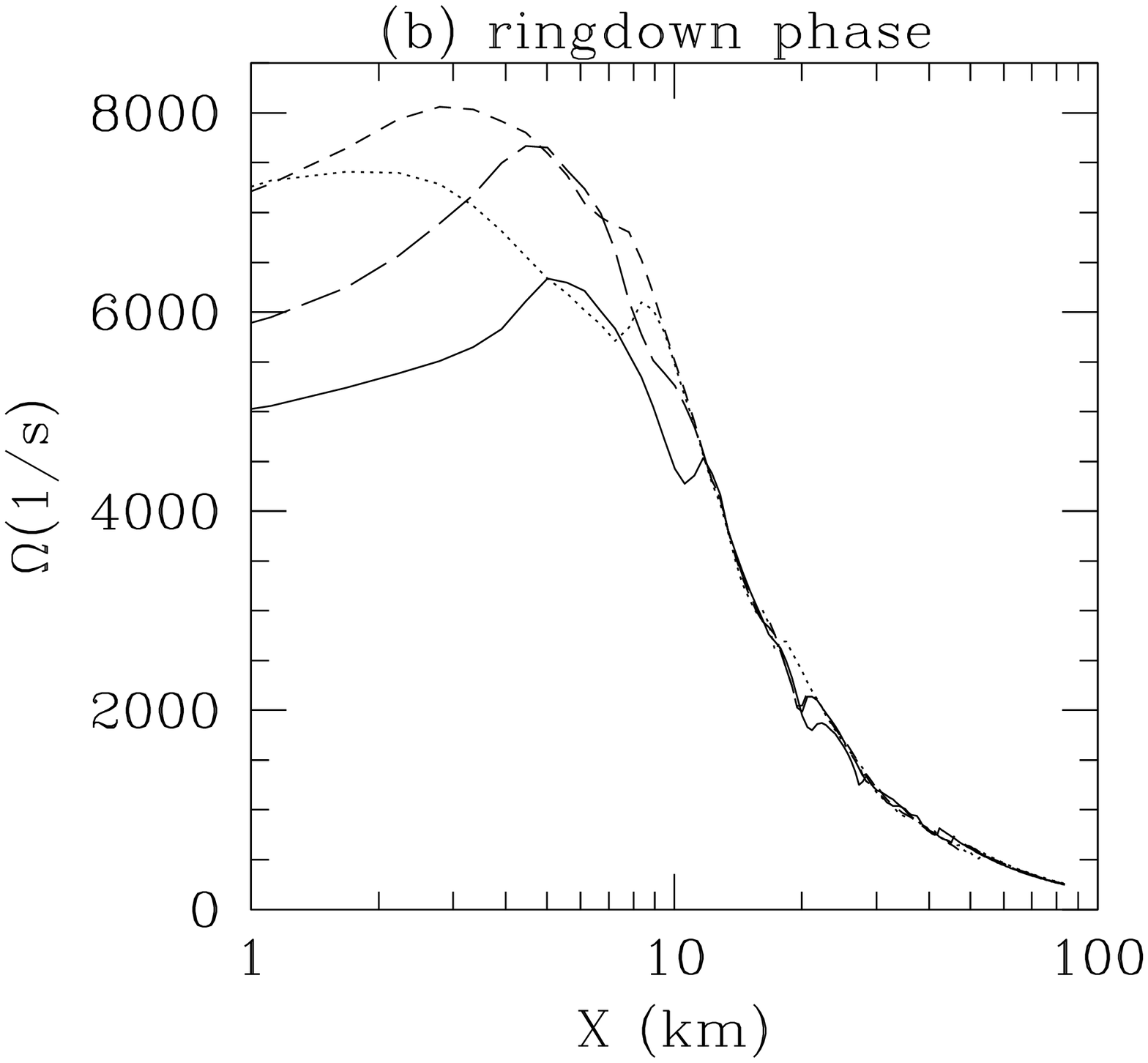}  
  \end{center}
  \vspace{-6mm}
  \caption{ The same as Fig. \ref{figure17} but for model D5d at
    selected time slices: (a)
    at the bounce phase of $t=83.78$ (solid curve), 83.91 (long dashed
    curve), 84.04 (dashed curve), and 84.15 ms (dotted curve); (b) at
    the ringdown phase of $t=86.10$ (solid curve), 86.32 (long dashed
    curve), 86.51 (dashed curve), and 86.71 ms (dotted curve). 
  }\label{figure20}
\end{figure}

In this subsection, we present the rotational profiles of the inner
region during the collapse and consider the possibility for the 
onset of the nonaxisymmetric instabilities.
Figures \ref{figure17}--\ref{figure20} show the angular
velocity profile along the $x$ axis at selected time 
slices for models D5a, D5b, D5c, and D5d, respectively. 
The first panels of Figs. \ref{figure17}-\ref{figure20} show the angular
velocity profile around the bounce phase. 
As the collapse proceeds, the angular velocity around the central
region increases toward the maximum value, $\Omega_{\rm max}$, achieved
at the bounce. 
The values are $\Omega_{\rm max} \approx 8000$, $10000$,
$6000$, and $7500$ s$^{-1}$ for models D5a, D5b, D5c, and D5d,
respectively (see dotted lines in the first panels). The inner core
at the bounce, when the maximum angular velocity is achieved, is
approximately rigidly rotating irrespective of the equations of state,
although the fluid elements outside the inner core is differentially
rotating as a whole. These results 
indicate that the differential rotation is not enhanced in the inner
core during the infall and bounce phases. 

The second panels of Figs. \ref{figure17}--\ref{figure20} 
show the angular velocity profiles in the ringdown phase, from which 
we find that differential rotation is enhanced in the ringdown
phase due to the oscillation of the protoneutron star and the infall
of the fluid elements of high specific angular momentum from the outer
region. However, the degree of the differential rotation is not very
high. This feature is also found in a Newtonian simulation \cite{Ott}. 
These panels also indicate that the amplitude of the
oscillation of the angular velocity is larger
for models with the equation of state 'c'. This is due to the fact
that the amplitude of the oscillation of the protoneutron star is
larger as described in Sec. \ref{dep_rhonuc}.

In the post-bounce oscillation phase,
the oscillation of the protoneutron star gradually damps. 
The damping is caused by the compressional work of the protoneutron
star to the infalling matter from the outer envelop. As a result of
damping, a quasistationary protoneutron star rotating rapidly at the
period  $P \equiv 2\pi / \Omega \sim 1$ ms is eventually formed.
Note that inner region of the formed protoneutron star is rigidly
rotating (see the third panel of Fig. \ref{figure18}).

To infer the nonaxisymmetric stabilities, we
estimate the value of $T_{\rm rot}/W$ at the bounce. 
Note that in general relativity, there is no unique definition
for $T_{\rm rot}/W$ for dynamical spacetimes. In this paper, we define
the rotational kinetic energy, $T_{\rm rot}$, and the gravitational
potential energy, $W$, for dynamical spacetime by
\beqn
&&T_{\rm rot} \equiv 
\frac{1}{2}\int d^{3}x \rho_{\ast}\hat{u}_{\varphi}v^{\varphi}, \\
&&W \equiv T_{\rm total} + U, 
{\rm ~~ where~} U \equiv \int d^{3}x\rho_{\ast} \varepsilon.  \label{def-W}
\eeqn
To define $W$, we use the fact that $M_{\rm ADM} \approx M_{\ast} - W
+ T_{\rm total} + U$ and $M_{\ast} \approx M_{\rm ADM}$. 
$T_{\rm total}$ denotes a total kinetic energy. 
Unfortunately, we do not know the appropriate definition for 
$T_{\rm total}$.
However, as far as configurations at the maximum
compression and at a final relaxed state are concerned, the rotational
kinetic energy $T_{\rm rot}$ is nearly equal to $T_{\rm total}$, and thus,
$T_{\rm total}$ in Eq. (\ref{def-W}) may be replaced by $T_{\rm rot}$.
In this definition, $T_{\rm rot}/W$ will give slightly
overestimated values in other stages such as the infall phase.
The value of $T_{\rm rot}/W$ at the bounce is $\approx 0.10$ and
$\approx 0.13$ for models D5b and D5a. 
These values are much smaller than the plausible
critical value of $T_{\rm rot}/W$ for the onset of the dynamical instability 
$T_{\rm rot}/W \approx 0.24$--0.25 for nearly rigidly rotating
stars \cite{Karino,SBS2,Ster}, and hence, the inner core at the bounce and
formed protoneutron star is unlikely to be unstable against
nonaxisymmetric perturbations in the dynamical time scale. 
However, these values are as large as the critical value
of $T_{\rm rot}/W$ for the secular instabilities driven by gravitational
wave emission \cite{Karino2,GW-secu,Ster}, or driven by viscosity
\cite{Vis-secu,Ster}. Therefore, the formed
protoneutron star may be unstable against nonaxisymmetric
perturbations in secular time scales $\gg 100$ ms. 

A sufficiently rapidly and differentially rotating iron core may 
collapse to form a protoneutron star of $T_{\rm rot}/W \agt 0.25$ and
become unstable against nonaxisymmetric perturbations.
Such a rotating nonaxisymmetric object will be a strong gravitational
wave emitter.
In a companion paper \cite{SS4}, we study conditions for the onset of
the dynamical instabilities in three-dimensional numerical simulation
in full general relativity. 
We find that the dynamical instabilities set in only for the
case that the progenitor is highly differentially rotating with $A
\alt 0.1$. Thus, the assumption of axial symmetry in this paper is
justified. 

\subsection{Gravitational waves}
\begin{figure}[htb]
  \vspace{-4mm}
  \begin{center}
    \epsfxsize=3.in
    \leavevmode
    \epsffile{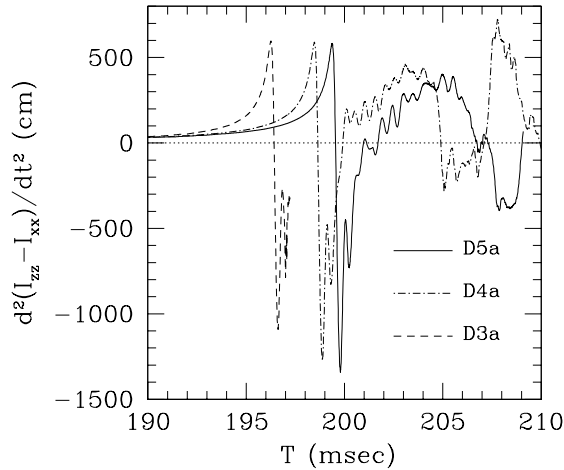} 
  \end{center}
  \vspace{-6mm}
  \caption{Gravitational waveforms computed by the quadrupole formula
    described in Sec. \ref{quad} for models D5a (solid curve) and D4a
    (dotted dashed curve), and D3a (dashed curve).
  }\label{figure21}
\end{figure}
\begin{figure}[htb]
  \vspace{-4mm}
  \begin{center}
    \epsfxsize=3.in
    \leavevmode
    \epsffile{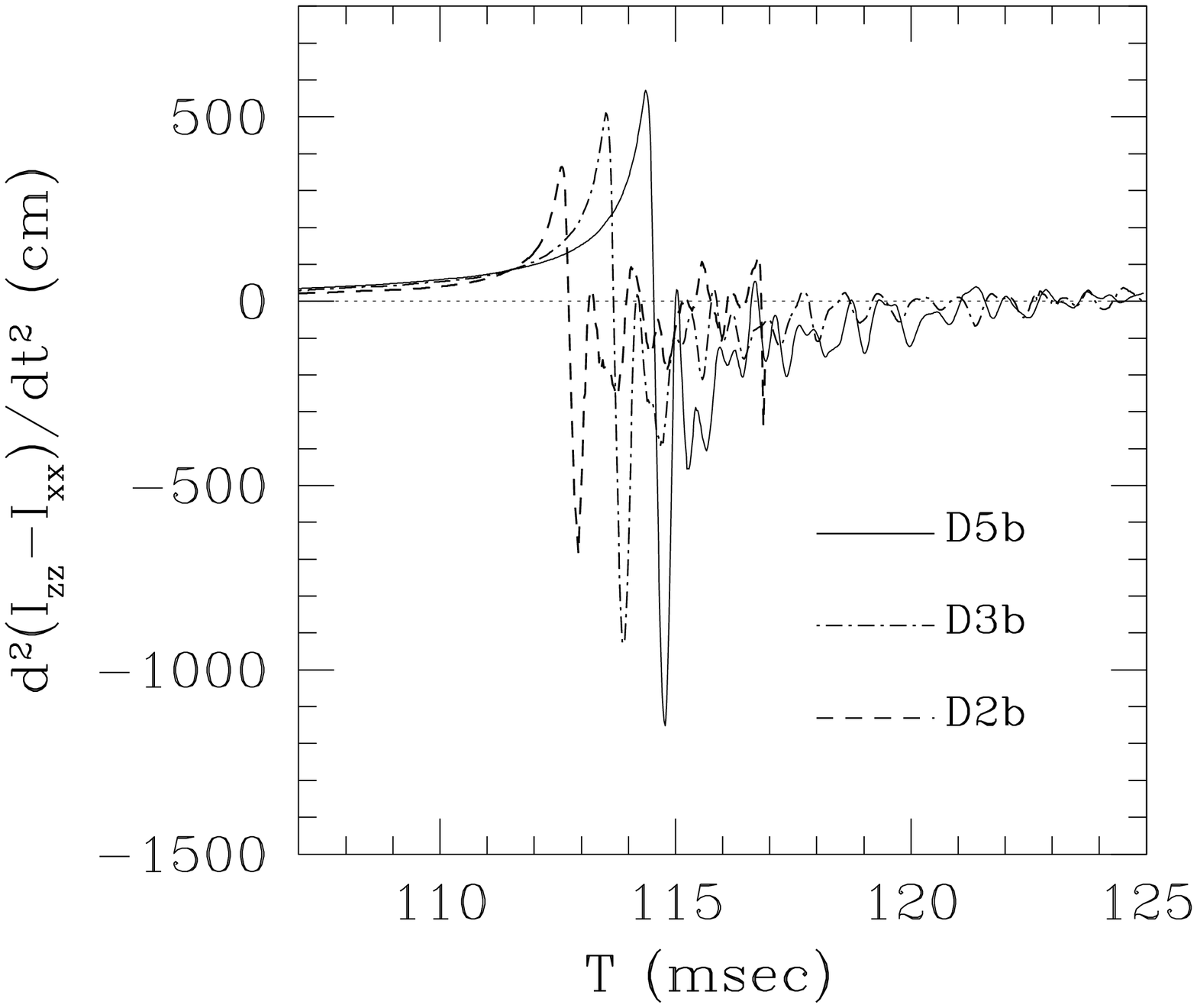} 
  \end{center}
  \vspace{-6mm}
  \caption{The same as Fig. \ref{figure21} but for 
    models D5b (solid curve), D3b (dotted dashed curve), and D2b (dashed
    curve). 
  }\label{figure22}
\end{figure}
\begin{figure}[htb]
  \vspace{-4mm}
  \begin{center}
    \epsfxsize=3.in
    \leavevmode
    \epsffile{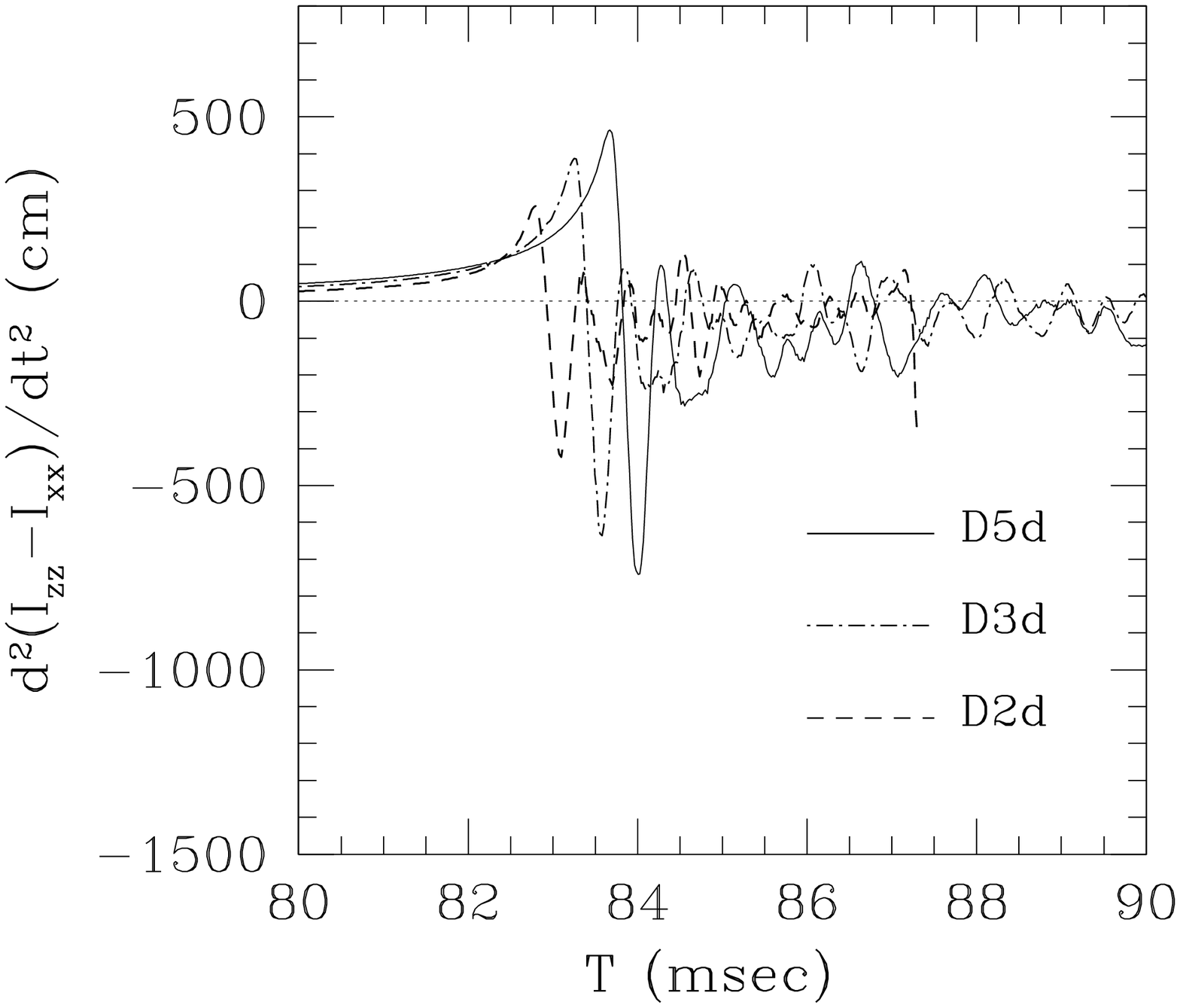} 
  \end{center}
  \vspace{-6mm}
  \caption{The same as Fig. \ref{figure21} but for models 
    D5d (solid curve), D3d (dotted dashed curve), and D2d (dashed
    curve).}\label{figure23}
\end{figure}
\begin{figure}[htb]
  \vspace{-4mm}
  \begin{center}
    \epsfxsize=3.in
    \leavevmode
    \epsffile{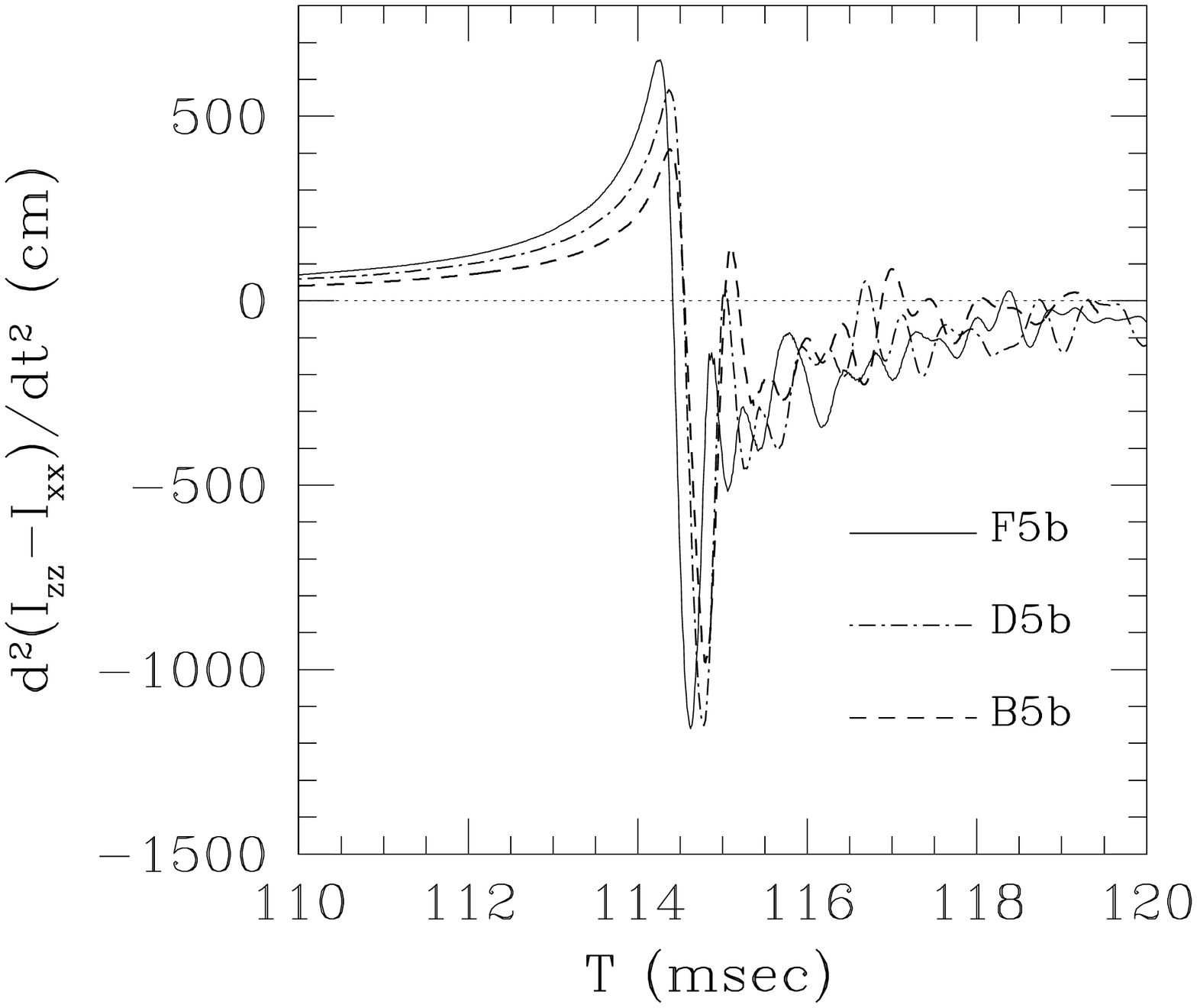} 
  \end{center}
  \vspace{-6mm}
  \caption{The same as Fig. \ref{figure21} but for models 
    F5b (solid curve), D5b (dotted dashed curve), and B5b (dashed
    curve). }\label{figure24}
\end{figure}
\begin{figure}[htb]
  \vspace{-4mm}
  \begin{center}
    \epsfxsize=3.in
    \leavevmode
    \epsffile{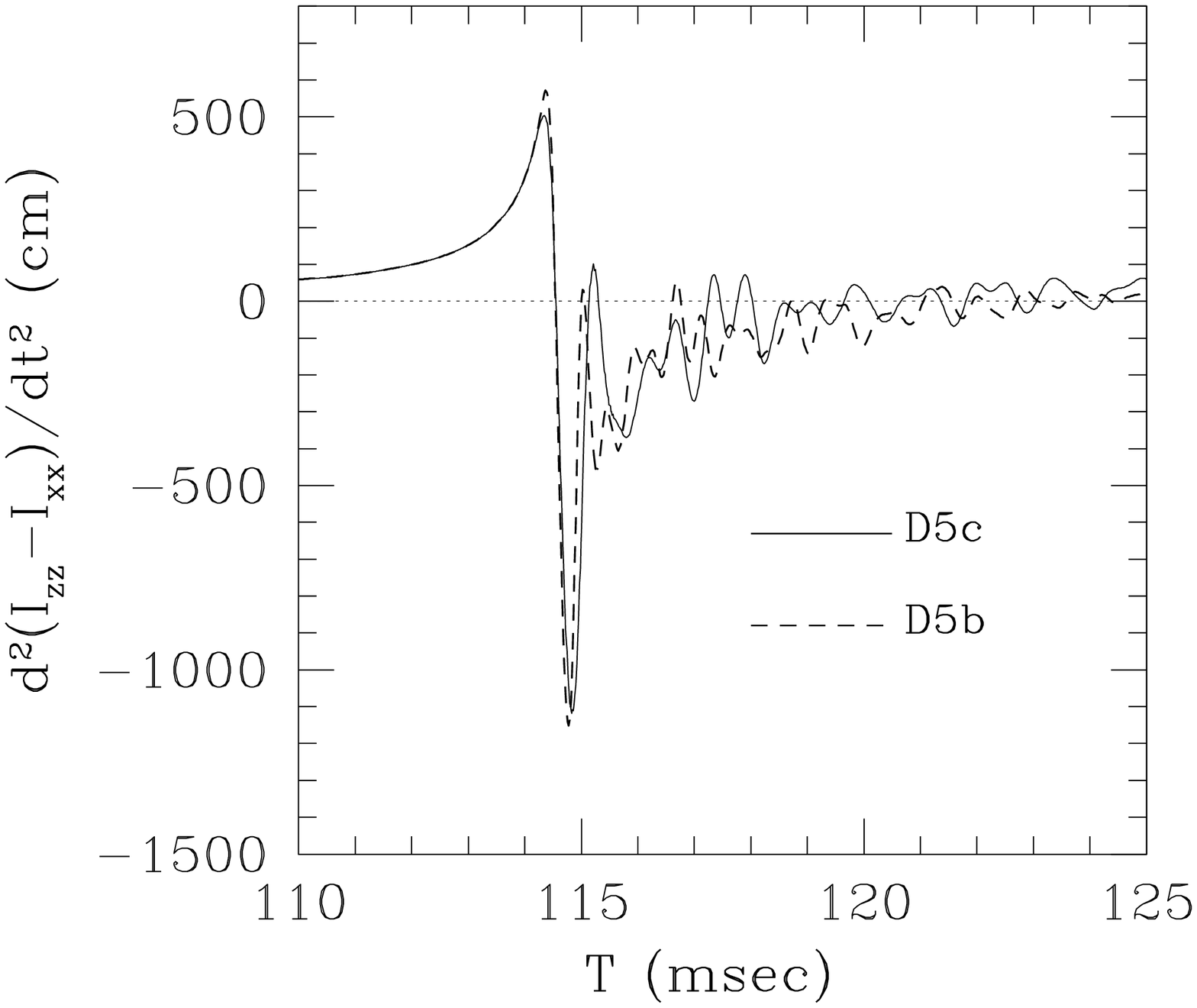} 
  \end{center}
  \vspace{-6mm}
  \caption{The same as Fig. \ref{figure21} but for models 
    D5b (dashed curve) and D5c (solid curve)}\label{figure25}
\end{figure}

\subsubsection{Gravitational waveforms}

Gravitational waveforms are computed in terms of the quadrupole
formula described in Sec. \ref{quad}. 
As illustrated in a previous paper \cite{SS}, approximate
gravitational waveforms can be computed even for highly relativistic,
highly oscillating, and rapidly rotating neutron stars using the
quadrupole formula 
except for systematic underestimate of the amplitudes of $O(M/R)$.
In the case of protoneutron star formation, gravitational waves are
emitted mainly by its oscillation. Thus, with the quadrupole formula,
it is possible to approximately compute gravitational waveforms
emitted during rotating iron core collapse to a protoneutron star. 

In Figs. \ref{figure21}--\ref{figure23}, gravitational waves for
models D2--D5 with equations of state 'a', 'b', and 'd' are shown. 
The gravitational waveforms during the black hole formation evaluated by
the quadrupole formula are also presented together for models D3a,
D2b, and D2d. We note that from $A_2=\ddot I_{zz}-\ddot I_{xx}$, 
the amplitude of gravitational waves at a distance of
$r$ from the source is calculated by
\beqn
h \approx 3 \times 10^{-20} \biggl( {A_{2} \over 1000~{\rm cm}}\biggr)
\biggl({10~{\rm kpc} \over r}\biggr) \sin^2\theta, \label{hamp}
\eeqn
where $\theta$ denotes the angle between the line of sight and the 
rotational axis, and $\langle \sin^2\theta \rangle=2/3$.
Thus, the typical amplitude is 
$\sim 2 \times 10^{-20}$ for a event at the galactic center. 

Figures \ref{figure21}--\ref{figure23} show
that the strong dependence of the dynamics of the collapse
on the values of $\Gamma_{1}$ is reflected in the amplitude of
gravitational waveforms. 
The following features are in particular worthy to note. 
First, the amplitudes of gravitational waves during the infall 
and bounce phases are smaller for the smaller value of $\Gamma_{1}$
(the larger value of $|\Gamma_{1}-4/3|$). 
The reason is that the smaller value of $\Gamma_{1}$ results in 
the smaller mass of the inner core at the bounce (cf. discussion in 
Sec. \ref{secGam}). This feature agrees with that found in previous
works \cite{Zweg,HD,SS2}. 

Second, the time-averaged amplitudes of the gravitational waves in the
ringdown phase are negative for models with equations of state 'b' and
'd' while positive for the model with 'a'.
This results from the difference of the dynamics in the infall
phase. For the equations of state 'a', the matter collapses and
bounces in a coherent manner, while for the equations of
state 'b' and 'd', in a less coherent manner.
In the case 'b' and 'd', therefore, the matter in the outer region
infalling toward the center suppresses the oscillation of the
protoneutron star. On the other hand, for models with 'a', 
such suppression dose not work effectively because of the
smaller mass fraction in the outer region.

Third, the gravitational waveforms for rapidly rotating models with
the equations of state 'a' (models D4a and D5a in Fig. \ref{figure21})
are qualitatively different from those for others. 
The modulation of the gravitational waveforms are quite 
remarkable. This reflects the bipolar explosion, in
which absolute value of $\ddot I_{zz}$ is much larger than that
of $\ddot I_{xx}$. 

Fourth, the so-called ``type-III'' gravitational waveforms \cite{Zweg}
are not found in any model.
Previous studies have indicated that type-III waveforms are generated
for a small value of $\Gamma_{1} \sim 1.28$ \cite{Zweg,HD,SS2}. In these 
waveforms, the amplitude of the first spike is significantly
suppressed to the value $|\ddot{I}_{zz}-\ddot{I}_{xx}| \sim 30$
cm. This is because the mass of the protoneutron star at 
the bounce phase is very small for such small value of $\Gamma_{1}$.
In the present case, the mass of the adopted iron core is much larger
than that for the previous studies \cite{Zweg,HD,SS2}. As a result,
the mass of the protoneutron star is not very small, and hence, the
type-III gravitational waveforms are not generated.

For a given mass, the amplitude of gravitational waves is increased
with the increase of the angular momentum in the present numerical
results. However, this is not trivial result because of
the following reasons.  First, recall that the amplitude of
gravitational waves is proportional to the quadrupole moment 
($|I_{zz}-I_{xx}|$) and the inverse square of the dynamical 
time scales, $\tau_{\rm char}$, of the system. The value of 
$|I_{zz}-I_{xx}|$ increases as the total angular momentum increases
since the radius and the degree of deformation of the inner core
become larger due to the increasing centrifugal force. On the other
hand, the characteristic time scale becomes longer as the angular
momentum of the inner core increases since the rotation effectively supplies
additional pressure to the inner core and the collapse is decelerated.
Thus, these two contrary effects may cancel each other with a certain
value of spin parameter $q$, resulting in the saturation of the
burst-like gravitational wave signals.  Indeed, such saturation was
found at $q \sim 1$ in previous Newtonian studies (e.g., \cite{Yamada})
since at such a large value of $q$, the iron core does not collapse
to a sufficiently compact state. However, saturation is not found in
the general relativistic studies \cite{HD,SS2}. This implies that the general
relativistic gravity is strong enough to overcome the centrifugal force
for the rigidly rotating case with $q \alt 1$. This effect also
suppresses to yield type-II gravitational waves \cite{HD} which
appear in the case that the centrifugal force is too strong
for the iron core to reach the nuclear density. 

In Fig. \ref{figure24}, we show gravitational waveforms for
models B5b, D5b, and F5b to illustrate the dependence of
them on the mass of the iron core. In the infall phase, 
the amplitude of gravitational waves is larger for the larger-mass model. 
This is quite natural since for the larger-mass model, 
the quadrupole moment of the inner cores are larger. 
On the other hand, the amplitude of the strong spike at 
the bounce saturates when the mass of the iron core reaches $\approx 2.5
M_{\odot}$. The plausible reason is that the inner core shrinks to be so
compact that the quadrupole moment is decreased significantly. 

The effect of mass on the amplitude of gravitational waves can be
analyzed by comparing the present results with the previous results 
in \cite{SS2}, in which collapse of the iron core of mass $M \approx
1.5M_{\odot}$ and rotational energy $T_{\rm rot}/W \approx 0.009$ are
studied. In \cite{SS2}, we found that 
the amplitudes of gravitational waves at the bounce are 
$|\ddot{I}_{zz}-\ddot{I}_{xx}|$ $\approx 650, 200$, and $30$ cm for
models with ($\Gamma_{1}$, $\Gamma_{2}$) = (1.32, 2.0), (1.31, 2.5),
and (1.28, 2.5). 
These results should be compared with the present results for D5a, D5b,
and D5d, in which $M \approx 2.5M_{\odot}$ and $T_{\rm rot}/W \approx
0.009$ and the amplitudes of gravitational waves are 
$|\ddot{I}_{zz}-\ddot{I}_{xx}| \approx$ 1350, 1200, and 750 cm,
respectively. These values are larger than those found in \cite{SS2}
by factors of $\approx 2$, 6, and 25. Thus, with the increase of the
mass, the amplitude is increased in a nonlinear manner.
For $\Gamma_1 =1.32$, the
increase factor is small. The reason is that for this case,
the inner core at the bounce becomes sufficiently large due to
the coherence collapse in the infall phase, and 
the mass is close to the value for the saturation as mentioned above.
For other cases, the increase factor is much larger. 
In particular, the increase factor for $\Gamma_1=1.28$ is
outstanding. This is
associated with the fact that the dynamics in the bounce phase
qualitatively changes with the increase of mass. 
For $\Gamma_1=1.28$ with a small mass $M \sim 1.5M_{\odot}$,
the mass of the inner core formed at the bounce is very small, and as a
result, type-III gravitational waves are emitted \cite{HD} with 
the small maximum amplitude as $|\ddot{I}_{zz}-\ddot{I}_{xx}| \sim 30$ cm.
On the other hand, for $\Gamma_1=1.28$ with a large mass
$M \sim 2.5M_{\odot}$, the mass of the inner core can be
sufficiently large due to its strong self-gravity. As a result,
type-I gravitational waves are emitted and 
the amplitude of gravitational waves is significantly increased. 

An interesting finding in gravitational waveforms for the larger-mass
model is that the first spike in the ringdown phase 
emitted just after the strong negative spike at the bounce 
decreases as the mass of the iron core increases (see Fig. \ref{figure24}). 
This spike is associated with the outward motion of the inner
core and usually positive for the smaller-mass models (such
as models of $M\approx1.5M_{\odot}$ studied in \cite{SS2,HD,Zweg}).
For models of larger mass such as studied in this paper, on the other
hand, the post-bounce first spike becomes negative due to suppression
of the outward oscillation by the infalling matter. The suppression is
larger in particular in the direction of the rotational axis
because of the absence of the centrifugal force. This decreases 
the oscillation amplitude to be negative. 

Gravitational waveforms for models D5b and D5c are compared to see the
dependence on $\rho_{\rm nuc}$ in Fig. \ref{figure25}. It shows that 
differences in the dynamics associated with the difference in
the value of $\rho_{\rm nuc}$ are reflected in gravitational waveforms.
For example, the amplitude
at the bounce for D5c is smaller than that for D5b. This is because the
central density at the bounce is smaller for D5c, and accordingly, the
characteristic time scale is longer.
On the other hand, the amplitude of the post-bounce first spike is
larger for D5c. This is because the amplitude of oscillation of the
inner core is larger for D5c as described in Sec. \ref{dep_rhonuc}.

Before closing this subsection, we comment on the
frequency of gravitational waves emitted in the case of 
black hole formation. After the formation of a black hole, 
the ambient matter falls into it. As a result, 
the so-called quasi-normal modes (e.g. \cite{Kokkotas}) of
the black hole will be excited and gravitational waves associated
with the spacetime oscillation will be emitted. 
According to Echeverria \cite{Eche}, semiempirical expression for the
real part of the quasi-normal mode frequency for $l=2$ is given by 
\beq
f \approx 12 \left( \frac{M}{M_\odot}\right)^{-1} 
\left( \frac{F(q)}{37/100} \right) ~{\rm kHz}, \label{qnm}
\eeq
where
\beq
F(q) \approx 1-\frac{63}{100}(1-q)^{3/10}.
\eeq
Equation (\ref{qnm}) indicates that
the frequency of gravitational waves associated
with the quasi-normal mode is $\sim 7.5$~kHz for models D3 and F5 for
which the mass and spin parameter of the formed black hole are
predicted to be $M \approx 2.3$--$2.4 M_{\odot}$ and $q \approx 0.7$ 
(see Sec. \ref{Prediction}). This value is far out of the
best-sensitive frequency range for detection of laser interferometric
gravitational wave detector \cite{KIP}. 

In reality, the mass of black hole will significantly increase in a
longterm evolution since the matter outside the iron core will fall 
into a black hole. After such accretion, the frequency of the
quasinormal mode will be decreased significantly. Gravitational waves 
emitted from a high-mass black hole $M \agt 20 M_{\odot}$ may have an
appropriate frequency for detection by the laser interferometric
detectors.
\subsubsection{Energy power spectra}

\begin{figure}[htb]
  \vspace{-8mm}
  \begin{center}
    \epsfxsize=2.8in
    \leavevmode
    (a)\epsffile{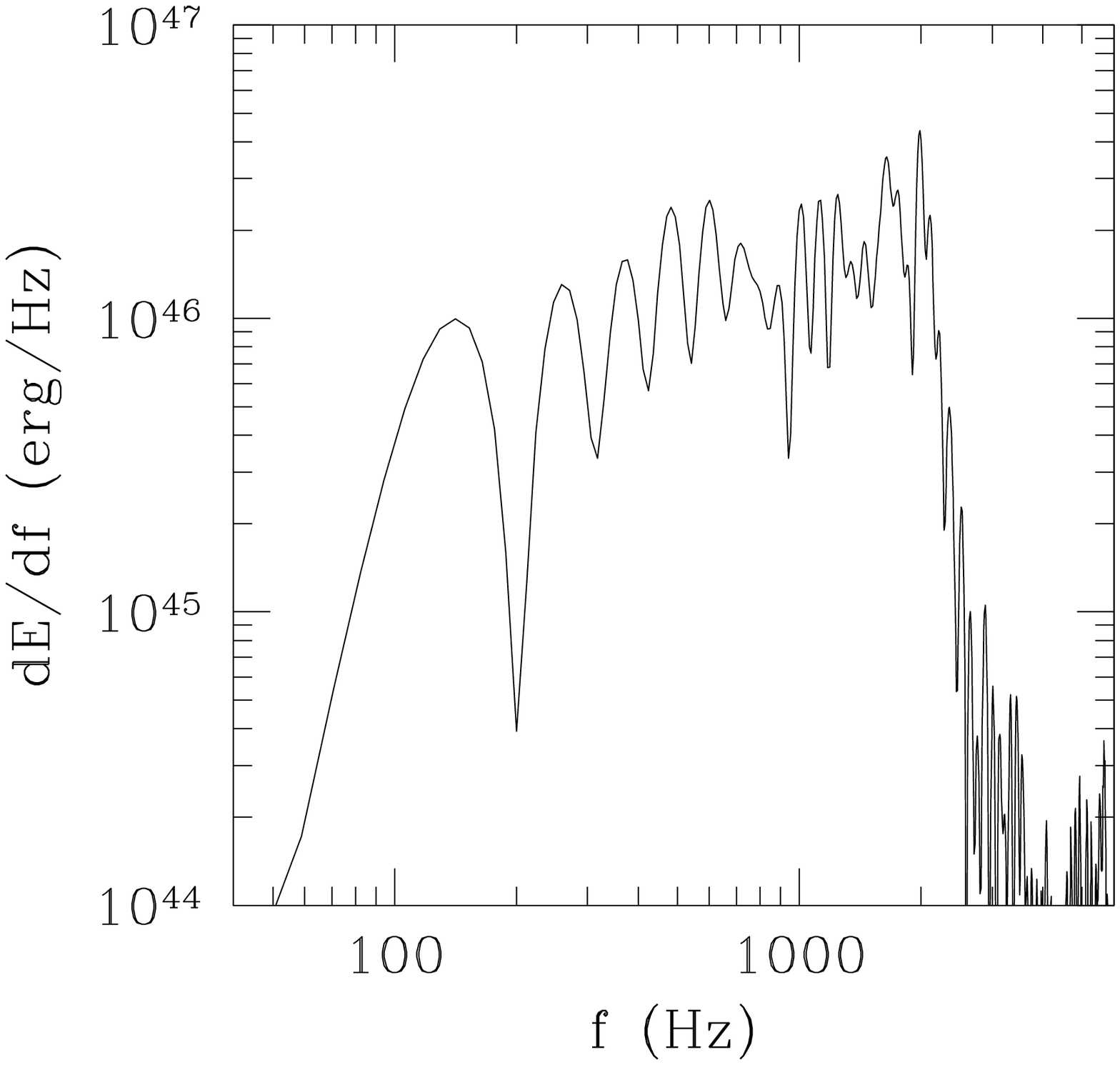}
    \epsfxsize=2.8in
    \leavevmode
    (b)\epsffile{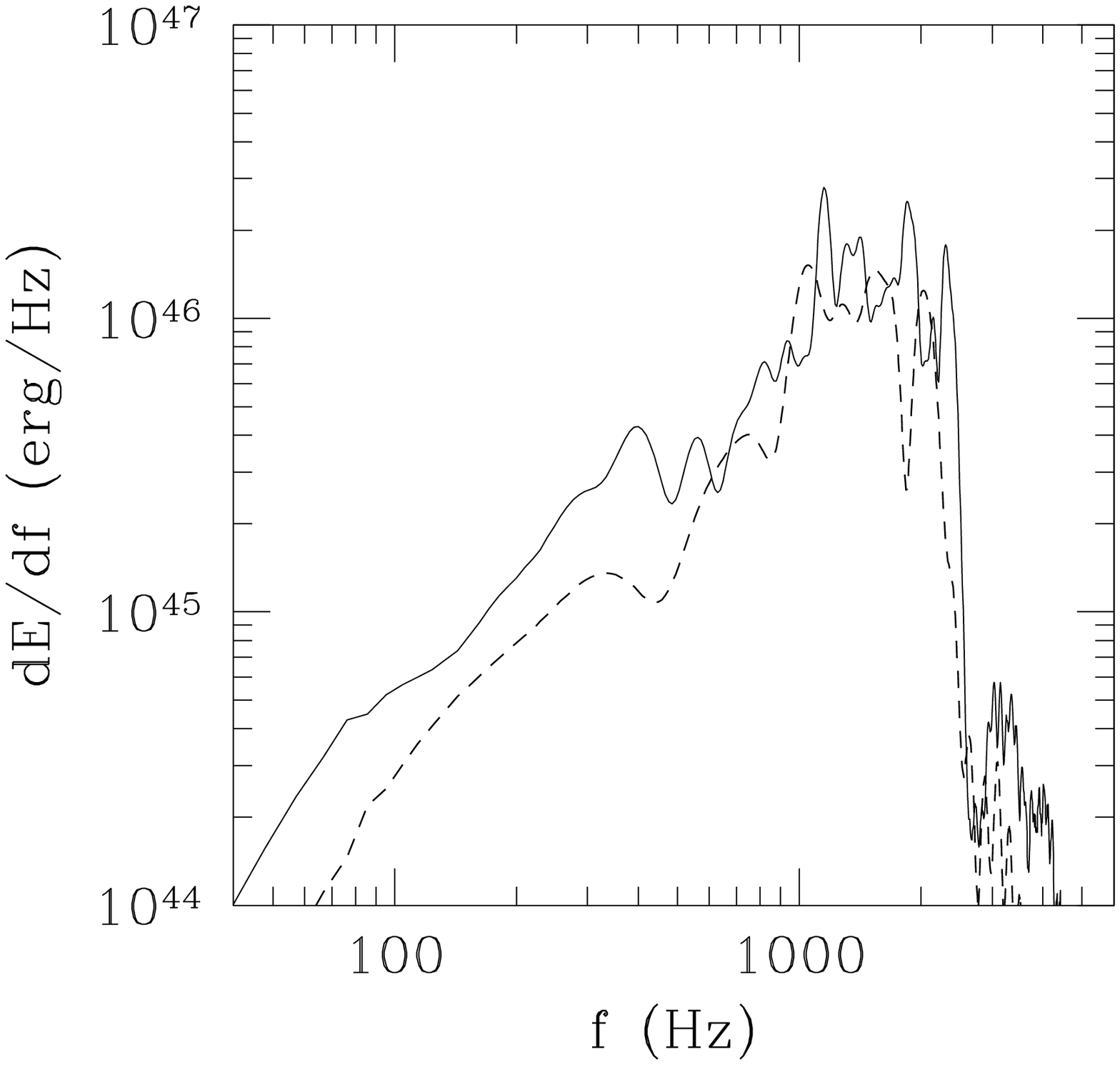} \\
    \epsfxsize=2.8in
    \leavevmode
    (c)\epsffile{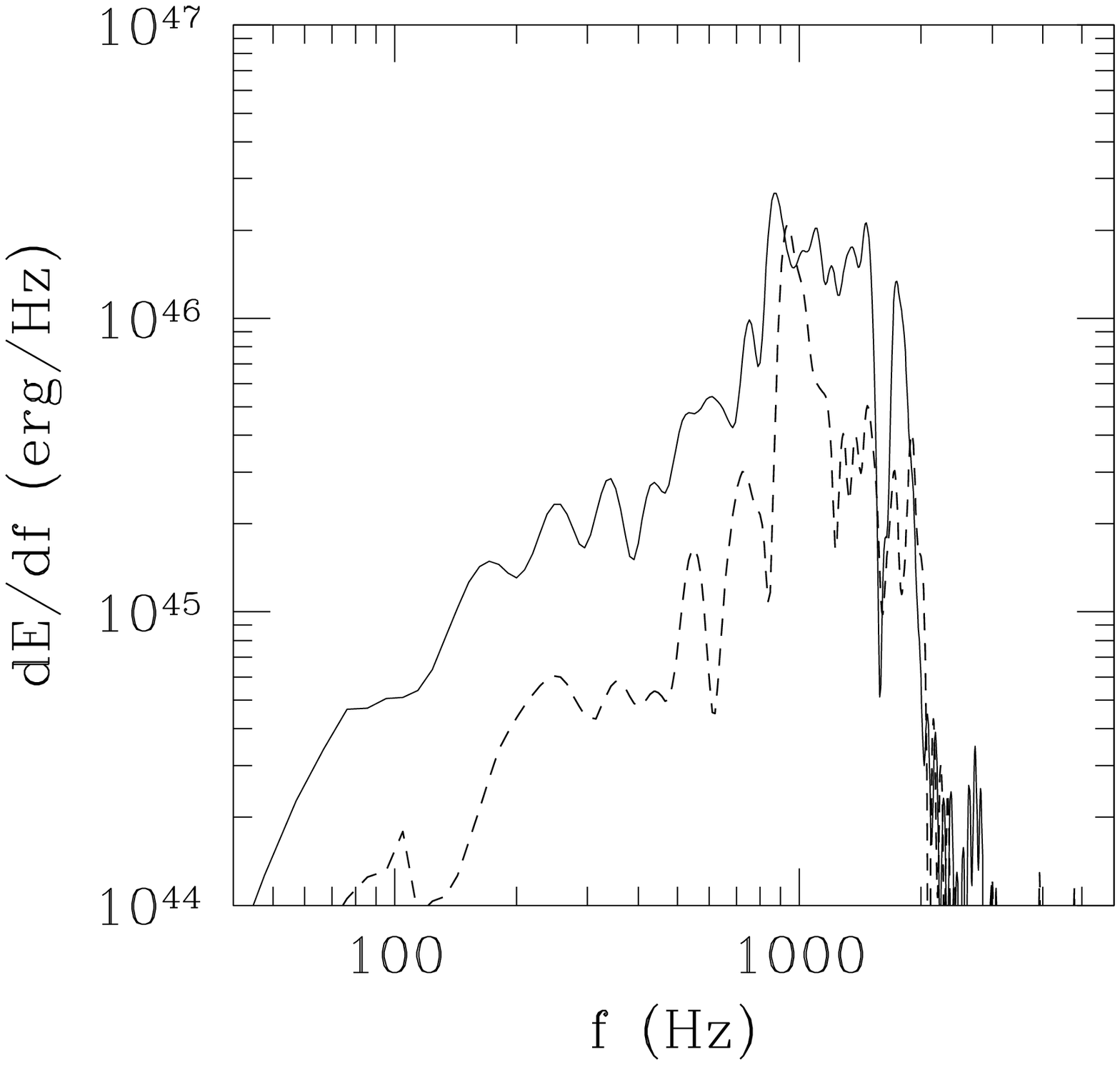}
    \epsfxsize=2.8in
    \leavevmode
    (d)\epsffile{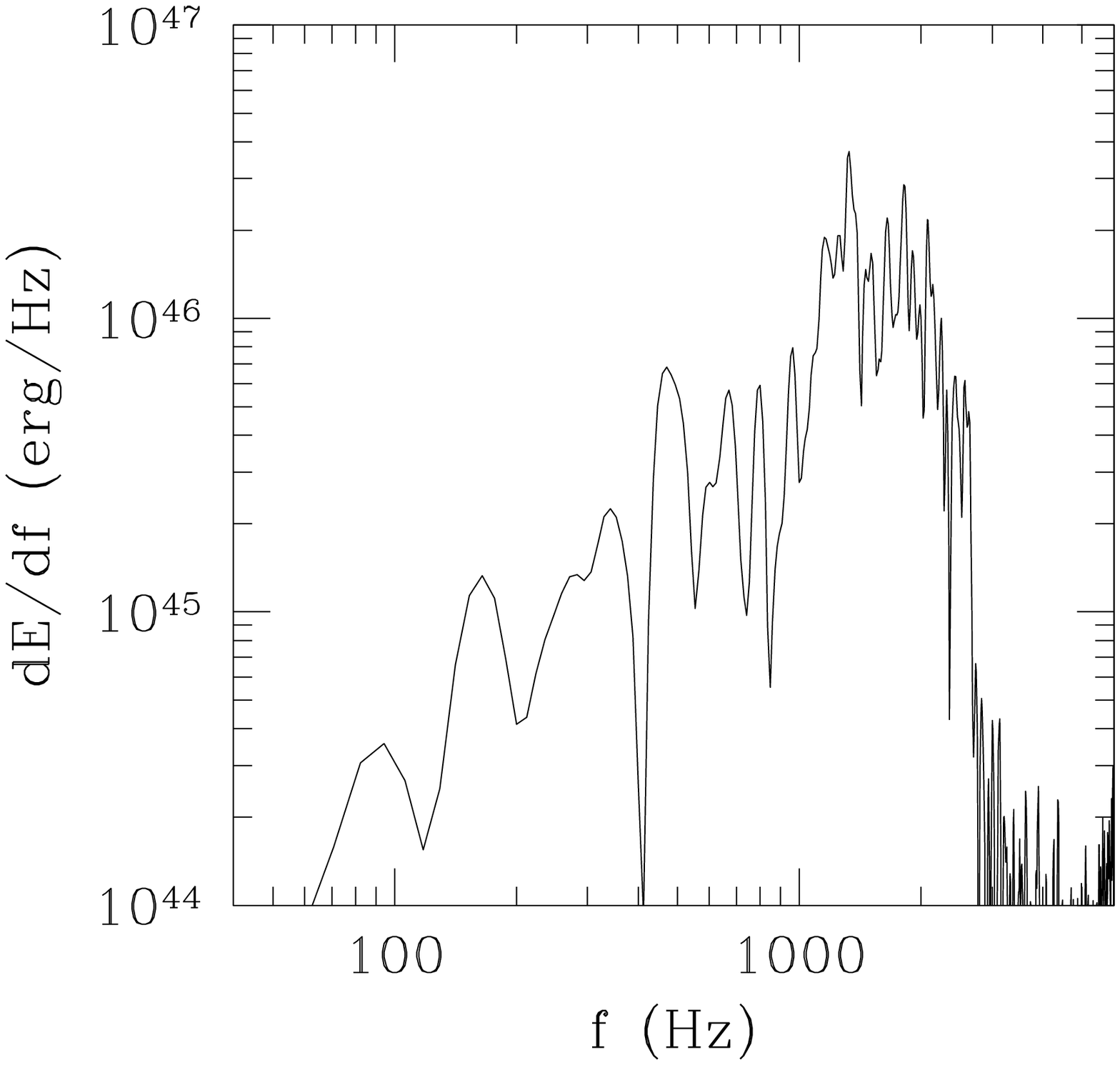}
  \end{center}
  \vspace{-4mm}
  \caption{The energy power spectra of $\tilde{A}_{2}(f)$ mode for
  (a) model D5a, (b) models D5b (solid curve) and A5b (dashed curve),
  (c) models D5c (solid curve) and D2c (dashed curve), and (d) model
  D5d. Note that gravitational waveforms for models D5a, D5b, D5c, and D5d
  are shown in Figs. \ref{figure21}, \ref{figure22}, \ref{figure25}, and
  \ref{figure23}, respectively}
\label{figure26} 
\end{figure}
In Fig. \ref{figure26}, we show the energy power spectra of 
gravitational waves with a mode of $l=2$ and $m=0$ 
for selected models. For all the models, the maximum values of the spectra are
located at $f_{\rm peak} \approx 1$--2 kHz, which are consistent with
the typical bounce time intervals of these models $\sim 0.5$--1 ms.
The value $f_{\rm peak} \approx 1$--2 kHz is 
by a factor of $\sim 2$ larger than the 
previous Newtonian results \cite{Zweg,Ott}.
This reflects the features in the dynamics of the collapse
that the central density and compactness of the
inner core at the bounce in general relativity
are larger than those in the Newtonian results due to 
the stronger attraction force \cite{HD}.
The peak frequency is slightly larger than
that found by Dimmelmeier et al. \cite{HD}. This is due to the fact that 
the mass of the iron core adopted in this work is larger than theirs. 
For $f \agt 2$ kHz, the spectra decline steeply for all the models.
This feature is consistent with that in \cite{HD,Ott}. 

The spectrum for model D5a (see Fig. \ref{figure26}(a)) is rather broad 
for a low frequency region ($100 \alt f \alt 1000$ Hz) and flatter 
than those for other models. Such characteristic spectrum 
results from the nature of the bipolar explosion. 
For models D5b--D5d (see Fig. \ref{figure26}(b)--(d)),
on the other hand, the spectra are quite 
similar to that of typical type-I burst-like gravitational waves
\cite{Monch,Zweg,HD}. In these cases, a few sharp peaks appear at
the frequency between 1 and 2 kHz. 
In Fig. \ref{figure26}(b), the energy power spectrum
for smaller-mass model A5b is shown for comparison with the spectrum for
model D5b. It is found that the shape of the spectra for these two models
is quite similar although the height of the  
peak for model A5b is smaller than that for model D5b 
and the peak slightly shifts to the low frequency side 
due to the fact that the mass is smaller. 

The spectrum for model D5c (see Fig. \ref{figure26}(c))
is similar to that for model D5b except for slight shift of the
spectrum to low frequency side. The smaller peak frequency results from 
the fact that the central density at the bounce and of the 
formed protoneutron star for model D5c is smaller than those
for model D5b (cf. Figs. \ref{figure3} and \ref{figure4}),
and hence, the dynamical time scale is longer. 
In Fig. \ref{figure26}(c), the energy power spectrum
for model D2c is also shown to see dependence of the spectra
on the angular momentum of the progenitor. 
The magnitude of the spectrum for model D2c is overall smaller than that
for model D5c, indicating the magnitude of the angular momentum
plays an important role for amplifying the gravitational wave amplitude. 
Also found is that the peak at $f \approx 1$ kHz is
dominant for model D2c. This indicates that for the smaller value of
the angular momentum, a single mode is dominantly excited. 
The reason is likely to be that the degree of deformation of the
inner core is smaller for the smaller value of the angular momentum, and
hence, the fundamental quadrupole mode of the formed protoneutron
star is dominantly excited. 

The spectrum for model D5d (see Fig. \ref{figure26}(d)) has a more
complicated shape than those for models D5b and D5c. In particular, 
several small peaks in the low frequency region are found 
between 100 and 1000 kHz, indicating that several oscillation modes
are excited simultaneously. This reflects the manner on the 
generation of the shocks at the surface of the protoneutron star
in the equation of state 'd':
As mentioned in Sec. \ref{NSform}, the mass of the formed protoneutron
star is initially small with the equation of state 'd', and as
a result, the accretion of the matter which gradually falls from the shock
layer subsequently proceeds. The accreted matter intermittently 
hits the protoneutron star and excites its oscillation modes 
in a complicated manner. As a result, gravitational waves of
several characteristic oscillation frequencies are generated. 
\section{Summary}\label{Summary}

We performed fully general relativistic simulations for black hole
formation through the collapse of rotating iron cores on assumption of
the axial symmetry for a wide variety of the mass, the angular
momentum, the velocity profile, and the equation of state. To
systematically study the dependence of the threshold mass for the prompt
black hole formation on the equations of state, we adopt a parametric
equation of state \cite{Janka,Zweg,HD}.  We choose the parameters so
that the maximum mass of the cold spherical neutron star becomes
$\approx 1.6M_{\odot}$.  The rotating iron cores before collapse are
simply modeled by the rotating $\Gamma =4/3$ polytrope in equilibrium. 
The mass of the iron core is set to be $\approx
2.0$--$3.0M_{\odot}$. The amount of rotational kinetic energy of the
core is changed in a wide range from $T_{\rm rot}/W = 0$ to $\approx
0.009$.

We have found that the threshold mass for the prompt black hole formation 
depends sensitively on the angular momentum, the rotational
velocity profile of the initial core, and the equations of state. 
The dependence of the threshold mass on the these elements
can be summarized as follows: 
(i) The thermal pressure generated by shocks increases the
threshold mass by 20--40$\%$. The magnitude of this factor
depends on the adopted equation of state: 
(ii) With the increase of the spin parameter $q$, the threshold mass
increased by $\sim 25 q^2~\%$ for the case that the progenitor is
rigidly rotating:  
(iii) Effect of differential rotation further increases the threshold mass, 
since the centrifugal force around the central region can be efficiently 
increased: 
(iv) The threshold mass depends sensitively on the equations of state
since the dynamics of the collapse does. 
For the smaller value of $|\Gamma_{1}-4/3|$ and for the larger value of
$\rho_{\rm nuc}$, prompt black hole formation becomes more liable
since the mass of the inner core at the bounce is larger for such
equations of state. 

About the dynamics of the collapse, we have found the following features:
(i) if $\Gamma_{1}$ is close to $4/3$, the collapse proceeds in an 
approximately homologous manner, and thus, the most part of the iron 
core collapses nearly simultaneously.
This implies that the mass of the inner core
formed at the bounce is larger for the smaller value of $|\Gamma_{1}-4/3|$,
that helps prompt black hole formation as mentioned above. 
In the case that the mass of the progenitor is not large enough,
not a black hole but a protoneutron star is formed. 
Since the mass of the inner core is larger for the small value of
$|\Gamma_{1}-4/3|$, the inner core shrinks more significantly 
resulting in a larger degree of the aspherical deformation and in 
a significant spin-up. 
As a result, the shape of the shock is highly aspherical and
the explosion proceeds in a strongly bipolar manner if the progenitor is
rapidly rotating as $T_{\rm rot}/W \approx 0.009$ (see Sec. \ref{Bipolar}): 
(ii) For $\Gamma_{1} \alt 1.3$, the collapse does not 
proceed in the homologous manner. Instead, only the central region
rapidly collapses and forms the inner core of a small mass at the bounce. 
As a result, prompt black hole formation is less liable. 
For the cases of protoneutron star formation, the aspherical
deformation of the inner core is not very remarkable since its mass at
the bounce is not sufficiently large, and hence, the effect of the 
centrifugal force is not very important. Consequently, 
the shape of the shock is only sightly aspherical even when 
the progenitor is rapidly rotating (see Sec. \ref{Bipolar}): 
(iii) For the equation of state 'c' in which the value of $\rho_{\rm nuc}$
is smaller than that for others, the pressure for 
$10^{14}~{\rm g/cm^3} \leq \rho \alt 2\times 10^{15}~{\rm g/cm^3}$ is larger
than that for the equations of state 'b' and 'd'.
As a result, strength of the shocks formed at the bounce is enhanced, 
and therefore, black holes are less liable to be formed. 

Gravitational waveforms are approximately computed in terms of a
quadrupole formula \cite{SS}. It is found that the amplitude of
gravitational waves at the bounce
increases monotonically as the mass and angular momentum of the iron
core increase as far as $M \alt 2.5 M_{\odot}$. 
In contrast to the previous results (e.g., \cite{Yamada}),
we do not find any tendency that 
the maximum amplitude saturates with the increase of the angular momentum 
for a given mass. This is due to the fact that the general
relativistic gravity is strong enough for the inner core
to form a compact protoneutron star overcoming the centrifugal force
in the present choice of the spin parameter ($0 \leq q \alt 1$) for 
the rigidly rotating case. 

Gravitational waveforms depend sensitively on the equation of state. 
For models with the equations of state 'b', 'c', and 'd', 
the so-called type-I gravitational waves are emitted
even with the mass $M \sim 2$--$2.5M_{\odot}$ 
as in the case of mass $M \approx 1.4$--$1.5M_{\odot}$ which
is studied in the previous papers \cite{HD,SS2}. With the increase
of the mass, the amplitude and the frequency of gravitational waves
become higher. Thus, the difference is only quantitative for the
equations of state 'b' and 'c'. 
A point worthy to note is that type-I gravitational waves
are emitted for $M \agt 2 M_{\odot}$ even for the equation of state 'd'
(with $\Gamma_1 =1.28$). For $M \approx 1.5M_{\odot}$, 
type-III gravitational waves are emitted for such small value of
$\Gamma_1$ since the mass of the inner core is very small at the bounce.
However, in the higher-mass case with $M \agt 2M_{\odot}$, 
the mass of the inner core formed at the bounce is large
enough to emit type-I gravitational waves. 
Gravitational waveforms in the collapse of rapidly rotating iron core with
the equation of state 'a' in which $\Gamma_{1} = 1.32$ are 
qualitatively different from others. The reason is that in such case, 
an outstanding bipolar explosion is induced along the rotational axis and 
gravitational waves associated with such extreme explosion 
and with the resulting oscillation of the inner core in the direction
of the rotational axis are emitted. Due to this change, 
the shape of the energy power spectrum is also qualitatively different
from others. 

In this paper, we assume that the collapse proceeds in the
axisymmetric manner. However, the rotating iron core may be
dynamically unstable against the nonaxisymmetric deformation
during the collapse if the spin is increased
significantly during the collapse, and as a result, 
the value of $T_{\rm rot}/W$ exceeds a critical value $\sim 0.27$.
To determine the criterion for the onset of nonaxisymmetric 
instabilities, we performed three-dimensional simulation for the
iron core collapse in a companion paper \cite{SS4}, which 
shows that in the collapse for rigidly rotating models, 
the value of $T_{\rm rot}/W$ for the formed protoneutron star
is far below the critical value for the onset of the dynamical
instabilities, and the nonaxisymmetric instability does not set in.
Therefore, the assumption of the axial symmetry adopted in this paper
is justified. 

Finally, we mention a direction of our next study.
In this paper, we have clarified a criterion for prompt black hole
formation in the iron core collapse adopting 
the parametric equations of state as a first step. We have shown that the
dynamics of the collapse and the criterion for the black hole formation 
depend sensitively on the equations of state. To obtain more realistic 
outputs that will be in nature, in the next step, 
it is necessary to adopt more realistic equations of state. 
We plan to perform such realistic simulations adopting
a realistic equation of state \cite{Shen} 
in the fully general relativistic framework. 

\vspace{4mm}

\begin{center}
{\bf Acknowledgments}
\end{center}

Numerical computations were performed 
on the FACOM VPP5000 machine in the data processing center of 
National Astronomical Observatory of Japan and NEC SX-6 in the data
processing center of ISAS in JAXA. 
This work is supported by JSPS Research Fellowship for Young
Scientists (No. 1611308) and by Monbukagakusho Grant (Nos. 15037204,
15740142, and 16029202).

\appendix

\section{Comparison with realistic equation of state}\label{compareEOS}
\begin{figure}[htb]
  \vspace{-4mm}
  \begin{center}
    \epsfxsize=3.5in
    \leavevmode
    \epsffile{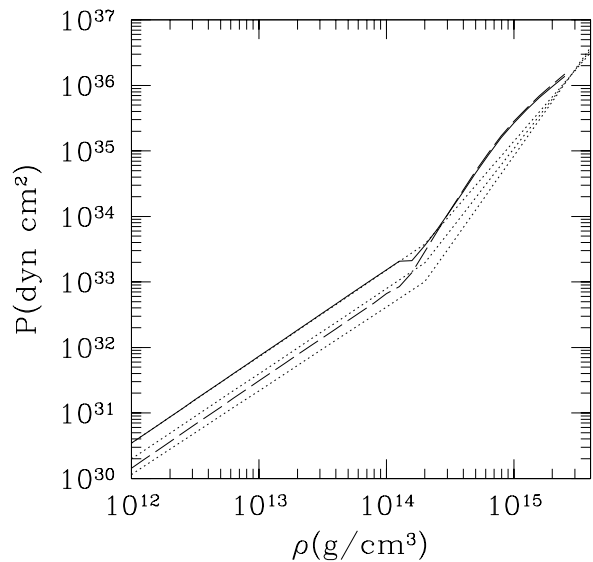} 
  \end{center}
  \vspace{-8mm}
  \caption{The pressure, $P$, as a function of the density, $\rho$,
  for the realistic equation of state at zero
  temperature. The solid curve corresponds to the case $Y_{e} = 0.5$
  and the long dashed to $Y_{e} = 0.2$. The three dotted curves denote
  the cold equations of state with the parameters listed in Table
  \ref{Table1}.}\label{figure27} 
\end{figure}
In this appendix, we compare the parametric equation of state (PEOS)
adopted in this paper (see Sec. \ref{secEOS}) with a realistic
equation of state proposed by Shen et al. \cite{Shen} (REOS). 
In Fig. \ref{figure27}, we show relations between the pressure and the
density of REOS at zero temperature
(adding the degenerate electron pressure)
together with those of PEOS. 
Since REOS depends on $Y_{e}$, we consider
the two values $Y_{e} = 0.5$ and $0.2$.
The $P$-$\rho$ relations of REOS and PEOS shows a good agreement, in
particular, in the subnuclear density. 
In the supranuclear density, on the other hand,
difference between the two equations of state becomes outstanding:
The REOS is stiffer than PEOS (see discussion about $\Gamma_{2}$
below).
This fact leads to the larger maximum mass of the spherical cold
neutron star for REOS. Indeed, the maximum ADM mass is
$\approx 2.2M_{\odot}$ \cite{Shen}, much larger than that for PEOS
($\approx 1.6M_{\odot}$). 

We first consider the validity of the adiabatic index of PEOS at
subnuclear density $\Gamma_{1}$.
The adiabatic index $\Gamma_{1}$ 
includes the effect of the electron capture. 
This implies that it corresponds to an average value 
$\langle \Gamma_{M} \rangle$ of the
'effective' adiabatic index $\Gamma_{M}$ \cite{CB,Monch} which
describes the change of the pressure along a collapse trajectory of a
given mass element: 
\beq
\Gamma_{M} = 
\left. \frac{\partial \log P}{\partial \log \rho} \right|_{S, Y_{e}}
+ \left. \frac{\partial \log P}{\partial Y_{e}} \right|_{\rho, S}
\left. \frac{\delta Y_{e}}{\delta \log \rho} \right|_{M}
+ \left. \frac{\partial \log P}{\partial S} \right|_{\rho, Y_{e}}
\left. \frac{\delta S}{\delta \log \rho} \right|_{M},
\eeq
where $\delta S = 0$ at zero temperature.
The crucial quantity for the dynamical behavior of the core is this
effective adiabatic index $\Gamma_{M}$ \cite{vRL,Monch}. 

On the other hand, the $P$-$\rho$ relations for REOS in
Fig. \ref{figure27} do not include the effect of the electron capture.
Since $Y_{e}$ decreases during the
collapse due to the electron capture, the $P$-$\rho$
relation will shift toward that of the smaller $Y_{e}$.
If the electron capture rate is small, the effective
adiabatic index will be close to $4/3$, while the rate is large, the
index will deviate from $4/3$.

A plausible value of $\langle \Gamma_{M} \rangle$ for REOS in
subnuclear density would be estimated as follows. 
First, previous studies suggest that the initial value of
$Y_{e}$ is $\approx 0.42$ and $-0.11 \alt \delta Y_{e} \alt -0.04$
during the collapse \cite{Bethe,ST}. 
Assuming these values and $\delta \log \rho \approx 4$, the average
value of $(\delta Y_{e}/ \delta \log \rho)|_{M}$ is $-0.01$--$-0.0275$.
On the other hand, the value of $(\partial \log P/\partial
Y_{e})|_{\rho, S}$ of REOS is 
$\approx 1.4$--$1.9$ for $10^{8} \alt \rho \alt
10^{14}$ and $0.31 \alt Y_{e} \alt 0.42$, and its averaged value is
$\approx 1.6$.
Thus, the average value for REOS is 
$1.29 \alt \langle \Gamma_{M} \rangle \alt 1.32$ depending the
electron capture rate.
Therefore, the range $1.28 \le \Gamma_{1} \le 1.32$ adopted in this
paper would be reasonable.

Then, let us consider the adiabatic index $\Gamma_{2}$ at supranuclear
density. 
The adiabatic index of REOS at supranuclear density is larger
($\Gamma \approx 3$) than that of PEOS $(2.25 \le \Gamma_{2} \le
2.75)$. This is due to a relatively larger value of incompressibility
of the REOS: $K_{s} = 281$ MeV \cite{Shen}. 
Although the real value of $K_{s}$ is uncertain at current status, a
recent study \cite{Colo} reported that a plausible range of the 
value of $K_{s}$ is $\approx 220$--$270$ MeV. 
Therefore, the adiabatic index can be much smaller
for a smaller value of $K_{s}$. 
For example, the adiabatic index is $\approx 2.2$ around the nuclear
matter density for a realistic equation of
state by Lattimer and Swesty with $K_{s} = 220$ MeV \cite{Latti}.
Thus, the range $2.25 \le \Gamma_{2} \le 2.75$ adopted in this paper
may be a plausible choice.

In summary, we conclude that the parameter
range adopted in the present paper is not far from predictions and
suggestion of realistic equations of state.

\end{document}